\documentclass[11pt,a4paper]{article}
\usepackage{float}
\usepackage{jheppub} % for details on the use of the package, please
\usepackage{braket,slashed,bm}
\usepackage{array,multirow}
\usepackage[normalem]{ulem}
\usepackage{cancel}
\usepackage[table]{xcolor}
\usepackage[vcentermath]{youngtab}

\usepackage[T1]{fontenc} % if needed

\usepackage{mathrsfs}

\usepackage{booktabs}
\usepackage{adjustbox}
\usepackage{mathtools}

\usepackage{soul}

\usepackage{arydshln}
\usepackage[utf8]{inputenc}

\def\nn{\nonumber\\ }
\def\hyp{\mathsf{y}}

\renewcommand{\O}{\mathcal{O}}
\newcommand{\hc}{\mathrm{h.c.}}

\begin{document}
%%----------------------------------------------------------------------------------------------------------------------------------------------------------------

\title{Dimension-8 Operators in the Standard Model Effective Field Theory}

\author{Christopher W.~Murphy}

\abstract{
We present a complete basis of dimension-8 operators in the Standard Model Effective Field Theory.
Attention is paid to operators that vanish in the absence of flavor structure.
There are dimension-8 SMEFT 44,807 operators.
We also briefly discuss a few aspects of phenomenology involving dimension-8 operators, including light-by-light scattering and electroweak precision data.
}
\maketitle

%%----------------------------------------------------------------------------------------------------------------------------------------------------------------

\section{Introduction}
\label{sec:intro}

The Standard Model (SM) is an extremely successful theory that has been rigorously tested at the Large Hadron Collider (LHC) and elsewhere.
Nevertheless it is widely expected that the SM is only an effective field theory (EFT), valid up to some cutoff scale $\Lambda$.
The Standard Model Effective Field Theory (SMEFT) generalizes the SM by adding a complete, but not over-complete basis of operators at every mass-dimension $d$ rather than stopping at $d=4$.\footnote{The SMEFT assumes there are no light hidden states such as sterile neutrinos or an axion, and that the Higgs boson form part of an $SU(2)_w$ doublet with hypercharge $\hyp = 1/2$. Other types of EFTs are possible where these assumptions are relaxed, but we do not consider them here.} 

The counting and classification of operators in the SMEFT has a long history.
Starting with dimension-5 there is a single type operator~\cite{Weinberg:1979sa}, $N_{\rm type} = 1$, and it violates lepton number.
At dimension-6, Ref.~\cite{Grzadkowski:2010es} classified the 76 baryon number preserving ($B$) Lagrangian terms; see~\cite{Buchmuller:1985jz} for earlier work.
The eight baryon number violating ($\slashed B$) terms were previously known~\cite{Abbott:1980zj}, yielding a total of $N_{\rm term} = 84$.
In terms of actual operators rather than terms in the Lagrangian, the counts explode when flavor structure is allowed.
For three generations of fermions, $n_g = 3$, there are $N_{\rm op} = 2499$ independent $B$ operators~\cite{Alonso:2013hga} and 546 $\slashed B$ operators~\cite{Alonso:2014zka}.
Hilbert series methods were applied to the SMEFT in Refs.~\cite{Lehman:2015via, Henning:2015daa, Lehman:2015coa, Henning:2015alf}, providing an elegant way to count the number of operators for arbitrary dimension $d$.
Computing tools $\mathtt{Sim2Int}$~\cite{Fonseca:2017lem}, $\mathtt{DEFT}$~\cite{Gripaios:2018zrz}, $\mathtt{BasisGen}$~\cite{Criado:2019ugp}, $\mathtt{ECO}$~\cite{Marinissen:2020jmb}, and $\mathtt{GrIP}$~\cite{Banerjee:2020bym} were subsequently developed, allowing for automated counting of operators.

Beyond counting operators, work has been done on their explicit forms as well.
Refs.~\cite{Lehman:2014jma, Liao:2016hru} classified the 18 dimension-7 operators.
So far only partial sets of dimension-8 operators exist in the literature.
This includes, however, all of the bosonic operators (in a basis where the number of derivatives is minimized)~\cite{Morozov:1985ef, Hays:2018zze, Remmen:2019cyz}.
Our goal in this work is to find a complete set of dimension-8 operators.
A subtlety in constructing the dimension-8 operator basis is that some operators vanish in the absence of flavor structure.
Our basis contains 231 types of operators.
For comparison, with $n_g = 1$ there are 993 operators, while for $n_g = 3$ there are instead 44807 operators~\cite{Henning:2015alf}.

Although the counting and classification of operators is certainly interesting in its own right, there is also a wide range of phenomenological implications of dimension-8 operators as well.
For some phenomena dimension-8 is the lowest dimension where the interactions become possible.
Most famous among these processes is light-by-light scattering.
Another area where dimension-8 effects have been studied is electroweak precision data (EWPD) where contributions to the $U$ parameter first arise at dimension-8~\cite{Grinstein:1991cd}.
Formally the dimension-6 operators are the leading terms in the EFT expansion.
However there are various scenarios in which this is not the case practically speaking.
Perhaps the most obvious among these is when the interference between the dimension-6 amplitude and the SM amplitude is suppressed or even vanishes.
Additionally there could be a difference in the experimental precision of the measurements being considered~\cite{Dawson:2017vgm}.
Finally, we comment on the structure the renormalization group evolution (RGE) equations of the dimension-8 operators. 

The rest of the paper is organized as follows.
Section~\ref{sec:note} lays down the notation and conventions we use, including the semantics of number of operators versus number of types of operators.
We then discuss how we performed the operator classification in Section~\ref{sec:cls_ops} with the results given in Sec.~\ref{sec:results}.
We briefly explore light-by-light scattering, EWPD, as well as models involving scalar $SU(2)_w$ quartets where there is interesting interplay between dimension-6 and dimension-8 effects in Section~\ref{sec:pheno}.
Additionally we comment on the renormalization group evolution (RGE) of the dimension-8 operators in Sec.~\ref{sec:rge} before concluding in Sec.~\ref{sec:con}.
For convenience we provide tables of dimension-6 and -7 operators in Appendix~\ref{sec:app}.

%%----------------------------------------------------------------------------------------------------------------------------------------------------------------

\section{Notation and Conventions}
\label{sec:note}

We start by considering the various uses of the word operator.
See Ref.~\cite{Fonseca:2019yya} for further discussion.
We a define operator to be a gauge and Lorentz invariant contraction of fields and derivatives with specific flavor indices.
A Lagrangian term, or just term or short, collects all of operators with the same gauge and Lorentz structure into a single unit, \textit{i.e.} a term collapses the flavor indices of otherwise identical operators.
By construction, a Lagrangian term of mass dimension $d \leq 8$ may contain no more than $n_g^4$ operators.\footnote{Starting at dimension-9 $n_g^6$ is possible.}
Ref.~\cite{Fonseca:2019yya} defines a type of operator as the collection of terms with the combination of fields (and derivatives) with conjugate counted separately.
In this work we use a broader definition of a type of operator where the conjugate fields are counted in unison with the un-conjugated fields.
Our types of operators are therefore supersets of those in~\cite{Fonseca:2019yya}, of which there are 541 to our 231.
This definition of a type of operator allows us systematically label the operators in a  phenomenologically friendly way.
The largest set of operators we consider is a class where the operators are grouped by the number of fields of a given spin as well as the number of derivatives.
It is useful to consider subclasses when discussing the RGE of the dimension-8 operators. 
Subclasses treat conjugate fields separately.
For example, class 1 has three subclasses, $\{X_L^4, X_L^2 X_R^2, X_R^4\} \in X^4$, and class 18 also has three subclasses, $\{\psi^4 H^2, \psi^2 \bar \psi^2 H^2, \bar \psi^4 H^2\} \in \psi^4 H^2$.
We tolerate a slight abuse of notation between classes and subclasses relying on context to distinguish which set is being discussed.

Moving onto physics conventions, the SM Lagrangian is given by
\begin{align}
\label{eq:sm}
\mathcal L_{\rm SM} &= - \frac{1}{4} \sum_X X_{\mu\nu} X^{\mu\nu} + (D_\mu H^\dag) (D^\mu H) + \sum_\psi \bar \psi i \slashed D \psi \\
&- \lambda \left(H^\dag H - \frac{v^2}{2}\right)^2 - \left[H^\dag_j \bar d\, Y_d q^j + \widetilde H^\dag_j \bar u\, Y_u q^j + H^\dag_j \bar e\, Y_e l^j + \hc \right] . \nonumber
\end{align}
In Eq~\eqref{eq:sm}, and throughout this work, we generically refer to field strengths as $X = \{G^A, W^I, B\}$, and to fermions as $\psi = \{l, e, q, u, d\}$.

The gauge covariant derivative is
\begin{equation}
(D_\mu q)^{j \alpha} = ((\partial_\mu + i g_1 \hyp B_\mu) \delta_\beta^\alpha \delta_k^j  + i g_2 (t^I)_k^j W^I_\mu \delta_\beta^\alpha + i g_3 (T^A)_\beta^\alpha A^A_\mu \delta_k^j) q^{k \beta} ,
\end{equation}
where the generators of $SU(3)_c$ and $SU(2)_w$ are $T^a$ and $t^I = \tau^I / 2$, respectively.
The $U(1)_y$ hypercharge is given by $\hyp$ with $Q = \tau^3 + \hyp$.
For $SU(3)_c$ fundamental and adjoint indices are denoted $\alpha, \beta, \gamma$ and $A, B, C$, respectively, while for $SU(2)_w$ the fundamental and adjoint indices are respectively labeled $j, k, m$ and $I, J, K$.

Anti-symmetrization of indices is denoted by a pair of square brackets, $[\mu\nu]$, and symmetrization is denoted by a pair of round brackets, $(\mu\nu)$.
The definition of $\widetilde H$ is
\begin{equation}
\widetilde H_j = \epsilon_{jk} H^{\dag k}
\end{equation}
where $\epsilon_{jk} = \epsilon_{[jk]}$ is the $SU(2)$ invariant tensor with $\epsilon_{12} = 1$.
The dual field strength is defined as
\begin{equation}
\widetilde X_{\mu\nu} = \tfrac{1}{2} \epsilon_{\mu\nu\rho\sigma} X^{\rho\sigma}
\end{equation}
with $\epsilon_{0123} = 1$. 

We will sometimes refer to the following combinations of field strength as they are typically what are used when counting coefficients,
\begin{equation}
X_{L,R}^{\mu\nu} = \tfrac{1}{2} (X^{\mu\nu} \mp i \widetilde X^{\mu\nu}) .
\end{equation}
These field strengths have simple Lorentz transformation properties, $X_L \sim (1, 0)$, $X_R \sim (0, 1)$ under $SU(2)_L \otimes SU(2)_R$.
Similarly $l$ and $q$ are left-handed fermion fields, whereas $e$, $u$, and $d$ are right-handed fields.
When necessary Lorentz indices in the fundamental representations are indicated by $a, b, \dot a, \dot b$, \textit{e.g.} $q_L \sim (q_L)_a$, $B_R \sim (B_R)_{(\dot a \dot b)}$.

The SMEFT extends the SM by adding all of the higher-dimensional operator that are gauge invariant under the SM with the caveat that redundant operators should not be included
\begin{equation}
\mathcal L_{\rm SMEFT} = \mathcal L_{\rm SM} + \sum_{d > 4} \mathcal{L}^{(d)} .
\end{equation}
For the dimension-6 operators we keep notation that has been well-established in the literature, see \textit{e.g.}~\cite{Alonso:2013hga}.
On the other hand, we use a systematic, if at times cumbersome, notation for labelling the operators of mass-dimension 7 and above.
For types of operators with a single Lagrangian term we label them as follows
\begin{equation}
\mathcal{L}^{(d)} \supset \sum_{\rm type} C_{\rm type} Q_{\rm type} , \quad n_{\rm term} = 1,
\end{equation}
where $n_{\rm term}$ is the number of terms of a given type.
The type of operator is denoted as the fields and the derivatives in the operator raised to the power of the number of times that type of object appears in the operator, \textit{e.g} the label $leBH^3$ indicates that this term has one left-handed lepton field, one right-handed electron field, one hypercharge field strength, and three Higgs fields.
If a type of operator has multiple terms, not counting Hermitian conjugates, we instead label the operators as
\begin{equation}
\mathcal{L}^{(d)} \supset \sum_{\rm type} \sum_{i = 1}^{n_{\rm term}} C_{\rm type}^{(i)} Q_{\rm type}^{(i)} , \quad n_{\rm term} > 1 .
\end{equation}
Consider as an explicit example,
\begin{equation}
\label{eq:notex}
\mathcal{L}^{(d=8)} \supset C_{\substack{u^2GH^2D \\ pr}}^{(1)} Q_{\substack{u^2GH^2D \\ pr}}^{(1)} + C_{\substack{u^2GH^2D \\ pr}}^{(2)} Q_{\substack{u^2GH^2D\\ pr}}^{(2)} + \left[C_{\substack{leBH^3 \\ pr}} Q_{\substack{leBH^3 \\ pr}} + \hc\right] ,
\end{equation}
with
\begin{align}
Q_{\substack{u^2GH^2D \\ pr}}^{(1)} &= (\bar u_p \gamma^\nu T^A u_r) D^\mu (H^\dag H) G^A_{\mu\nu} , \nn
Q_{\substack{u^2GH^2D\\ pr}}^{(2)} &= (\bar u_p \gamma^\nu T^A u_r) D^\mu (H^\dag H) \widetilde G^A_{\mu\nu} , \nn
Q_{\substack{leBH^3 \\ pr}} &= (\bar l_p \sigma^{\mu\nu} e_r) H (H^\dag H) B_{\mu\nu} .
\end{align}
Flavor indices explicitly appear in Eq.~\eqref{eq:notex}.
Fermion fields have a flavor index $p, r, s, t$ that runs over 1, 2, 3 for three generations.
The fermion fields themselves are in the weak eigenstate basis.
The Yukawa matrices, $Y_{e, u, d}$, in Eq.~\eqref{eq:sm} are matrices in flavor space.

Note that we do not explicitly label the transpose of a spinor in fermion bilinears involving a charge conjugation operator, \textit{e.g.} $\psi_1 C \psi_2 \equiv \psi_1^T C \psi_2$, $\psi_1 C \sigma_{\mu\nu} \psi_2 \equiv \psi_1^\top C \sigma_{\mu\nu} \psi_2$.
Finally, it is convenient to define Hermitian derivatives \textit{e.g.}
\begin{align}
i H^\dag \overleftrightarrow{D}_\mu H &= i H^\dag (D_\mu H) - i (D_\mu H^\dag) H , \nn
i H^\dag \overleftrightarrow{D}^I_\mu H &= i H^\dag \tau^I (D_\mu H) - i (D_\mu H^\dag) \tau^I H .
\end{align}

%%----------------------------------------------------------------------------------------------------------------------------------------------------------------

\section{Operator Classification}
\label{sec:cls_ops}

Having defined our notation in Sec.~\ref{sec:note} we can more precisely state our goal.
We are trying to find the minimum number of Lagrangian terms needed to give all of the operators at dimension-8 subject to the constraint that no term may contain more than $n_g^4$ operators.
For bosonic and two-fermion operators this constraint is trivially satisfied as those terms always contain one and $n_g^2$ operators, respectively.
Four-fermion operators where two or more of the fields are identical constitute the interesting cases.

We use existing results from the literature when they are available.
All of the bosonic operators have been classified previously~\cite{Morozov:1985ef, Hays:2018zze, Remmen:2019cyz}.
Ref.~\cite{Hays:2018zze} also gave partial results for three of the two-fermion classes that were sufficient to allow us to deduce the remaining operators in those classes.
As this work was being finalized Refs.~\cite{Alioli:2020kez, Remmen:2020vts} appeared, which classified a subset of four-fermion operators with two derivatives.
However there are still non-trivial results for us to work out in that class.

When classifying the dimension-8 operators we exploit the fact that not only are the types of operators known, but the number of operators is also known, see Ref.~\cite{Henning:2015alf}.
In particular, we leverage the Python package $\mathtt{BasisGen}$~\cite{Criado:2019ugp}, which we use to get the number of operators for each type of operator.
Additionally, we use the Mathematica program $\mathtt{Sym2Int}$~\cite{Fonseca:2017lem}, which not only gives the number of operators per type, but also the flavor representations when there are identical particles in the operator.
Furthermore, $\mathtt{Sym2Int}$ gives the number of Lagrangian terms per type of operator except when the operator contains both derivatives and identical particles. 
In that case a range is given because the permutation symmetry of operators with derivatives is ambiguous due to integration by parts (IBP) redundancies.

Beyond getting the number of terms correct we need to ensure that the operators in our basis are independent.
Operators with derivatives can be related through integration by parts.
When there are multiple derivatives care must be taken to select operators for the basis that span the entire space of possible operators for that class.
See the discussion of class 16 below for an example of this.
Operators can also be related to each other through the equations of motion (EOM).
We use the EOM the remove redundant terms, trading them for basis operators in the same class, operators with fewer derivatives, and sometimes operators of lower mass dimension.
See the discussion of class 17 below for an example of this, and see \textit{e.g.}~\cite{Jenkins:2013zja} for the SM equations of motion.
Our basis does not explicitly contain an EOM.
Operators with derivatives can be IBP, and some of the resulting terms contain an EOM.
However it is never the case that all of the resulting terms have an EOM.
Lastly, there are various tensor and spinor identities that relate operators to each other.
There are the Fierz identities, for example for $SU(2)$
\begin{equation}
\label{eq:fierz}
(\tau^I)_j^k (\tau^I)_m^n = 2 \delta_j^n \delta_m^k - \delta_j^k \delta_m^n .
\end{equation}
There are identities involving the Levi-Civita symbol, \textit{e.g.} in two-dimensions
\begin{equation}
\label{eq:lc}
\epsilon_{jk} \epsilon_{mn} + \epsilon_{jm} \epsilon_{nk} + \epsilon_{jn} \epsilon_{km} = 0
\end{equation}
There are identities for products of Dirac matrices, \textit{e.g.} the anti-symmetric Dirac tensor is self-dual
\begin{equation}
\label{eq:dsd}
\epsilon^{\rho\tau\mu\nu} \sigma_{\mu\nu} P_R = 2 i \sigma^{\rho\tau} P_R .
\end{equation}

%--------------------------------------------------------------------------------------------------------------
\subsection{Bosonic Operators}

\begin{enumerate}
%--------------------------------------------------------------------------------------------------------------
\item $X^4$ \\
The $X^4$ operators for a single Yang-Mills field were classified in Ref.~\cite{Morozov:1985ef}.
Ref.~\cite{Remmen:2019cyz} generalized this result to the SM field content.
Note that dimension-8 is the lowest dimension where a subclass of operators contains both $X_L$ and $X_R$.

%--------------------------------------------------------------------------------------------------------------
\item $H^8$ \\
$(H^\dag H)^4$ is the only possibility.

%--------------------------------------------------------------------------------------------------------------
\item $H^6D^2$ \\
The $H^6D^2$ operators were classified in Ref.~\cite{Hays:2018zze}.

%--------------------------------------------------------------------------------------------------------------
\item $H^4D^4$ \\
Both Refs.~\cite{Hays:2018zze} and~\cite{Remmen:2019cyz} classified the $H^4D^4$ operators.

%--------------------------------------------------------------------------------------------------------------
\item $X^3H^2$ \\
The $X^3H^2$ operators were classified in Ref.~\cite{Hays:2018zze}.
Note that the two terms in $Q_{W^2BH^2}^{(2)}$ are equivalent via the identity
\begin{equation}
\widetilde X_{\mu\rho} Y^{\rho\nu} = - X^{\nu\rho} \widetilde Y_{\rho\mu} - \frac{1}{2} X_{\alpha\beta} \widetilde Y^{\alpha\beta} \delta_\mu^\nu .
\end{equation}
For backwards compatibility we construct our basis with $Q_{W^2BH^2}^{(2)}$ as originally defined by Ref.~\cite{Hays:2018zze} as opposed to, say, only keeping the second term.

%--------------------------------------------------------------------------------------------------------------
\item $X^2H^4$ \\
The $X^2H^4$ operators were classified in Ref.~\cite{Hays:2018zze}.

%--------------------------------------------------------------------------------------------------------------
\item $X^2H^2D^2$ \\
Both Refs.~\cite{Hays:2018zze} and~\cite{Remmen:2019cyz} classified the $X^2H^2D^2$ operators.

%--------------------------------------------------------------------------------------------------------------
\item $XH^4D^2$ \\
The $XH^4D^2$ operators were classified in Ref.~\cite{Hays:2018zze}.

%--------------------------------------------------------------------------------------------------------------
\subsection{Two-Fermion Operators}

%--------------------------------------------------------------------------------------------------------------
\item $\psi^2X^2H$ \\
For the dimension-8 class $\psi^2X^2H$, 24 terms arise from joining a field strength to a dimension-6 operator of the form $\psi^2XH$, whereas 48 terms come from the product of two field strengths and a Yukawa interaction, $(X^2) (\psi^2H)$.
See Table~\ref{tab:smeft6ops} for the dimension-6 operators.

%--------------------------------------------------------------------------------------------------------------
\item $\psi^2XH^3$ \\
In the class $\psi^2XH^3$, 16 of the 22 terms are identical to the dimension-6 terms $\psi^2XH$ up to an extra factor of $(H^{\dag} H)$.
The remaining six terms, all involving $W^I_{\mu\nu}$, instead have the dimension-2 covariant $(H^{\dag} \tau^I H)$.

%--------------------------------------------------------------------------------------------------------------
\item $\psi^2H^2D^3$ \\
Ref.~\cite{Hays:2018zze} classified the four terms involving $q^2H^2D^3$.
The remaining 12 terms in the class can be deduced from the results of Ref.~\cite{Hays:2018zze}.

%--------------------------------------------------------------------------------------------------------------
\item $\psi^2H^5$ \\
The class $\psi^2H^5$ is identical to the dimension-6 class $\psi^2H^3$ up to an extra factor of $(H^{\dag} H)$.

%--------------------------------------------------------------------------------------------------------------
\item $\psi^2H^4D$ \\
Ref.~\cite{Hays:2018zze} classified the four operators involving $q^2H^4D$.
The term $Q_{q^2H^4D}^{(2)}$ contains a sum
\begin{equation}
\label{eq:q2H4D}
Q_{q^2H^4D}^{(2)} = Q_{q^2H^4D}^{(2l)} + Q_{q^2H^4D}^{(2r)} .
\end{equation}

The term $Q_{q^2H^4D}^{(3)}$ is related to the righthand side of Eq.~\eqref{eq:q2H4D} as follows
\begin{equation}
\label{eq:Q3}
i Q_{q^2H^4D}^{(3)} = - Q_{q^2H^4D}^{(2l)} + Q_{q^2H^4D}^{(2r)} .
\end{equation}
This can be seen using the following variation of Eq.~\eqref{eq:fierz}
\begin{equation}
\label{eq:fierz_ep}
\delta_j^k (\tau^I)_m^n - (\tau^I)_j^k \delta_m^n = i \epsilon^{IJK} (\tau^J)_j^n (\tau^K)_m^k .
\end{equation}
As was the case with the class 5 operator $Q_{W^2BH^2}^{(2)}$ we choose to keep $Q_{q^2H^4D}^{(2)}$ and $Q_{q^2H^4D}^{(3)}$ is our basis as opposed to their summands for backwards compatibility.
The remaining nine terms in the class can be deduced from the results of Ref.~\cite{Hays:2018zze}.

%--------------------------------------------------------------------------------------------------------------
\item $\psi^2X^2D$ \\
We use integration by parts to place the derivative on a fermion field.
Then in order for the operator to not be ``reduced'' to a class with fewer derivatives through the use of the equations of motion the fermionic component of the operator must not be Lorentz invariant.
Class 7 also contains two field strengths, see above, and a subset of the operators in class 7 have covariants formed from Higgs fields and derivatives that transform as $(1, 1)$ under $SU(2)_L \otimes SU(2)_R$.
We take the field strength components of that subset of class 7 operators and use them for the class 14 operators, contracting them with fermionic covariants of the form $\bar \psi \gamma^\mu D^\nu \psi$.
In particular, we use $Q_{G^2H^2D^2}^{(1)}$ as the template for when a fermion is not charged under a gauge group, $Q_{W^2H^2D^2}^{(1, 4-6)}$ when it is charged under a gauge group, and $Q_{WBH^2D^2}^{(1, 4-6)}$ when it is charged under two gauge groups.
Three terms are not covered by this procedure.
They involve quarks and two gluon field strength where the $SU(3)_c$ adjoint indices are contracted with the symmetric $d^{ABC}$ symbol.

%--------------------------------------------------------------------------------------------------------------
\item $\psi^2XH^2D$ \\
Ref.~\cite{Hays:2018zze} classified the 12 terms involving $q^2WH^2D$.
Of the remaining 74 terms, 68 of them have a form analogous to those classified by Ref.~\cite{Hays:2018zze}.
The final six terms in the class are instead analogous to the dimension-6 operator $Q_{Hud}$ with the addition of a field strength.

%--------------------------------------------------------------------------------------------------------------
\item $\psi^2XHD^2$ \\
Things become more complicated when there are two or more derivatives in the operator.
As such it is useful to introduce some additional machinery to classify the operators.
We use the procedure given in Ref.~\cite{Lehman:2015via} for removing terms that are reducible through the use of the equations of motion.
In a nutshell, the procedure says Lorentz indices should be symmetrized for representations that are triplets or higher under either $SU(2)_L$ or $SU(2)_R$.
In terms of Lorentz indices we have for example
\begin{equation}
\label{eq:eom}
D \psi_L \sim (D \psi_L)_{(a b), \dot a}, \quad D X_R \sim (D X_R)_{a, (\dot a \dot b \dot c)}, \quad D^2 H \sim (D^2 H)_{(a b), (\dot a \dot b)} .
\end{equation} 

We now work through a representative example with field content $\bar l, e, H, B_L$ following the procedure laid out in Ref.~\cite{Hays:2018zze}.
Using relations like those in~\eqref{eq:eom} we see that operators with a derivative acting on the field strength or two derivatives acting on a fermion can be reduced using the EOM. Ignoring for the time being constraints from IBP this leaves us with four possibilities
\begin{align}
\label{eq:cand}
x_1 &= (D \bar l)_{a, (\dot a \dot c)} e_{\dot d} (D H)_{b, \dot b} B_{(c d)} \epsilon^{a c} \epsilon^{b d} \epsilon^{\dot a \dot d} \epsilon^{\dot c \dot b}, \nn
x_2 &= \bar l_{\dot c} (D e)_{a, (\dot a \dot d)} (D H)_{b, \dot b} B_{(c d)} \epsilon^{a c} \epsilon^{b d} \epsilon^{\dot a \dot c} \epsilon^{\dot d \dot b}, \nn
x_3 &= (D \bar l)_{a, (\dot a \dot c)} (D e)_{b, (\dot b \dot d)} H B_{(c d)} \epsilon^{a c} \epsilon^{b d} \epsilon^{\dot a \dot b} \epsilon^{\dot c \dot d}, \nn
x_4 &= \bar l_{\dot c} e_{\dot d} (D^2 H)_{(a b), (\dot a \dot b)} B_{(c d)} \epsilon^{a c} \epsilon^{b d} \epsilon^{\dot a \dot c} \epsilon^{\dot b \dot d},
\end{align} 
where we have not shown the $SU(2)_w$ contraction as it is trivial.

To determine redundancies coming from integration by parts we need operators transforming as $(\tfrac{1}{2}, \tfrac{1}{2})$ under the Lorentz group with one fewer derivative than the operators of interest.
There are three possibilities in this example
\begin{align}
\label{eq:ibp}
y_1 &= (D \bar l)_{a, (\dot a \dot c)} e_{\dot d} H B_{(c d)} \epsilon^{a c} \epsilon^{\dot a \dot d}, \nn
y_2 &= \bar l_{\dot c} (D e)_{a, (\dot a \dot d)} H B_{(c d)} \epsilon^{a c} \epsilon^{\dot a \dot c}, \nn
y_3 &= \bar l_{\dot c} e_{\dot d} (D H)_{a, \dot a} B_{(c d)} \tfrac{1}{2} \epsilon^{a c} (\epsilon^{\dot a \dot c} + \epsilon^{\dot a \dot d}) . 
\end{align}
The derivatives of the $y_i$ show which of the $x_i$ are related by IBP
\begin{align}
\label{eq:const}
D_{b, \dot b}\, y_1 &= x_1 + x_3 = 0 ,\nn
D_{b, \dot b}\, y_2 &= x_2 + x_3 = 0 ,\nn
D_{b, \dot b}\, y_3 &= \tfrac{1}{2} x_1 + \tfrac{1}{2} x_2 + x_4 = 0 ,
\end{align}
where the appropriate contraction of the remaining Lorentz indices in the leftmost terms is understood.
By inspection of~\eqref{eq:const} we see that any of the four candidate operator can be transformed into any of the remaining three through the use of IBP.

We repeat the same procedure, omitting the details here, to find the class 17 operators with field content $\bar l, e, H, B_R$.
In this case there are eight candidate operators, $x_i$, and six operators in the Lorentz four-vector representation, $y_i$.
Given the larger number of operators in this case we use Mathematica to solve the system of constraints, yielding two operators.

We are left with a total of three $Q_{leHBD^2}$ terms ($+ \hc$)
\begin{align}
&\epsilon^{a c} \epsilon^{b d} \epsilon^{\dot a \dot b} \epsilon^{\dot c \dot d} \bar l_{\dot a} (D e)_{a, (\dot b \dot c)} (D H)_{b, \dot d} (B_L)_{c d} , \nn
&\epsilon^{ab} \epsilon^{\dot a \dot b} \epsilon^{\dot c \dot e} \epsilon^{\dot d \dot f} \bar l_{\dot a} (D e)_{a, (\dot b \dot c)} (D H)_{b, \dot d} (B_R)_{(\dot e \dot f)} , \nn
&\epsilon^{ab} \epsilon^{\dot a \dot d} \epsilon^{\dot b \dot e} \epsilon^{\dot c \dot f} \bar l_{\dot a} e_{\dot b} (D H)_{a, \dot c} (D B_R)_{b, (\dot d \dot e \dot f)} ,
\end{align}
which matches the counting we found using $\mathtt{BasisGen}$.
Translating the Lorentz contractions from $SU(2)_L \otimes SU(2)_R$ to $SO(3, 1)$ and translating $B_L$ and $B_R$ to $B$ and $\widetilde B$ we find
\begin{align}
\label{eq:lebhd2}
Q_{leBHD^2}^{(1)} &= (\bar l_p \sigma^{\mu\nu} D^\rho e_r) (D_\nu H) B_{\rho\mu} \nn
Q_{leBHD^2}^{(2)} &= (\bar l_p D^\rho e_r) (D^\nu H) \widetilde B_{\rho\nu} \nn
Q_{leBHD^2}^{(3)} &= (\bar l_p \sigma^{\mu\nu} e_r) (D^\rho H) (D_\rho B_{\mu\nu}) .
\end{align}
The remaining 42 terms in class 17 can be deduced from the $Q_{leHBD^2}$ operators in~\eqref{eq:lebhd2} $+ \hc$.

%--------------------------------------------------------------------------------------------------------------
\item $\psi^2H^3D^2$ \\
Starting with relations like those in~\eqref{eq:eom} we see that any operator in this class with two derivatives acting on the same field can be reduced using the EOM.
Therefore an operator with one derivative on each of the two fermion fields can be traded for an operator with one derivative on a Higgs field, one derivative on a fermion plus operators with fewer derivatives.
From here we extend the results of Ref.~\cite{Grzadkowski:2010es} to move any remaining derivatives acting on fermions onto Higgs fields, again plus operators with fewer derivatives.
The first relation is
\begin{align}
\label{eq:ten}
H^\dag H (D_\mu H) \bar \psi \sigma^{\mu\nu} D_\nu \psi &= \tfrac{i}{2} H^\dag H (D_\mu H) \bar \psi (\gamma^\mu \slashed D - \slashed D \gamma^\mu) \psi \nn
&= i H^\dag H (D_\mu H) \bar \psi \gamma^\mu \slashed D \psi - i H^\dag H (D_\mu H) \bar \psi D^\mu \psi \nn
&= - i H^\dag H (D_\mu H) \bar \psi D^\mu \psi + \fbox{$\psi^2H^4D$} + \fbox{$E$} ,
\end{align}
where \fbox{$E$} represents operators that vanish via the EOM.
The other relation we need is
\begin{align}
2 H^\dag H (D_\mu H) \bar \psi D^\mu \psi &= H^\dag H (D_\mu H) \bar \psi (\gamma^\mu \slashed D + \slashed D \gamma^\mu) \psi \nn
&= \left(H^\dag H (D_\mu H) \bar \psi \gamma^\mu \slashed D \psi - H^\dag H (D_\mu H) (\slashed D \bar \psi) \gamma^\mu \psi \right. \nn
&\left. - D_\nu [H^\dag H (D_\mu H)] \bar \psi \gamma^\nu \gamma^\mu \psi + \fbox{$T$}\right) \nn
&= - D_\nu [H^\dag H (D_\mu H)] \bar \psi \gamma^\nu \gamma^\mu \psi + \fbox{$E$} + \fbox{$T$}
\end{align}
where \fbox{$T$} stands for a total derivative.

After all this we are left with six terms $+ \hc$ where the derivatives act only on Higgs fields.
In particular, having already established that operators in this class cannot have two derivatives acting on the same field, the derivatives can either act on $H$ and $H^\dag$ or there can be one derivative on each of the $H$ fields.
For each of these cases the fermion pair can either be in the $(0, 0)$ or $(0, 1)$ representation of the Lorentz group.
(The classification for the Hermitian conjugate operators proceeds in an identical fashion.)
Finally when the derivatives act on $H$ and $H^\dag$ the covariants can either be $SU(2)_w$ singlets or adjoints.
The same logic applies for all three choices for the pair of fermions.

%--------------------------------------------------------------------------------------------------------------
\subsection{Four-Fermion Operators}

%--------------------------------------------------------------------------------------------------------------
\item $\psi^4H^2$ \\
All 38 of the $\psi^4$ dimension-6 terms can be multiplied by $(H^{\dag} H)$.
Focusing on $B$ preserving operators, an additional 23 terms are formed by inserting a $\tau^I$ into a $\psi^4$ operator (with at least two left-handed fermions) and joining it to the dimension-2 covariant $(H^{\dag} \tau^I H)$.
Operators of the type $Q_{l^2q^2H^2}$ provide three of the these terms, whereas all other types of operators provide one term $(+ \hc)$. 
There are also 22 $\psi^4H^2$ terms that, schematically, are products of Yukawa interactions, $(\bar L R H) (\bar L R H)$ or $(\bar L R H) (H^\dag \bar R L)$.
Among these there is one new type of operator, $Q_{l^2udH^2}$.
Bi-Yukawa terms with four fields that are fundamentals under either $SU(2)_w$ or $SU(3)_c$ where two of those fields are identical can be contracted either as singlets or adjoints.
There are some redundant operators involving identical left-handed fermions.
For example, the terms
\begin{align}
\label{eq:4lredun}
Q_{l^4H^2}^{(3)} &= (\bar l_p \gamma^\mu \tau^I l_r) (\bar l_s \gamma_\mu \tau^I l_t) (H^\dag H) , \nn
Q_{l^4H^2}^{(4)} &= (\bar l_p \gamma^\mu \tau^I l_r) (\bar l_s \gamma_\mu l_t) (H^\dag \tau^I H) , \nn
Q_{l^4H^2}^{(5)} &= \epsilon^{IJK} (\bar l_p \gamma^\mu \tau^I l_r) (\bar l_s \gamma_\mu \tau^J \tau^I l_t) (H^\dag \tau^K H) ,
\end{align}
are related to the operators in our basis 
\begin{align}
\label{eq:l5}
Q_{\substack{l^4H^2 \\ prst}}^{(3)} &= 2 Q_{\substack{l^4H^2 \\ ptsr}}^{(1)} - Q_{\substack{l^4H^2 \\ prst}}^{(1)} , \nn
Q_{\substack{l^4H^2 \\ prst}}^{(4)} &= Q_{\substack{l^4H^2 \\ stpr}}^{(2)}, \nn
i Q_{\substack{l^4H^2 \\ prst}}^{(5)} &= Q_{\substack{l^4H^2 \\ ptsr}}^{(2)} - Q_{\substack{l^4H^2 \\ srpt}}^{(2)} .
\end{align}
This can be seen using Eqs.~\eqref{eq:fierz} and~\eqref{eq:fierz_ep}.
On the other hand, the four-quark terms 
\begin{align}
Q_{q^4H^2}^{(3)} &= (\bar q_p \gamma^\mu \tau^I q_r) (\bar q_s \gamma_\mu \tau^I q_t) (H^\dag H) , \nn
Q_{q^4H^2}^{(5)} &= \epsilon^{IJK} (\bar q_p \gamma^\mu \tau^I q_r) (\bar q_s \gamma_\mu \tau^J q_t) (H^\dag \tau^K H) ,
\end{align}
whose four-lepton analogs can be found in~\eqref{eq:4lredun}, are not redundant due to the non-trivial color representations of the quarks.
Therefore these terms must be retained in our basis.
The flavor structures of the four-quark, two-Higgs operators are
\begin{align}
\label{eq:Qflav}
Q_{\substack{q^4H^2 \\ prst}}^{(1-3)} &\sim \left(\young(ps) \otimes \young(rt)\right) \oplus \left(\young(p,s) \otimes \young(r,t)\right) , \nn
Q_{\substack{q^4H^2 \\ prst}}^{(4, 5)} &\sim \left(\young(ps) \otimes \young(r,t)\right) \oplus \left(\young(p,s) \otimes \young(rt)\right) .
\end{align}
In the last line of~\eqref{eq:Qflav} we define a particular combination of operators
\begin{equation}
Q_{\substack{q^4H^2 \\ prst}}^{(4)} = Q_{\substack{q^4H^2 \\ prst}}^{(1)} + Q_{\substack{q^4H^2 \\ prst}}^{(2)} + Q_{\substack{q^4H^2 \\ stpr}}^{(2)} + Q_{\substack{q^4H^2 \\ prst}}^{(3)} .
\end{equation}
The operators $Q_{q^4H^2}^{(4, 5)}$ vanish when there is only one generation of quarks.

The terms for baryon number violating operators with dimension-6 analogs follow the logic laid out for the $B$ operators.
However there are two interesting flavor structures.
The dimension-6 operator $Q_{qque}$ is symmetric in its $q$ flavor indices~\cite{Abbott:1980zj}.
This constraint is broken by the additional $SU(2)_w$ in $Q_{eq^2uH^2}$, giving it full flavor rank, $n_g^4$.
The dimension-6 operator $Q_{qqql}$ also has a flavor constraint~\cite{Abbott:1980zj}
\begin{equation}
\label{eq:qqql}
Q_{\substack{qqql \\ prst}} + Q_{\substack{qqql \\ rpst}} = Q_{\substack{qqql \\ sprt}} + Q_{\substack{qqql \\ srpt}},
\end{equation}
which can be derived using Eq.~\eqref{eq:lc}.
The operators $Q_{lq^3H^2}^{(1, 2)}$ respect this constraint, leading to each of their Lagrangian terms $(+ \hc)$ containing $\tfrac{2}{3} n_g^2 (2 n_g^2 + 1)$ operators.
On the other hand, $Q_{lq^3H^2}^{(3)}$ has mixed symmetry.
It is symmetric in $p$ and $r$ and antisymmetric in $r$ and $s$, which causes it to vanish when there is only when generation of fermions.
As a result of the six fundamental $SU(2)_w$ indices there are eight redundant operators
\begin{align}
\label{eq:lq3h2}
Q_{lq^3H^2}^{(1a)} &= \epsilon_{\alpha\beta\gamma} \epsilon_{mj} \epsilon_{kn} (q_p^{m\alpha} C q_r^{j\beta}) (q_s^{k\gamma} C l_t^n) (H^\dag H) , \nn
Q_{lq^3H^2}^{(1b)} &= \epsilon_{\alpha\beta\gamma} (\tau^I \epsilon)_{mj} (\tau^I \epsilon)_{kn} (q_p^{m\alpha} C q_r^{j\beta}) (q_s^{k\gamma} C l_t^n) (H^\dag H) , \nn 
Q_{lq^3H^2}^{(2a)} &= \epsilon_{\alpha\beta\gamma} \epsilon_{mj} (\tau^I \epsilon)_{kn} (q_p^{m\alpha} C q_r^{j\beta}) (q_s^{k\gamma} C l_t^n) (H^\dag \tau^I H) , \nn
Q_{lq^3H^2}^{(2b)} &= \epsilon_{\alpha\beta\gamma} (\tau^I \epsilon)_{jn} \epsilon_{km}  (q_p^{m\alpha} C q_r^{j\beta}) (q_s^{k\gamma} C l_t^n) (H^\dag \tau^I H) , \nn
Q_{lq^3H^2}^{(3a)} &= \epsilon_{\alpha\beta\gamma} (\tau^I \epsilon)_{mj} \epsilon_{kn} (q_p^{m\alpha} C q_r^{j\beta}) (q_s^{k\gamma} C l_t^n) (H^\dag \tau^I H) , \nn
Q_{lq^3H^2}^{(3b)} &= \epsilon_{\alpha\beta\gamma} \epsilon_{jn} (\tau^I \epsilon)_{km} (q_p^{m\alpha} C q_r^{j\beta}) (q_s^{k\gamma} C l_t^n) (H^\dag \tau^I H) , \nn
Q_{lq^3H^2}^{(4a)} &= \epsilon_{\alpha\beta\gamma} \epsilon^{IJK} (\tau^I \epsilon)_{mn} (\tau^J \epsilon)_{jk} (q_p^{m\alpha} C q_r^{j\beta}) (q_s^{k\gamma} C l_t^n) (H^\dag \tau^K H) , \nn
Q_{lq^3H^2}^{(4b)} &= \epsilon_{\alpha\beta\gamma} \epsilon^{IJK} (\tau^I \epsilon)_{mj} (\tau^J \epsilon)_{kn} (q_p^{m\alpha} C q_r^{j\beta}) (q_s^{k\gamma} C l_t^n) (H^\dag \tau^K H) .
\end{align}
The operators in~\eqref{eq:lq3h2} can be written in terms of the operators in our basis using Eq.~\eqref{eq:fierz}, Eq.~\eqref{eq:fierz_ep}, and the following relations obtained from Eq.~\eqref{eq:fierz}
\begin{align}
\epsilon_{mj} (\tau^I \epsilon)_{kn} + (\tau^I \epsilon)_{mj} \epsilon_{kn} &= \epsilon_{mn} (\tau^I \epsilon)_{jk} - (\tau^I \epsilon)_{mn} \epsilon_{jk} , \nn
i \epsilon^{IJK} [(\tau^J \epsilon)_{mn} (\tau^K \epsilon)_{jk} - \tfrac{1}{2} (\tau^J \epsilon)_{mj} (\tau^K \epsilon)_{kn}] &= \epsilon_{mn} (\tau^I \epsilon)_{jk} + (\tau^I \epsilon)_{mn} \epsilon_{jk} .
\end{align} 
In particular, the relations are
\begin{align}
- Q_{\substack{lq^3H^2 \\ prst}}^{(1a)} &= Q_{\substack{lq^3H^2 \\ prst}}^{(1)} + Q_{\substack{lq^3H^2 \\ rpst}}^{(1)} , \nn
- Q_{\substack{lq^3H^2 \\ prst}}^{(1b)} &= Q_{\substack{lq^3H^2 \\ prst}}^{(1)} - Q_{\substack{lq^3H^2 \\ rpst}}^{(1)} , \nn
Q_{\substack{lq^3H^2 \\ prst}}^{(2b)} &= Q_{\substack{lq^3H^2 \\ rpst}}^{(2)} , \nn
Q_{\substack{lq^3H^2 \\ prst}}^{(3b)} &= - Q_{\substack{lq^3H^2 \\ rpst}}^{(3)} , \nn
2 Q_{\substack{lq^3H^2 \\ prst}}^{(2a)} &= (Q_{\substack{lq^3H^2 \\ prst}}^{(2)} + Q_{\substack{lq^3H^2 \\ rpst}}^{(2)}) - (Q_{\substack{lq^3H^2 \\ prst}}^{(3)} + Q_{\substack{lq^3H^2 \\ rpst}}^{(3)}) , \nn
2 Q_{\substack{lq^3H^2 \\ prst}}^{(3a)} &= (Q_{\substack{lq^3H^2 \\ prst}}^{(2)} - Q_{\substack{lq^3H^2 \\ rpst}}^{(2)}) - (Q_{\substack{lq^3H^2 \\ prst}}^{(3)} - Q_{\substack{lq^3H^2 \\ rpst}}^{(3)}) , \nn
Q_{\substack{lq^3H^2 \\ prst}}^{(4a)} &= Q_{\substack{lq^3H^2 \\ rpst}}^{(2)} - Q_{\substack{lq^3H^2 \\ rpst}}^{(3)} , \nn
\tfrac{1}{2}Q_{\substack{lq^3H^2 \\ prst}}^{(4b)} &= - (Q_{\substack{lq^3H^2 \\ prst}}^{(2)} - Q_{\substack{lq^3H^2 \\ rpst}}^{(2)}) - (Q_{\substack{lq^3H^2 \\ prst}}^{(3)} + Q_{\substack{lq^3H^2 \\ rpst}}^{(3)}) .
\end{align}

In addition to $\slashed B$ operators with dimension-6 analogs, three new types of operators appear.
These were three of the types of operators identified by Ref.~\cite{Henning:2015alf} as types that vanish in the absence of flavor structure.
All three contain a Lorentz singlet pair of quarks in the antisymmetric $\bar 3$ representation of $SU(3)_c$, yielding $n_g^3 (n_g -1)$ independent operators.
The operators of type $Q_{lq^3H^2}$ are different from these three (and others identified by~\cite{Henning:2015alf}) in that there is at least one Lagrangian term in the absence of flavor.
However not all of the terms are present in the absence of flavor structure. 
Dimension-8 is the lowest mass dimension where this happens.
The vanishing of operators in the absence of flavor structure first occurs at dimension-7.

%--------------------------------------------------------------------------------------------------------------
\item $\psi^4X$ \\
For a pair of currents there are 114 terms formed an operator by contracting the currents with a field strength, and inserting $SU(2)_w$ and $SU(3)_c$ generators and invariants as necessary.
There are at least two terms per $J J X$ operator type, one from $X_L$ and one from $X_R$.
The largest number of terms is eight, which occurs for operator types $Q_{u^2d^2G}$, $Q_{q^2u^2G}$, and $Q_{q^2d^2G}$, where the $SU(3)_c$ combinations are $(8 \otimes 1 \otimes 8)$, $(1 \otimes 8 \otimes 8)$, $(8 \otimes 8 \otimes 8)_A$, and $(8 \otimes 8 \otimes 8)_S$ for the, say, $u$ current, $d$ current and $G$, respectively.

There are five types of operators that were identified in~\cite{Henning:2015alf} involving identical currents contracted with the hypercharge field strength, \textit{e.g.} $(\bar l \gamma^\mu l) (\bar l \gamma^\nu l) B_{\mu\nu}$, which vanish in the absence of flavor structure as the contraction forces the fermions into an antisymmetric flavor representation.
For each Lagrangian term the number of operators is
\begin{equation}
\label{eq:4f}
\left(\overline{\yng(1)} \otimes \yng(1) \otimes \overline{\yng(1)} \otimes \yng(1)\right)_A = adj \oplus adj \oplus \bar a s \oplus \bar s a = \frac{1}{2} n_g^2 (n_g^2 - 1)
\end{equation}
The relevant group theory results can be found in \textit{e.g.}~\cite{Dashen:1994qi}.
The electron is a special case as it does not have $SU(2)_w$ or $SU(3)_c$ indices.
As such only half of the operators of this class 19 type are independent with the rest being related through a Fierz identity.

For operators with fermion chirality $(\bar L R) (\bar R L)$ there are two possibilities per field strength, one with the left-handed field strength and the other with its right-handed counterpart.
For operators with fermion chirality $(\bar L R) (\bar L R)$ there are instead three choices for the Lorentz contractions, all with $X_R$.
Two of these terms involve a tensor bilinear and a scalar bilinear while the third has two tensor bilinears.
Additionally when all the fermions are quarks there are two choices for the $SU(3)_c$ contractions.

For the baryon number violating operators with two left-handed and two right-handed fermions there are two possible Lorentz contractions, one with $X_L$ and one with $X_R$.
When these operators involve a gluon field strength there are also two possible arrangements of the $SU(3)_c$ indices, $8 \otimes 3 \otimes 3 \otimes 3 = (3 \oplus \bar 6 \ldots) \otimes 3 \otimes 3 = 1 \oplus 1 \ldots$.
The operators of type $Q_{eq^2uX}$ with $X = W_R$ or $B_L$ are in antisymmetric flavor representations for the $q$ pair, and as such vanish in the absence of flavor structure.
Instead $Q_{eq^2uX}$ with $X = W_L$ or $B_R$ are symmetric in $p$ and $r$.
The types involving gluons have full flavor rank as the $q$ fields can either be a color $\bar 3$ or $6$, compensating for other (anti)symmetries of the operator.

On the other hand, when the fermions all have the same chirality three Lorentz contractions are possible.
For operators of the type $Q_{eu^2dB}$ the gauge contractions are fixed and it is the Lorentz contractions that dictate the flavor representation of the $u$ pair is, leading to one symmetric and one full rank term $+ \hc$.
Instead for $Q_{eu^2dG}$ type operators, the additional freedom coming from the color indices of the gluon, allowing for full flavor rank, $n_g^4$, in all the Lagrangian terms.
In the case of $Q_{lq^3X}$ only two of the three Lorentz contractions are independent due to the identical $q$ fields.
There are two terms of $Q_{lq^3G}$ operators ($+ \hc$) that have the same flavor representations as the their dimension-6 analog, $Q_{qqql}$, along with one term of $Q_{lq^3W}$ and one term of $Q_{lq^3B}$, again $+ \hc$.
For each of these types of operators there are an equal number of operator types that have mixed symmetry, $\tfrac{1}{3} n_g^2 (n_g^2 - 1)$, that vanish in the absence of flavor structure similar to $Q_{lq^3H^2}^{(3)}$.
Finally there is a third term $+ \hc$ for the operators involving $W_{\mu\nu}^I$, $Q_{lq^3W}^{(2)}$, that is in a symmetric plus mixed flavor representation.

Operators of the type $Q_{lq^3W}$ are another complicated case that deserve further discussion.
Here the quarks can be in either the 2 or the 4 representation of both $SU(2)_w$ and the $SU(2)_L$ of the Lorentz group.
Naively there are four terms ($+ \hc$) to consider with redundant operators handled in a similar fashion as the $Q_{lq^3H^2}$ case.
However we can unambiguously combine the two terms that are Lorentz quartets into a single term, reducing the number of terms to three $+ \hc$.
To see this consider the quark flavor symmetries of the four cases.
Following Ref.~\cite{Fonseca:2019yya}, specifically its Table 2, we decompose the product of the gauge and Lorentz representations into irreducible representations of the permutation group of three objects, $S_3$, which gives us the flavor representations of the three quarks.
As we have been seen before, when the quarks are in the 2 of both $SU(2)_w$ and $SU(2)_L$ they have symmetric, mixed, and antisymmetric flavor representations.
When one of the two representations is a 2 and the other is a 4 there is only a mixed flavor representation.
Finally, when both representations are the 4 there is only the symmetric representation.
We can combine the two terms that are in the 4 of $SU(2)_L$ into $Q_{lq^3W}^{(2)}$ as they contain distinct flavor representations.
This is not unambiguously possible for $Q_{lq^3W}^{(1)}$ and $Q_{lq^3W}^{(3)}$, or $Q_{lq^3H^2}^{(2)}$ and $Q_{lq^3H^2}^{(3)}$ as each of those terms contain a mixed representation.
In equations, the naive terms that are in the 4 of $SU(2)_L$ are
\begin{align}
Q_{lq^3W}^{(2a)} &= \epsilon_{\alpha\beta\gamma} (\tau^I \epsilon)_{mn} \epsilon_{jk} (q_p^{m\alpha} C  \sigma^{\mu\nu}q_r^{j\beta}) (q_s^{k\gamma} C l_t^n) W^I_{\mu\nu} ,\nn
Q_{lq^3W}^{(2b)} &= \epsilon_{\alpha\beta\gamma} \epsilon_{mn} (\tau^I \epsilon)_{jk} (q_p^{m\alpha} C \sigma^{\mu\nu}q_r^{j\beta}) (q_s^{k\gamma} C  l_t^n) W^I_{\mu\nu} .
\end{align}
They can be combined as 
\begin{equation}
2 Q_{\substack{lq^3W \\ prst}}^{(2)} = (Q_{\substack{lq^3W \\ prst}}^{(2a)} - Q_{\substack{lq^3W \\ rpst}}^{(2a)}) - (Q_{\substack{lq^3W \\ prst}}^{(2b)} - Q_{\substack{lq^3W \\ rpst}}^{(2b)}) .
\end{equation}
Other combinations are possible of course, but this combination makes it clear there is a symmetric and a mixed flavor representation.

%--------------------------------------------------------------------------------------------------------------
\item $\psi^4HD$ \\

In class 20 the operators either have one fermion transforming as $(\tfrac{1}{2}, 0)$ and three transforming as $(0, \tfrac{1}{2})$ under the Lorentz group, or vice versa.
When describing the classification of this class we assume the former case.
Then, from~\eqref{eq:eom} the derivative cannot act on the left-handed fermion.
Otherwise the operator would be reduced by the EOM.

We start with the baryon number conserving operators.
Consider the case when the four fermion fields are distinguishable, \textit{e.g.} $\bar d, d, \bar l, e$ (with conjugate fields are counted separately).
Here there are two independent Lorentz contractions when the derivative acts on the Higgs field.
A third Lorentz structure is related to the first two by Eq.~\eqref{eq:lc}.
When the derivative acts on a fermion there is instead only one possible Lorentz contraction.
We use type $Q_{led^2HD}$ as an example.
It has the five aforementioned candidate terms
\begin{align}
x_1 &= \bar d_a d_{\dot a} \bar l_{\dot b} e_{\dot c} (D H)_{b \dot d} \epsilon^{ab} \epsilon^{\dot a \dot b} \epsilon^{\dot c \dot d} , \nn
x_2 &= \bar d_a d_{\dot a} \bar l_{\dot b} e_{\dot c} (D H)_{b \dot d} \epsilon^{ab} \epsilon^{\dot a \dot d} \epsilon^{\dot c \dot b} , \nn
x_3 &= \bar d_a (D d)_{b (\dot a \dot d)} \bar l_{\dot b} e_{\dot c} H \tfrac{1}{2} \epsilon^{ab} (2 \epsilon^{\dot a \dot b} \epsilon^{\dot c \dot d} - \epsilon^{\dot a \dot d} \epsilon^{\dot c \dot b}) , \nn
x_4 &= \bar d_a d_{\dot a} (D \bar l)_{a (\dot b \dot d)} e_{\dot c} H \tfrac{1}{2} \epsilon^{ab} (\epsilon^{\dot a \dot b} \epsilon^{\dot c \dot d} + \epsilon^{\dot a \dot d} \epsilon^{\dot c \dot b}) , \nn
x_5 &= \bar d_a d_{\dot a} l_{\dot b} (D e)_{a (\dot c \dot d)} H \tfrac{1}{2} \epsilon^{ab} \epsilon^{\dot a \dot b} \epsilon^{\dot c \dot d} ,
\end{align}
where Eq.~\eqref{eq:lc} is used to remove redundant Lorentz structures.
There are two constraint equations
\begin{align}
D y_1 &= x_1 + x_3 + \tfrac{1}{2} x_4 + \tfrac{1}{2} x_5 = 0, \nn
D y_2 &= x_2 - x_3 + \tfrac{1}{2} x_4 = 0,
\end{align}
and we choose to keep $x_1$, $x_2$, and $x_4$ in our basis.
For other types of operators where all four fermions are distinguishable we keep the analogs of $x_1$, $x_2$, and $x_4$ as well.
In addition, if there are four fields in the operator that are fundamentals under $SU(2)_w$ or $SU(3)_c$, including the Higgs, then there are two possible contractions of those gauge indices for each possible Lorentz contraction.

The other possibility is that only three of the fermion fields are unique, \textit{e.g.} $\bar d, d, \bar q, d$.
In this case there is only one way to contract the Lorentz indices when the derivative acts on the Higgs field, and only two possible ways to assign the derivative to fermions.
When the derivative acts on the Higgs field or the repeated fermion there are four, two, and one possible gauge contractions when the repeated fermion is $q$, one of $\{l, u, d\}$, or $e$, respectively.
Instead when the derivative acts on the fermion is not repeated, \textit{e.g.} $\bar d, d, (D \bar q), d$, there are two possible gauge contractions if the repeated fermion is $q$ and only one otherwise.
Here the repeated fermion is in a symmetric Lorentz representation and so the gauge contractions must be antisymmetric, eliminating half the possibilities.

The baryon number violating operators follow the same rules.
There are a couple of non-trivial flavor cases.
When there are duplicate right-handed fermions they form a symmetric flavor representation if the derivative acts on the Higgs field, and have full flavor rank if the derivative acts on the one of the duplicate fermions.
For the term $Q_{eq^3HD}$, the derivative acts on one of the $q$ fields, breaking the flavor constrain, Eq.~\eqref{eq:qqql}, giving it full flavor rank.

%--------------------------------------------------------------------------------------------------------------
\item $\psi^4D^2$ \\

We start with operators with equal numbers of left- and right-handed fermions, both $B$ and $\slashed B$.
There are two ways to assign the derivatives to the fields that are not related by IBP.
The first is to assign the derivatives to fermions that have the same Lorentz representation, and second is to assign the derivatives to fermions with conjugate Lorentz representations.
The only exception to this is when both currents are electron currents in which case there is only one independent term.
As previously mentioned this is due to the fact that the electron is a singlet both $SU(2)_w$ and $SU(3)_c$.
As this work was being completed Ref.~\cite{Alioli:2020kez} appeared, which classified nine terms from class 19.
Some of their operators subsume both derivatives into the d'Alembertian operator.
Our logic is consistent with the results of Ref.~\cite{Alioli:2020kez}.
The difference is we use relations like~\eqref{eq:eom} to reduce operators with two derivatives acting on the same fermion to classes with fewer derivatives.
As such the $\psi^4D^2$ operators in our basis where derivatives act symmetrically take the form $D_\mu (\psi_1 \Gamma \psi_2) D^\mu (\psi_3 \Gamma \psi_4)$ where $\Gamma$ is some, possibly scalar, combination of gamma matrices. 
The remaining 30 current-current terms and the four terms with chirality $(\bar L R) (\bar R L)$ in class 19 have a form analogous to the operators classified by Ref.~\cite{Alioli:2020kez}.
Also, the derivatives in $Q_{eq^2uD^2}$ break the flavor constraint present in its dimension-6 analog $Q_{qque}$.

For operators with either all left-handed or all right-handed fermions there are three possible Lorentz contractions, two where the fermions without derivatives form a scalar and a third where they form a tensor.
An example of an IBP constraint equation for these types of operators is
\begin{equation}
Q_{lequD^2}^{(1)} + 2 (D_\mu \bar l_p^j D^\mu e_r) \epsilon_{jk} (\bar q_s^k u_t) =  \fbox{$E$} .
\end{equation}
There is a single term ($+ \hc$) for the type $Q_{lq^3D^2}$.
As was the case with $Q_{eq^2uD^2}$, the derivatives in $Q_{lq^3D^2}$ break the flavor constraint present in some other operators of type $Q_{lq^3\ldots}$, allowing the term to have full flavor rank.
On the other hand, for the first time we encounter a term of the type $Q_{eu^2d\ldots}$ that vanishes in the absence of flavor structure.
There is also a second type of $Q_{eu^2dD^2}$ operator ($+ \hc$) that has full flavor rank.

%%%
\end{enumerate}

%%----------------------------------------------------------------------------------------------------------------------------------------------------------------

\section{The Complete Set of Dimension-8 Operators}
\label{sec:results}

Having gone through our classification of the dimension-8 operators in the previous Section we are now ready to tabulate the results.
Table~\ref{tab:summary} summarizes the results tables that follow it.
The links in the rightmost column point to the table(s) of results for a given class.
The number of types of operators and the number of Lagrangian terms in the class are given in the third and fourth columns from the left, respectively.
For comparison the number of operators from Ref.~\cite{Henning:2015alf} is given in the second column from the right.
Lines separate the classes based on the number of fermions in the class. 
Four-fermion operators are further divided into subclasses either preserving or violating baryon number.
Additionally the number to the right of the $+$ sign in $N_{\rm type, term}$ for the four-fermion operators is the number of types or terms that vanish in the absence of flavor structure
Tables~\ref{tab:smeft8class_1_2_3_4} and \ref{tab:smeft8class_5_6_7_8} contain bosonic operators.
Tables~\ref{tab:smeft8class_9}, \ref{tab:smeft8class_10_11_12_13}, \ref{tab:smeft8class_14qud}, \ref{tab:smeft8class_14le_15RR}, \ref{tab:smeft8class_15LL}, and \ref{tab:smeft8class_16_17} contain two-fermion operators.
Tables~\ref{tab:smeft8class_18_21}, \ref{tab:smeft8class_18},  \ref{tab:smeft8class_19_LL_RR}, \ref{tab:smeft8class_19_LLRR}, \ref{tab:smeft8class_19_LRRL_LRLR}, \ref{tab:smeft8class_19_slashedB}, \ref{tab:smeft8class_20_le_qu}, \ref{tab:smeft8class_20_qd_slashedB}, and \ref{tab:smeft8class_21} contain four-fermions operators.

%--------------------------------------------------------------------------------------------------------------
\begin{table}[H]
\begin{center}
\renewcommand{\arraystretch}{1.2}
\begin{tabular}[t]{c | c | c | c | c | c}
$\#$ & Class & $N_{\rm type}$ & $N_{\rm term}$ & $N_{\rm op}$~\cite{Henning:2015alf} & Table(s) \\
\hline
%--------------------------------------------------------
1    & $X^4$                & 7 & 43 & 43    & \ref{tab:smeft8class_1_2_3_4} \\
2    & $H^8$                & 1  & 1   & 1       &  \ref{tab:smeft8class_1_2_3_4} \\
3    & $H^6D^2$          & 1  & 2   & 2     &  \ref{tab:smeft8class_1_2_3_4} \\
4    & $H^4D^4$           & 1  & 3   & 3    &  \ref{tab:smeft8class_1_2_3_4} \\
5    & $X^3H^2$          & 3  & 6   & 6       & \ref{tab:smeft8class_5_6_7_8} \\
6    & $X^2H^4$          & 5 & 10 & 10    & \ref{tab:smeft8class_5_6_7_8} \\
7    & $X^2H^2D^2$    & 4 & 18 & 18   & \ref{tab:smeft8class_5_6_7_8} \\
8    & $XH^4D^2$        &  2 & 6   & 6     & \ref{tab:smeft8class_5_6_7_8} \\ \hline
%--------------------------------------------------------
9    & $\psi^2X^2H$     & 16 & 96 & $96 n_g^2$ & \ref{tab:smeft8class_9} \\
10  & $\psi^2XH^3$     & 8 & 22 & $22 n_g^2$  &  \ref{tab:smeft8class_10_11_12_13} \\
11  & $\psi^2H^2D^3$  & 6 & 16 & $16 n_g^2$ &  \ref{tab:smeft8class_10_11_12_13} \\
12  & $\psi^2H^5$       & 3   & 6   & $6 n_g^2$    &  \ref{tab:smeft8class_10_11_12_13} \\
13  & $\psi^2H^4D$    & 6 & 13 & $13 n_g^2$  &  \ref{tab:smeft8class_10_11_12_13} \\
14  & $\psi^2X^2D$     & 21  & 57 & $57 n_g^2$  &  \ref{tab:smeft8class_14qud}, \ref{tab:smeft8class_14le_15RR} \\
15  & $\psi^2XH^2D$  & 16 & 92  & $92 n_g^2$ &  \ref{tab:smeft8class_14le_15RR}, \ref{tab:smeft8class_15LL} \\
16  & $\psi^2XHD^2$  & 8  & 48  & $48 n_g^2$ &  \ref{tab:smeft8class_16_17} \\
17  & $\psi^2H^3D^2$  & 3 & 36  & $36 n_g^2$ &  \ref{tab:smeft8class_16_17} \\ \hline 
%--------------------------------------------------------
18$(B)$  & \multirow{2}{*}{$\psi^4H^2$}  & 19  & 75 + 1 & $n_g^2 (67 n_g^2 + n_g +7)$  &  \ref{tab:smeft8class_18_21}, \ref{tab:smeft8class_18} \\
18$(\slashed B)$  &                                 & $4+3$  & $12+8$ & $\tfrac{1}{3} n_g^2 (43 n_g^2 - 9 n_g + 2)$  &  \ref{tab:smeft8class_18_21} \\ 
19$(B)$  & \multirow{2}{*}{$\psi^4X$}  & $40+5$  & $156+12$ & $4 n_g^2 (40 n_g^2 - 1)$ &  \ref{tab:smeft8class_19_LL_RR}, \ref{tab:smeft8class_19_LLRR}, \ref{tab:smeft8class_19_LRRL_LRLR} \\
19$(\slashed B)$  &                             & 4  & $44+12$   & $2 n_g^3 (21 n_g + 1)$    &  \ref{tab:smeft8class_19_slashedB} \\
20$(B)$  & \multirow{2}{*}{$\psi^4HD$}  & 16  & $134+2$ & $n_g^3 (135 n_g - 1)$ &  \ref{tab:smeft8class_20_le_qu}, \ref{tab:smeft8class_20_qd_slashedB} \\
20$(\slashed B)$  &                                 & 7 & 32 & $n_g^3 (29 n_g + 3)$    &  \ref{tab:smeft8class_20_qd_slashedB} \\ 
21$(B)$  & \multirow{2}{*}{$\psi^4D^2$}  & 18  & 55 & $\tfrac{11}{2} n_g^2 (9 n_g^2 + 1)$ &  \ref{tab:smeft8class_18_21}, \ref{tab:smeft8class_21} \\
21$(\slashed B)$  &                                 & 4  & $10+2$ & $n_g^3 (11 n_g - 1)$   &  \ref{tab:smeft8class_18_21} \\ \hline \hline
%--------------------------------------------------------
  & $B$                & $204+5$  & $895+15$ & 895(36971), $n_g = 1 (3)$ & \\
  & $\slashed B$  & $19+3$     & $98+22$   & 98(7836), $n_g = 1 (3)$   & \\
  & Total               & $223+8$  & $993+37$ & 993(44807), $n_g = 1 (3)$ &
\end{tabular}
\end{center}
\caption{Summary of the contents of the tables to follow. 
The links in the rightmost column point to the table(s) of results for a given class.
The number of types of operators and the number of Lagrangian terms in the class are given in the third and fourth columns from the left, respectively.
For comparison the number of operators from Ref.~\cite{Henning:2015alf} is given in the second column from the right.
Lines separate the classes based on the number of fermions in the class. 
Four-fermion operators are further divided into subclasses either preserving or violating baryon number.
Additionally the number to the right of the $+$ sign in $N_{\rm type(term)}$ for the four-fermion operators is the number of types(terms) that vanish in the absence of flavor structure.}
\label{tab:summary}
\end{table}

%--------------------------------------------------------------------------------------------------------------
\subsection{Results for Bosonic Operators}

See Tables~\ref{tab:smeft8class_1_2_3_4} and \ref{tab:smeft8class_5_6_7_8}.

%%%%%%%%%%%%%%%%%%%%%%%%%%%%%%%%%%%%%%%%%%
% SMEFT d=8 Classes 1, 2, 3, 4
%%%%%%%%%%%%%%%%%%%%%%%%%%%%%%%%%%%%%%%%%%
\begin{table}[H]
\begin{center}
\begin{adjustbox}{width=0.8\textwidth,center}
\small
%%%%%%%%%%%%
\begin{minipage}[t]{6cm}
\renewcommand{\arraystretch}{1.5}
\begin{tabular}[t]{c|c}
\multicolumn{2}{c}{\boldmath$1:X^4,\, X^3 X^{\prime}$} \\
\hline
$Q_{G^4}^{(1)}$  &  $(G_{\mu\nu}^A G^{A\mu\nu}) (G_{\rho\sigma}^B G^{B\rho\sigma})$ \\
$Q_{G^4}^{(2)}$  &  $(G_{\mu\nu}^A \widetilde{G}^{A\mu\nu}) (G_{\rho\sigma}^B \widetilde{G}^{B\rho\sigma})$ \\
$Q_{G^4}^{(3)}$  &  $(G_{\mu\nu}^A G^{B\mu\nu}) (G_{\rho\sigma}^A G^{B\rho\sigma})$ \\
$Q_{G^4}^{(4)}$  &  $(G_{\mu\nu}^A \widetilde{G}^{B\mu\nu}) (G_{\rho\sigma}^A \widetilde{G}^{B\rho\sigma})$ \\
$Q_{G^4}^{(5)}$  &  $(G_{\mu\nu}^A G^{A\mu\nu}) (G_{\rho\sigma}^B \widetilde{G}^{B\rho\sigma})$ \\
$Q_{G^4}^{(6)}$  &  $(G_{\mu\nu}^A G^{B\mu\nu}) (G_{\rho\sigma}^A \widetilde{G}^{B\rho\sigma})$ \\
$Q_{G^4}^{(7)}$  &  $d^{ABE} d^{CDE} (G_{\mu\nu}^A G^{B\mu\nu}) (G_{\rho\sigma}^C G^{D\rho\sigma})$ \\
$Q_{G^4}^{(8)}$  &  $d^{ABE} d^{CDE} (G_{\mu\nu}^A \widetilde{G}^{B\mu\nu}) (G_{\rho\sigma}^C \widetilde{G}^{D\rho\sigma})$ \\
$Q_{G^4}^{(9)}$  &  $d^{ABE} d^{CDE} (G_{\mu\nu}^A G^{B\mu\nu}) (G_{\rho\sigma}^C \widetilde{G}^{D\rho\sigma})$ \\
$Q_{W^4}^{(1)}$  &  $(W_{\mu\nu}^I W^{I\mu\nu}) (W_{\rho\sigma}^J W^{J\rho\sigma})$ \\
$Q_{W^4}^{(2)}$  &  $(W_{\mu\nu}^I \widetilde{W}^{I\mu\nu}) (W_{\rho\sigma}^J \widetilde{W}^{J\rho\sigma})$ \\
$Q_{W^4}^{(3)}$  &  $(W_{\mu\nu}^I W^{J\mu\nu}) (W_{\rho\sigma}^I W^{J\rho\sigma})$ \\
$Q_{W^4}^{(4)}$  &  $(W_{\mu\nu}^I \widetilde{W}^{J\mu\nu}) (W_{\rho\sigma}^I \widetilde{W}^{J\rho\sigma})$ \\
$Q_{W^4}^{(5)}$  &  $(W_{\mu\nu}^I W^{I\mu\nu}) (W_{\rho\sigma}^J \widetilde{W}^{J\rho\sigma})$ \\
$Q_{W^4}^{(6)}$  &  $(W_{\mu\nu}^I W^{J\mu\nu}) (W_{\rho\sigma}^I \widetilde{W}^{J\rho\sigma})$ \\
$Q_{B^4}^{(1)}$  &  $(B_{\mu\nu} B^{\mu\nu}) (B_{\rho\sigma} B^{\rho\sigma})$ \\
$Q_{B^4}^{(2)}$  &  $(B_{\mu\nu} \widetilde{B}^{\mu\nu}) (B_{\rho\sigma} \widetilde{B}^{\rho\sigma})$ \\
$Q_{B^4}^{(3)}$  &  $(B_{\mu\nu} B^{\mu\nu}) (B_{\rho\sigma} \widetilde{B}^{\rho\sigma})$ \\
$Q_{G^3B}^{(1)}$  &  $d^{ABC} (B_{\mu\nu} G^{A\mu\nu}) (G_{\rho\sigma}^B G^{C\rho\sigma})$ \\
$Q_{G^3B}^{(2)}$  &  $d^{ABC} (B_{\mu\nu} \widetilde{G}^{A\mu\nu}) (G_{\rho\sigma}^B \widetilde{G}^{C\rho\sigma})$ \\
$Q_{G^3B}^{(3)}$  &  $d^{ABC} (B_{\mu\nu} \widetilde{G}^{A\mu\nu}) (G_{\rho\sigma}^B G^{C\rho\sigma})$ \\
$Q_{G^3B}^{(4)}$  &  $d^{ABC} (B_{\mu\nu} G^{A\mu\nu}) (G_{\rho\sigma}^B \widetilde{G}^{C\rho\sigma})$
\end{tabular}
\end{minipage}
\hspace{1cm}
%%%%%%%%%%%%
\begin{minipage}[t]{5cm}
\renewcommand{\arraystretch}{1.5}
\begin{tabular}[t]{c|c}
\multicolumn{2}{c}{\boldmath$1:X^2 X^{\prime 2}$} \\
\hline
$Q_{G^2W^2}^{(1)}$  &  $(W_{\mu\nu}^I W^{I\mu\nu}) (G_{\rho\sigma}^A G^{A\rho\sigma})$ \\
$Q_{G^2W^2}^{(2)}$  &  $(W_{\mu\nu}^I \widetilde{W}^{I\mu\nu}) (G_{\rho\sigma}^A \widetilde{G}^{A\rho\sigma})$ \\
$Q_{G^2W^2}^{(3)}$  &  $(W_{\mu\nu}^I G^{A\mu\nu}) (W_{\rho\sigma}^I G^{A\rho\sigma})$ \\
$Q_{G^2W^2}^{(4)}$  &  $(W_{\mu\nu}^I \widetilde{G}^{A\mu\nu}) (W_{\rho\sigma}^I \widetilde{G}^{A\rho\sigma})$ \\
$Q_{G^2W^2}^{(5)}$  &  $(W_{\mu\nu}^I \widetilde{W}^{I\mu\nu}) (G_{\rho\sigma}^A G^{A\rho\sigma})$ \\
$Q_{G^2W^2}^{(6)}$  &  $(W_{\mu\nu}^I W^{I\mu\nu}) (G_{\rho\sigma}^A \widetilde{G}^{A\rho\sigma})$ \\
$Q_{G^2W^2}^{(7)}$  &  $(W_{\mu\nu}^I G^{A\mu\nu}) (W_{\rho\sigma}^I \widetilde{G}^{A\rho\sigma})$ \\
$Q_{G^2B^2}^{(1)}$  &  $(B_{\mu\nu} B^{\mu\nu}) (G_{\rho\sigma}^A G^{A\rho\sigma})$ \\
$Q_{G^2B^2}^{(2)}$  &  $(B_{\mu\nu} \widetilde{B}^{\mu\nu}) (G_{\rho\sigma}^A \widetilde{G}^{A\rho\sigma})$ \\
$Q_{G^2B^2}^{(3)}$  &  $(B_{\mu\nu} G^{A\mu\nu}) (B_{\rho\sigma} G^{A\rho\sigma})$ \\
$Q_{G^2B^2}^{(4)}$  &  $(B_{\mu\nu} \widetilde{G}^{A\mu\nu}) (B_{\rho\sigma} \widetilde{G}^{A\rho\sigma})$ \\
$Q_{G^2B^2}^{(5)}$  &  $(B_{\mu\nu} \widetilde{B}^{\mu\nu}) (G_{\rho\sigma}^A G^{A\rho\sigma})$ \\
$Q_{G^2B^2}^{(6)}$  &  $(B_{\mu\nu} B^{\mu\nu}) (G_{\rho\sigma}^A \widetilde{G}^{A\rho\sigma})$ \\
$Q_{G^2B^2}^{(7)}$  &  $(B_{\mu\nu} G^{A\mu\nu}) (B_{\rho\sigma} \widetilde{G}^{A\rho\sigma})$ \\
$Q_{W^2B^2}^{(1)}$  &  $(B_{\mu\nu} B^{\mu\nu}) (W_{\rho\sigma}^I W^{I\rho\sigma})$ \\
$Q_{W^2B^2}^{(2)}$  &  $(B_{\mu\nu} \widetilde{B}^{\mu\nu}) (W_{\rho\sigma}^I \widetilde{W}^{I\rho\sigma})$ \\
$Q_{W^2B^2}^{(3)}$  &  $(B_{\mu\nu} W^{I\mu\nu}) (B_{\rho\sigma} W^{I\rho\sigma})$ \\
$Q_{W^2B^2}^{(4)}$  &  $(B_{\mu\nu} \widetilde{W}^{I\mu\nu}) (B_{\rho\sigma} \widetilde{W}^{I\rho\sigma})$ \\
$Q_{W^2B^2}^{(5)}$  &  $(B_{\mu\nu} \widetilde{B}^{\mu\nu}) (W_{\rho\sigma}^I W^{I\rho\sigma})$ \\
$Q_{W^2B^2}^{(6)}$  &  $(B_{\mu\nu} B^{\mu\nu}) (W_{\rho\sigma}^I \widetilde{W}^{I\rho\sigma})$ \\
$Q_{W^2B^2}^{(7)}$  &  $(B_{\mu\nu} W^{I\mu\nu}) (B_{\rho\sigma} \widetilde{W}^{I\rho\sigma})$
\end{tabular}
\end{minipage}
%%%%%%%%
\end{adjustbox}
%%%%%%%%%%%%%%%%%%%%%%
\begin{adjustbox}{width=1.0\textwidth,center}
\small
%%%%%%%%%%%%
\begin{minipage}[t]{2.4cm}
\renewcommand{\arraystretch}{1.5}
\begin{tabular}[t]{c|c}
\multicolumn{2}{c}{\boldmath$2:H^8$} \\
\hline
$Q_{H^8}$  &  $(H^\dag H)^4$ 
\end{tabular}
\end{minipage}
\hspace{1cm}
%%%%%%%%%%%%
\begin{minipage}[t]{5.6cm}
\renewcommand{\arraystretch}{1.5}
\begin{tabular}[t]{c|c}
\multicolumn{2}{c}{\boldmath$3:H^6D^2$} \\
\hline
$Q_{H^6}^{(1)}$  & $(H^{\dag} H)^2 (D_{\mu} H^{\dag} D^{\mu} H)$ \\
$Q_{H^6}^{(2)}$  & $(H^{\dag} H) (H^{\dag} \tau^I H) (D_{\mu} H^{\dag} \tau^I D^{\mu} H)$
\end{tabular}
\end{minipage}
\hspace{1cm}
%%%%%%%%%%%%
\begin{minipage}[t]{4.9cm}
\renewcommand{\arraystretch}{1.5}
\begin{tabular}[t]{c|c}
\multicolumn{2}{c}{\boldmath$4:H^4D^4$} \\
\hline
$Q_{H^4}^{(1)}$  &  $(D_{\mu} H^{\dag} D_{\nu} H) (D^{\nu} H^{\dag} D^{\mu} H)$ \\ 
$Q_{H^4}^{(2)}$  &  $(D_{\mu} H^{\dag} D_{\nu} H) (D^{\mu} H^{\dag} D^{\nu} H)$ \\ 
$Q_{H^4}^{(3)}$  &  $(D^{\mu} H^{\dag} D_{\mu} H) (D^{\nu} H^{\dag} D_{\nu} H)$
\end{tabular}
\end{minipage}
%%%%%%%%
\end{adjustbox}
\end{center}
\caption{The dimension-eight operators in the SMEFT whose field content is either entirely gauge field strengths or Higgs boson fields.}
\label{tab:smeft8class_1_2_3_4}
\end{table}

%%%%%%%%%%%%%%%%%%%%%%%%%%%%%%%%%%%%%%%%%%
% SMEFT d=8 Classes 5, 6, 7, 8
%%%%%%%%%%%%%%%%%%%%%%%%%%%%%%%%%%%%%%%%%%
\begin{table}[H]
\begin{center}
%%%%%%%%%%%%%%%%%%%%%%
\begin{adjustbox}{width=1.0\textwidth,center}
\small
%%%%%%%%%%%%
\begin{minipage}[t]{8cm}
\renewcommand{\arraystretch}{1.5}
\begin{tabular}[t]{c|c}
\multicolumn{2}{c}{\boldmath$5:X^3H^2$} \\
\hline
$Q_{G^3H^2}^{(1)}$  &  $f^{ABC} (H^\dag H) G_{\mu}^{A\nu} G_{\nu}^{B\rho} G_{\rho}^{C\mu}$ \\
$Q_{G^3H^2}^{(2)}$  &  $f^{ABC} (H^\dag H) G_{\mu}^{A\nu} G_{\nu}^{B\rho} \widetilde{G}_{\rho}^{C\mu}$ \\
$Q_{W^3H^2}^{(1)}$  &  $\epsilon^{IJK} (H^\dag H) W_{\mu}^{I\nu} W_{\nu}^{J\rho} W_{\rho}^{K\mu}$ \\
$Q_{W^3H^2}^{(2)}$  &  $\epsilon^{IJK} (H^\dag H) W_{\mu}^{I\nu} W_{\nu}^{J\rho} \widetilde{W}_{\rho}^{K\mu}$ \\
$Q_{W^2BH^2}^{(1)}$  &  $\epsilon^{IJK} (H^\dag \tau^I H) B_{\mu}^{\,\nu} W_{\nu}^{J\rho} W_{\rho}^{K\mu}$ \\
$Q_{W^2BH^2}^{(2)}$  &  $\epsilon^{IJK} (H^\dag \tau^I H) (\widetilde{B}^{\mu\nu} W_{\nu\rho}^J W_{\mu}^{K\rho} + B^{\mu\nu} W_{\nu\rho}^J \widetilde{W}_{\mu}^{K\rho})$
\end{tabular}
\end{minipage}
\hspace{1cm}
%%%%%%%%%%%%
\begin{minipage}[t]{6cm}
\renewcommand{\arraystretch}{1.5}
\begin{tabular}[t]{c|c}
\multicolumn{2}{c}{\boldmath$6:X^2H^4$} \\
\hline
$Q_{G^2H^4}^{(1)}$  & $(H^\dag H)^2 G^A_{\mu\nu} G^{A\mu\nu}$ \\
$Q_{G^2H^4}^{(2)}$  & $(H^\dag H)^2 \widetilde G^A_{\mu\nu} G^{A\mu\nu}$ \\
$Q_{W^2H^4}^{(1)}$  & $(H^\dag H)^2 W^I_{\mu\nu} W^{I\mu\nu}$ \\
$Q_{W^2H^4}^{(2)}$  & $(H^\dag H)^2 \widetilde W^I_{\mu\nu} W^{I\mu\nu}$ \\
$Q_{W^2H^4}^{(3)}$  & $(H^\dag \tau^I H) (H^\dag \tau^J H) W^I_{\mu\nu} W^{J\mu\nu}$ \\
$Q_{W^2H^4}^{(4)}$  & $(H^\dag \tau^I H) (H^\dag \tau^J H) \widetilde W^I_{\mu\nu} W^{J\mu\nu}$ \\
$Q_{WBH^4}^{(1)}$  & $ (H^\dag H) (H^\dag \tau^I H) W^I_{\mu\nu} B^{\mu\nu}$ \\
$Q_{WBH^4}^{(2)}$  & $(H^\dag H) (H^\dag \tau^I H) \widetilde W^I_{\mu\nu} B^{\mu\nu}$ \\
$Q_{B^2H^4}^{(1)}$  & $ (H^\dag H)^2 B_{\mu\nu} B^{\mu\nu}$ \\
$Q_{B^2H^4}^{(2)}$  & $(H^\dag H)^2 \widetilde B_{\mu\nu} B^{\mu\nu}$ \\
\end{tabular}
\end{minipage}
%%%%%%%
\end{adjustbox}
%%%%%%%%%%%%%%%%%%%%%%
\begin{adjustbox}{width=1.0\textwidth,center}
\small
%%%%%%%%%%%%
\begin{minipage}[t]{7.8cm}
\renewcommand{\arraystretch}{1.5}
\begin{tabular}[t]{c|c}
\multicolumn{2}{c}{\boldmath$7:X^2H^2D^2$} \\
\hline
$Q_{G^2H^2D^2}^{(1)}$  &  $(D^{\mu} H^{\dag} D^{\nu} H) G_{\mu\rho}^A G_{\nu}^{A \rho}$ \\
$Q_{G^2H^2D^2}^{(2)}$  &  $(D^{\mu} H^{\dag} D_{\mu} H) G_{\nu\rho}^A G^{A \nu\rho}$ \\
$Q_{G^2H^2D^2}^{(3)}$  &  $(D^{\mu} H^{\dag} D_{\mu} H) G_{\nu\rho}^A \widetilde{G}^{A \nu\rho}$ \\
$Q_{W^2H^2D^2}^{(1)}$  &  $(D^{\mu} H^{\dag} D^{\nu} H) W_{\mu\rho}^I W_{\nu}^{I \rho}$ \\
$Q_{W^2H^2D^2}^{(2)}$  &  $(D^{\mu} H^{\dag} D_{\mu} H) W_{\nu\rho}^I W^{I \nu\rho}$ \\
$Q_{W^2H^2D^2}^{(3)}$  &  $(D^{\mu} H^{\dag} D_{\mu} H) W_{\nu\rho}^I \widetilde{W}^{I \nu\rho}$ \\
$Q_{W^2H^2D^2}^{(4)}$  &  $i \epsilon^{IJK} (D^{\mu} H^{\dag} \tau^I D^{\nu} H) W_{\mu\rho}^J W_{\nu}^{K \rho}$ \\
$Q_{W^2H^2D^2}^{(5)}$  &  $\epsilon^{IJK} (D^{\mu} H^{\dag} \tau^I D^{\nu} H) (W_{\mu\rho}^J \widetilde{W}_{\nu}^{K \rho} - \widetilde{W}_{\mu\rho}^J W_{\nu}^{K \rho})$ \\
$Q_{W^2H^2D^2}^{(6)}$  &  $i \epsilon^{IJK} (D^{\mu} H^{\dag} \tau^I D^{\nu} H) (W_{\mu\rho}^J \widetilde{W}_{\nu}^{K \rho} + \widetilde{W}_{\mu\rho}^J W_{\nu}^{K \rho})$ \\
$Q_{WBH^2D^2}^{(1)}$  &  $(D^{\mu} H^{\dag} \tau^I D_{\mu} H) B_{\nu\rho} W^{I \nu\rho}$ \\
$Q_{WBH^2D^2}^{(2)}$  &  $(D^{\mu} H^{\dag} \tau^I D_{\mu} H) B_{\nu\rho} \widetilde{W}^{I \nu\rho}$ \\
$Q_{WBH^2D^2}^{(3)}$  &  $i (D^{\mu} H^{\dag} \tau^I D^{\nu} H) (B_{\mu\rho} W_{\nu}^{I \rho} - B_{\nu\rho} W_{\mu}^{I\rho})$ \\
$Q_{WBH^2D^2}^{(4)}$  &  $(D^{\mu} H^{\dag} \tau^I D^{\nu} H) (B_{\mu\rho} W_{\nu}^{I \rho} + B_{\nu\rho} W_{\mu}^{I\rho})$ \\
$Q_{WBH^2D^2}^{(5)}$  &  $i (D^{\mu} H^{\dag} \tau^I D^{\nu} H) (B_{\mu\rho} \widetilde{W}_\nu^{^I \rho} - B_{\nu\rho} \widetilde{W}_\mu^{^I \rho})$ \\
$Q_{WBH^2D^2}^{(6)}$  &  $(D^{\mu} H^{\dag} \tau^I D^{\nu} H) (B_{\mu\rho} \widetilde{W}_\nu^{^I \rho} + B_{\nu\rho} \widetilde{W}_\mu^{^I \rho})$ \\
$Q_{B^2H^2D^2}^{(1)}$  &  $(D^{\mu} H^{\dag} D^{\nu} H) B_{\mu\rho} B_{\nu}^{\,\,\,\rho}$ \\
$Q_{B^2H^2D^2}^{(2)}$  &  $(D^{\mu} H^{\dag} D_{\mu} H) B_{\nu\rho} B^{\nu\rho}$ \\
$Q_{B^2H^2D^2}^{(3)}$  &  $(D^{\mu} H^{\dag} D_{\mu} H) B_{\nu\rho} \widetilde{B}^{\nu\rho}$
\end{tabular}
\end{minipage}
\hspace{0.8cm}
%%%%%%%%%%%%
\begin{minipage}[t]{6.4cm}
\renewcommand{\arraystretch}{1.5}
\begin{tabular}[t]{c|c}
\multicolumn{2}{c}{\boldmath$8:XH^4D^2$} \\
\hline
$Q_{WH^4D^2}^{(1)}$  & $(H^{\dag} H) (D^{\mu} H^{\dag} \tau^I D^{\nu} H) W_{\mu\nu}^I$ \\
$Q_{WH^4D^2}^{(2)}$  & $(H^{\dag} H) (D^{\mu} H^{\dag} \tau^I D^{\nu} H) \widetilde{W}_{\mu\nu}^I$ \\
$Q_{WH^4D^2}^{(3)}$  & $\epsilon^{IJK} (H^{\dag} \tau^I H) (D^{\mu} H^{\dag} \tau^J D^{\nu} H) W_{\mu\nu}^K$ \\
$Q_{WH^4D^2}^{(4)}$  & $\epsilon^{IJK} (H^{\dag} \tau^I H) (D^{\mu} H^{\dag} \tau^J D^{\nu} H) \widetilde{W}_{\mu\nu}^K$ \\
$Q_{BH^4D^2}^{(1)}$  & $(H^{\dag} H) (D^{\mu} H^{\dag} D^{\nu} H) B_{\mu\nu}$ \\
$Q_{BH^4D^2}^{(2)}$  & $(H^{\dag} H) (D^{\mu} H^{\dag} D^{\nu} H) \widetilde{B}_{\mu\nu}$
\end{tabular}
\end{minipage}
%%%%%%%
\end{adjustbox}
\end{center}
\caption{Bosonic dimension-eight operators in the SMEFT containing both gauge field strengths and Higgs boson fields.}
\label{tab:smeft8class_5_6_7_8}
\end{table}

%--------------------------------------------------------------------------------------------------------------
\subsection{Results for Two-Fermion Operators}

See Tables~\ref{tab:smeft8class_9}, \ref{tab:smeft8class_10_11_12_13}, \ref{tab:smeft8class_14qud}, \ref{tab:smeft8class_14le_15RR}, \ref{tab:smeft8class_15LL}, and \ref{tab:smeft8class_16_17}.

%%%%%%%%%%%%%%%%%%%%%%%%%%%%%%%%%%%%%%%%%%
% SMEFT d=8 Class 9 Operators
%%%%%%%%%%%%%%%%%%%%%%%%%%%%%%%%%%%%%%%%%%
\begin{table}[H]
\begin{center}
\begin{adjustbox}{width=0.81\textwidth,center}
\small
%%%%%%%%%%%%
\begin{minipage}[t]{5.6cm}
\renewcommand{\arraystretch}{1.5}
\begin{tabular}[t]{c|c}
\multicolumn{2}{c}{\boldmath$9:\psi^2X^2H + \hc$} \\
\hline
$Q_{leG^2H}^{(1)}$  &  $(\bar l_p e_r) H G^A_{\mu\nu} G^{A\mu\nu}$ \\
$Q_{leG^2H}^{(2)}$  &  $(\bar l_p e_r) H \widetilde G^A_{\mu\nu} G^{A\mu\nu}$ \\
$Q_{leW^2H}^{(1)}$  &  $(\bar l_p e_r) H W^I_{\mu\nu} W^{I\mu\nu} $ \\
$Q_{leW^2H}^{(2)}$  &  $(\bar l_p e_r) H \widetilde W^I_{\mu\nu} W^{I\mu\nu}$ \\
$Q_{leW^2H}^{(3)}$  &  $\epsilon^{IJK} (\bar l_p \sigma^{\mu\nu} e_r) \tau^I H W_{\mu\rho}^J W_\nu^{K\rho}$ \\
$Q_{quG^2H}^{(1)}$  &  $(\bar q_p u_r) \widetilde H G^A_{\mu\nu} G^{A\mu\nu} $ \\
$Q_{quG^2H}^{(2)}$  &  $(\bar q_p u_r) \widetilde H \widetilde G^A_{\mu\nu} G^{A\mu\nu} $ \\
$Q_{quG^2H}^{(3)}$  &  $d^{ABC} (\bar q_p T^A u_r) \widetilde H G^B_{\mu\nu} G^{C\mu\nu} $ \\
$Q_{quG^2H}^{(4)}$  &  $d^{ABC} (\bar q_p T^A u_r) \widetilde H \widetilde G^B_{\mu\nu} G^{C\mu\nu} $ \\
$Q_{quG^2H}^{(5)}$  &  $f^{ABC} (\bar q_p \sigma^{\mu\nu} T^A u_r) \widetilde H G_{\mu\rho}^B G_\nu^{C\rho}$ \\
$Q_{quGWH}^{(1)}$  &  $(\bar q_p T^A u_r) \tau^I \widetilde H G^A_{\mu\nu} W^{I\mu\nu} $ \\
$Q_{quGWH}^{(2)}$  &  $(\bar q_p T^A u_r) \tau^I \widetilde H \widetilde G^A_{\mu\nu} W^{I\mu\nu} $ \\
$Q_{quGWH}^{(3)}$  &  $(\bar q_p \sigma^{\mu\nu} T^A u_r) \tau^I \widetilde H G_{\mu\rho}^A W_\nu^{I\rho}$ \\
$Q_{quGBH}^{(1)}$  &  $(\bar q_p T^A u_r) \widetilde H G^A_{\mu\nu} B^{\mu\nu} $ \\
$Q_{quGBH}^{(2)}$  &  $(\bar q_p T^A u_r) \widetilde H \widetilde G^A_{\mu\nu} B^{\mu\nu} $ \\
$Q_{quGBH}^{(3)}$  &  $(\bar q_p \sigma^{\mu\nu} T^A u_r) \widetilde H G_{\mu\rho}^A B_\nu^{\,\,\,\rho}$ \\
$Q_{quW^2H}^{(1)}$  &  $(\bar q_p u_r) \widetilde H W^I_{\mu\nu} W^{I\mu\nu} $ \\
$Q_{quW^2H}^{(2)}$  &  $(\bar q_p u_r) \widetilde H \widetilde W^I_{\mu\nu} W^{I\mu\nu} $ \\
$Q_{quW^2H}^{(3)}$  &  $\epsilon^{IJK} (\bar q_p \sigma^{\mu\nu} u_r) \tau^I \widetilde H W_{\mu\rho}^J W_\nu^{K\rho}$ \\
$Q_{quWBH}^{(3)}$  &  $(\bar q_p \sigma^{\mu\nu} u_r) \tau^I \widetilde H W_{\mu\rho}^I B_\nu^{\,\,\,\rho}$ \\
$Q_{quWBH}^{(1)}$  &  $(\bar q_p u_r) \tau^I \widetilde H W^I_{\mu\nu} B^{\mu\nu}$ \\
$Q_{quWBH}^{(2)}$  &  $(\bar q_p u_r) \tau^I \widetilde H \widetilde W^I_{\mu\nu} B^{\mu\nu} $ \\
$Q_{quB^2H}^{(1)}$  &  $(\bar q_p u_r) \widetilde H B_{\mu\nu} B^{\mu\nu} $ \\
$Q_{quB^2H}^{(2)}$  &  $(\bar q_p u_r) \widetilde H \widetilde B_{\mu\nu} B^{\mu\nu}$
\end{tabular}
\end{minipage}
%%%%%%%%%%%%
\hspace{1cm}
\begin{minipage}[t]{5.6cm}
\renewcommand{\arraystretch}{1.5}
\begin{tabular}[t]{c|c}
\multicolumn{2}{c}{\boldmath$9:\psi^2X^2H + \hc$} \\
\hline
$Q_{leWBH}^{(1)}$  &  $(\bar l_p e_r) \tau^I H W^I_{\mu\nu} B^{\mu\nu} $ \\
$Q_{leWBH}^{(2)}$  &  $(\bar l_p e_r) \tau^I H \widetilde W^I_{\mu\nu} B^{\mu\nu} $ \\
$Q_{leWBH}^{(3)}$  &  $(\bar l_p \sigma^{\mu\nu} e_r) \tau^I H W_{\mu\rho}^I B_\nu^{\,\,\,\rho}$ \\
$Q_{leB^2H}^{(1)}$  &  $(\bar l_p e_r) H B_{\mu\nu} B^{\mu\nu} $ \\
$Q_{leB^2H}^{(2)}$  &  $(\bar l_p e_r) H \widetilde B_{\mu\nu} B^{\mu\nu} $ \\
$Q_{qdG^2H}^{(1)}$  &  $(\bar q_p d_r) H G^A_{\mu\nu} G^{A\mu\nu} $ \\
$Q_{qdG^2H}^{(2)}$  &  $(\bar q_p d_r) H \widetilde G^A_{\mu\nu} G^{A\mu\nu} $ \\
$Q_{qdG^2H}^{(3)}$  &  $d^{ABC} (\bar q_p T^A d_r) H G^B_{\mu\nu} G^{C\mu\nu} $ \\
$Q_{qdG^2H}^{(4)}$  &  $d^{ABC} (\bar q_p T^A d_r) H \widetilde G^B_{\mu\nu} G^{C\mu\nu} $ \\
$Q_{qdG^2H}^{(5)}$  &  $f^{ABC} (\bar q_p \sigma^{\mu\nu} T^A d_r) H G_{\mu\rho}^B G_\nu^{C\rho}$ \\
$Q_{qdGWH}^{(1)}$  &  $(\bar q_p T^A d_r) \tau^I H G^A_{\mu\nu} W^{I\mu\nu} $ \\
$Q_{qdGWH}^{(2)}$  &  $(\bar q_p T^A d_r) \tau^I H \widetilde G^A_{\mu\nu} W^{I\mu\nu} $ \\
$Q_{qdGWH}^{(3)}$  &  $(\bar q_p \sigma^{\mu\nu} T^A d_r) \tau^I H G_{\mu\rho}^A W_\nu^{I\rho}$ \\
$Q_{qdGBH}^{(1)}$  &  $(\bar q_p T^A d_r) H G^A_{\mu\nu} B^{\mu\nu} $ \\
$Q_{qdGBH}^{(2)}$  &  $(\bar q_p T^A d_r) H \widetilde G^A_{\mu\nu} B^{\mu\nu} $ \\
$Q_{qdGBH}^{(3)}$  &  $(\bar q_p \sigma^{\mu\nu} T^A d_r) H G_{\mu\rho}^A B_\nu^{\,\,\,\rho}$ \\
$Q_{qdW^2H}^{(1)}$  &  $(\bar q_p d_r) H W^I_{\mu\nu} W^{I\mu\nu} $ \\
$Q_{qdW^2H}^{(2)}$  &  $(\bar q_p d_r) H \widetilde W^I_{\mu\nu} W^{I\mu\nu} $ \\
$Q_{qdW^2H}^{(3)}$  &  $\epsilon^{IJK} (\bar q_p \sigma^{\mu\nu} d_r) \tau^I H W_{\mu\rho}^J W_\nu^{K\rho}$ \\
$Q_{qdWBH}^{(1)}$  &  $(\bar q_p d_r) \tau^I H W^I_{\mu\nu} B^{\mu\nu} $ \\
$Q_{qdWBH}^{(2)}$  &  $(\bar q_p d_r) \tau^I H \widetilde W^I_{\mu\nu} B^{\mu\nu} $ \\
$Q_{qdWBH}^{(3)}$  &  $(\bar q_p \sigma^{\mu\nu} d_r) \tau^I H W_{\mu\rho}^I B_\nu^{\,\,\,\rho}$ \\
$Q_{qdB^2H}^{(1)}$  &  $(\bar q_p d_r) H B_{\mu\nu} B^{\mu\nu} $ \\
$Q_{qdB^2H}^{(2)}$  &  $(\bar q_p d_r) H \widetilde B_{\mu\nu} B^{\mu\nu} $
\end{tabular}
\end{minipage}
%%%%%%%
\end{adjustbox}
\end{center}
\caption{The dimension-eight operators in the SMEFT of class-9 with field content $\psi^2X^2H$. All of the operators have Hermitian conjugates. The subscripts $p, r$ are weak-eigenstate indices.}
\label{tab:smeft8class_9}
\end{table}

%%%%%%%%%%%%%%%%%%%%%%%%%%%%%%%%%%%%%%%%%%
% SMEFT d=8 Classes 10, 11, 12 ,13
%%%%%%%%%%%%%%%%%%%%%%%%%%%%%%%%%%%%%%%%%%
\begin{table}[H]
\begin{center}
%%%%%%%%%%%%%%%%%%%%%%
\begin{adjustbox}{width=0.89\textwidth,center}
\small
%%%%%%%%%%%%
\begin{minipage}[t]{5.4cm}
\renewcommand{\arraystretch}{1.5}
\begin{tabular}[t]{c|c}
\multicolumn{2}{c}{\boldmath$10:\psi^2XH^3 + \hc$} \\
\hline
$Q_{leWH^3}^{(1)}$  & $(\bar l_p \sigma^{\mu\nu} e_r) \tau^I H (H^\dag H) W_{\mu\nu}^I$ \\
$Q_{leWH^3}^{(2)}$  & $(\bar l_p \sigma^{\mu\nu} e_r) H (H^\dag \tau^I H) W_{\mu\nu}^I$ \\
$Q_{leBH^3}$  & $(\bar l_p \sigma^{\mu\nu} e_r) H (H^\dag H) B_{\mu\nu}$ \\
$Q_{quGH^3}$  & $(\bar q_p \sigma^{\mu\nu} T^A u_r) \widetilde H (H^\dag H) G_{\mu\nu}^A$ \\
$Q_{quWH^3}^{(1)}$  & $(\bar q_p \sigma^{\mu\nu} u_r) \tau^I \widetilde H (H^\dag H) W_{\mu\nu}^I$ \\
$Q_{quWH^3}^{(2)}$  & $(\bar q_p \sigma^{\mu\nu} u_r) \widetilde H (H^\dag \tau^I H) W_{\mu\nu}^I$ \\
$Q_{quBH^3}$  & $(\bar q_p \sigma^{\mu\nu} u_r) \widetilde H (H^\dag H) B_{\mu\nu}$ \\
$Q_{qdGH^3}$  & $(\bar q_p \sigma^{\mu\nu} T^A d_r) H (H^\dag H) G_{\mu\nu}^A$ \\
$Q_{qdWH^3}^{(1)}$  & $(\bar q_p \sigma^{\mu\nu} d_r) \tau^I H (H^\dag H) W_{\mu\nu}^I$ \\
$Q_{qdWH^3}^{(2)}$  & $(\bar q_p \sigma^{\mu\nu} d_r) H (H^\dag \tau^I H) W_{\mu\nu}^I$ \\
$Q_{qdBH^3}$  & $(\bar q_p \sigma^{\mu\nu} d_r) H (H^\dag H) B_{\mu\nu}$ 
\end{tabular}
\end{minipage}
%%%%%%%%%%%%
\hspace{1cm}
\begin{minipage}[t]{6.9cm}
\renewcommand{\arraystretch}{1.5}
\begin{tabular}[t]{c|c}
\multicolumn{2}{c}{\boldmath$11:\psi^2H^2D^3$} \\
\hline
$Q_{l^2H^2D^3}^{(1)}$  &  $i (\bar{l}_p \gamma^{\mu} D^{\nu} l_r) (D_{(\mu}D_{\nu)}H^{\dag} H)$ \\
$Q_{l^2H^2D^3}^{(2)}$  &  $i (\bar{l}_p \gamma^{\mu} D^{\nu} l_r) (H^{\dag} D_{(\mu}D_{\nu)} H)$ \\
$Q_{l^2H^2D^3}^{(3)}$  &  $i (\bar{l}_p \gamma^{\mu} \tau^I D^{\nu} l_r) (D_{(\mu}D_{\nu)}H^{\dag} \tau^I H)$ \\
$Q_{l^2H^2D^3}^{(4)}$  &  $i (\bar{l}_p \gamma^{\mu} \tau^I D^{\nu} l_r) (H^{\dag} \tau^I D_{(\mu}D_{\nu)} H)$ \\
$Q_{e^2H^2D^3}^{(1)}$  &  $i (\bar{e}_p \gamma^{\mu} D^{\nu} e_r) (D_{(\mu}D_{\nu)}H^{\dag} H)$ \\
$Q_{e^2H^2D^3}^{(2)}$  &  $i (\bar{e}_p \gamma^{\mu} D^{\nu} e_r) (H^{\dag} D_{(\mu}D_{\nu)} H)$ \\
$Q_{q^2H^2D^3}^{(1)}$  &  $i (\bar{q}_p \gamma^{\mu} D^{\nu} q_r) (D_{(\mu}D_{\nu)}H^{\dag} H)$ \\
$Q_{q^2H^2D^3}^{(2)}$  &  $i (\bar{q}_p \gamma^{\mu} D^{\nu} q_r) (H^{\dag} D_{(\mu}D_{\nu)} H)$ \\
$Q_{q^2H^2D^3}^{(3)}$  &  $i (\bar{q}_p \gamma^{\mu} \tau^I D^{\nu} q_r) (D_{(\mu}D_{\nu)}H^{\dag} \tau^I H)$ \\
$Q_{q^2H^2D^2}^{(4)}$  &  $i (\bar{q}_p \gamma^{\mu} \tau^I D^{\nu} q_r) (H^{\dag} \tau^I D_{(\mu}D_{\nu)} H)$ \\
$Q_{u^2H^2D^3}^{(1)}$  &  $i (\bar{u}_p \gamma^{\mu} D^{\nu} u_r) (D_{(\mu}D_{\nu)}H^{\dag} H)$ \\
$Q_{u^2H^2D^3}^{(2)}$  &  $i (\bar{u}_p \gamma^{\mu} D^{\nu} u_r) (H^{\dag} D_{(\mu}D_{\nu)} H)$ \\
$Q_{d^2H^2D^3}^{(1)}$  &  $i (\bar{d}_p \gamma^{\mu} D^{\nu} d_r) (D_{(\mu}D_{\nu)}H^{\dag} H)$ \\
$Q_{d^2H^2D^3}^{(2)}$  &  $i (\bar{d}_p \gamma^{\mu} D^{\nu} d_r) (H^{\dag} D_{(\mu}D_{\nu)} H)$ \\
$Q_{udH^2D^3} + \hc$  &  $i (\bar{u}_p \gamma^{\mu} D^{\nu} d_r) (\widetilde H^{\dag} D_{(\mu}D_{\nu)} H)$
\end{tabular}
\end{minipage}
\end{adjustbox}
%%%%%%%%%%%%%%%%%%%%%%
\begin{adjustbox}{width=0.97\textwidth,center}
\small
%%%%%%%%%%%%
\begin{minipage}[t]{3.7cm}
\renewcommand{\arraystretch}{1.5}
\begin{tabular}[t]{c|c}
\multicolumn{2}{c}{\boldmath$12:\psi^2H^5 + \hc$} \\
\hline
$Q_{leH^5}$  & $(H^\dag H)^2 (\bar l_p e_r H)$ \\
$Q_{quH^5}$  & $(H^\dag H)^2 (\bar q_p u_r \widetilde H )$ \\
$Q_{qdH^5}$  & $(H^\dag H)^2 (\bar q_p d_r H)$
\end{tabular}
\end{minipage}
%%%%%%%%%%%%
\hspace{1cm}
\begin{minipage}[t]{9.9cm}
\renewcommand{\arraystretch}{1.5}
\begin{tabular}[t]{c|c}
\multicolumn{2}{c}{\boldmath$13:\psi^2H^4D$} \\
\hline
$Q_{l^2H^4D}^{(1)}$  &  $ i (l_p \gamma^{\mu} l_r) (H^{\dag} \overleftrightarrow{D}_{\mu} H) (H^{\dag} H)$ \\
$Q_{l^2H^4D}^{(2)}$  &  $ i (l_p \gamma^{\mu} \tau^I l_r) [(H^{\dag} \overleftrightarrow{D}_{\mu}^I H) (H^{\dag} H) + (H^{\dag} \overleftrightarrow{D}_{\mu} H) (H^{\dag} \tau^I H)]$ \\
$Q_{l^2H^4D}^{(3)}$  &  $ i \epsilon^{IJK} (l_p \gamma^{\mu} \tau^I l_r) (H^{\dag} \overleftrightarrow{D}_{\mu}^J H) (H^{\dag} \tau^K H)$ \\
$Q_{l^2H^4D}^{(4)}$  &  $ \epsilon^{IJK} (l_p \gamma^{\mu} \tau^I l_r) (H^{\dag} \tau^J H) D_{\mu} (H^{\dag} \tau^K H)$ \\
$Q_{e^2H^4D}$  &  $ i (e_p \gamma^{\mu} e_r) (H^{\dag} \overleftrightarrow{D}_{\mu} H) (H^{\dag} H)$ \\
$Q_{q^2H^4D}^{(1)}$  &  $ i (q_p \gamma^{\mu} q_r) (H^{\dag} \overleftrightarrow{D}_{\mu} H) (H^{\dag} H)$ \\
$Q_{q^2H^4D}^{(2)}$  &  $ i (q_p \gamma^{\mu} \tau^I q_r) [(H^{\dag} \overleftrightarrow{D}_{\mu}^I H) (H^{\dag} H) + (H^{\dag} \overleftrightarrow{D}_{\mu} H) (H^{\dag} \tau^I H)]$ \\
$Q_{q^2H^4D}^{(3)}$  &  $ i \epsilon^{IJK} (q_p \gamma^{\mu} \tau^I q_r) (H^{\dag} \overleftrightarrow{D}_{\mu}^J H) (H^{\dag} \tau^K H)$ \\
$Q_{q^2H^4D}^{(4)}$  &  $ \epsilon^{IJK} (q_p \gamma^{\mu} \tau^I q_r) (H^{\dag} \tau^J H) D_{\mu} (H^{\dag} \tau^K H)$ \\
$Q_{u^2H^4D}$  &  $ i (u_p \gamma^{\mu} u_r) (H^{\dag} \overleftrightarrow{D}_{\mu} H) (H^{\dag} H)$ \\
$Q_{d^2H^4D}$  &  $ i (d_p \gamma^{\mu} d_r) (H^{\dag} \overleftrightarrow{D}_{\mu} H) (H^{\dag} H)$ \\
$Q_{udH^4D} + \hc$  &  $ i (u_p \gamma^{\mu} d_r) (\widetilde H^{\dag} \overleftrightarrow{D}_{\mu} H) (H^{\dag} H)$
\end{tabular}
\end{minipage}
\end{adjustbox}
\end{center}
\caption{The dimension-eight operators in the SMEFT of classes-10, through -13, all of which have two fermions. The operators $Q_{udH^4D}$ and $Q_{udH^2D^3}$ have Hermitian conjugates. The subscripts $p, r$ are weak-eigenstate indices.}
\label{tab:smeft8class_10_11_12_13}
\end{table}

%%%%%%%%%%%%%%%%%%%%%%%%%%%%%%%%%%%%%%%%%%
% SMEFT d=8 Class 14 q
%%%%%%%%%%%%%%%%%%%%%%%%%%%%%%%%%%%%%%%%%%
\begin{table}[H]
\begin{center}
%%%%%%%%%%%%%%%%%%%%%%
\begin{adjustbox}{width=0.95\textwidth,center}
\small
%%%%%%%%%%%%
\begin{minipage}[t]{5.9cm}
\renewcommand{\arraystretch}{1.5}
\begin{tabular}[t]{c|c}
\multicolumn{2}{c}{\boldmath$14:\psi^2X^2D$} \\
\hline
$Q_{q^2G^2D}^{(1)}$  &  $i (\bar q_p \gamma^\mu \overleftrightarrow{D}^\nu q_r) G_{\mu\rho}^A G_{\nu}^{A\rho}$ \\
$Q_{q^2G^2D}^{(2)}$  &  $f^{ABC} (\bar q_p \gamma^\mu T^A \overleftrightarrow{D}^\nu q_r) G_{\mu\rho}^B G_{\nu}^{C\rho}$ \\
$Q_{q^2G^2D}^{(3)}$  &  $i d^{ABC} (\bar q_p \gamma^\mu T^A \overleftrightarrow{D}^\nu q_r) G_{\mu\rho}^B G_{\nu}^{C\rho}$ \\
$Q_{q^2W^2D}^{(1)}$  &  $i (\bar q_p \gamma^\mu \overleftrightarrow{D}^\nu q_r) W_{\mu\rho}^I W_{\nu}^{I\rho}$ \\
$Q_{q^2W^2D}^{(2)}$  &  $\epsilon^{IJK} (\bar q_p \gamma^\mu \tau^I \overleftrightarrow{D}^\nu q_r) W_{\mu\rho}^J W_{\nu}^{K\rho}$ \\
$Q_{q^2B^2D}$  &  $i (\bar q_p \gamma^\mu \overleftrightarrow{D}^\nu q_r) B_{\mu\rho} B_\nu^{\,\,\,\rho}$  \\
$Q_{u^2G^2D}^{(1)}$  &  $i (\bar u_p \gamma^\mu \overleftrightarrow{D}^\nu u_r) G_{\mu\rho}^A G_{\nu}^{A\rho}$ \\
$Q_{u^2G^2D}^{(2)}$  &  $f^{ABC} (\bar u_p \gamma^\mu T^A \overleftrightarrow{D}^\nu u_r) G_{\mu\rho}^B G_{\nu}^{C\rho}$ \\
$Q_{u^2G^2D}^{(3)}$  &  $i d^{ABC} (\bar u_p \gamma^\mu T^A \overleftrightarrow{D}^\nu u_r) G_{\mu\rho}^B G_{\nu}^{C\rho}$ \\
$Q_{u^2W^2D}$  &  $i (\bar u_p \gamma^\mu \overleftrightarrow{D}^\nu u_r) W_{\mu\rho}^I W_{\nu}^{I\rho}$ \\
$Q_{u^2B^2D}$  &  $i (\bar u_p \gamma^\mu \overleftrightarrow{D}^\nu u_r) B_{\mu\rho} B_\nu^{\,\,\,\rho}$ \\
$Q_{d^2G^2D}^{(1)}$  &  $i (\bar d_p \gamma^\mu \overleftrightarrow{D}^\nu d_r) G_{\mu\rho}^A G_{\nu}^{A\rho}$ \\
$Q_{d^2G^2D}^{(2)}$  &  $f^{ABC} (\bar d_p \gamma^\mu T^A \overleftrightarrow{D}^\nu d_r) G_{\mu\rho}^B G_{\nu}^{C\rho}$ \\
$Q_{d^2G^2D}^{(3)}$  &  $i d^{ABC} (\bar d_p \gamma^\mu T^A \overleftrightarrow{D}^\nu d_r) G_{\mu\rho}^B G_{\nu}^{C\rho}$ \\
$Q_{d^2W^2D}$  &  $i (\bar d_p \gamma^\mu \overleftrightarrow{D}^\nu d_r) W_{\mu\rho}^I W_{\nu}^{I\rho}$ \\
$Q_{d^2B^2D}$  &  $i (\bar d_p \gamma^\mu \overleftrightarrow{D}^\nu d_r) B_{\mu\rho} B_\nu^{\,\,\,\rho}$ 
\end{tabular}
\end{minipage}
\hspace{1cm}
%%%%%%%%%%%%
\begin{minipage}[t]{7.4cm}
\renewcommand{\arraystretch}{1.5}
\begin{tabular}[t]{c|c}
\multicolumn{2}{c}{\boldmath$14:\psi^2X^2D$} \\
\hline
$Q_{q^2G^2D}^{(4)}$  &  $i f^{ABC} (\bar q_p \gamma^\mu T^A \overleftrightarrow{D}^\nu q_r) (G_{\mu\rho}^B \widetilde{G}_{\nu}^{C\rho} - \widetilde{G}_{\mu\rho}^B G_\nu^{C \rho})$ \\
$Q_{q^2G^2D}^{(5)}$  &  $f^{ABC} (\bar q_p \gamma^\mu T^A \overleftrightarrow{D}^\nu q_r) (G_{\mu\rho}^B \widetilde{G}_{\nu}^{C\rho} + \widetilde{G}_{\mu\rho}^B G_\nu^{C \rho})$ \\
$Q_{q^2GWD}^{(1)}$  &  $(\bar q_p \gamma^\mu T^A \tau^I \overleftrightarrow{D}^\nu q_r) (G^A_{\mu\rho} W_{\nu}^{I\rho} - G^A_{\nu\rho} W_{\mu}^{I \rho})$ \\
$Q_{q^2GWD}^{(2)}$  &  $i (\bar q_p \gamma^\mu T^A \tau^I \overleftrightarrow{D}^\nu q_r) (G^A_{\mu\rho} W_{\nu}^{I\rho} + G^A_{\nu\rho} W_{\mu}^{I \rho})$ \\
$Q_{q^2GWD}^{(3)}$  &  $(\bar q_p \gamma^\mu T^A \tau^I \overleftrightarrow{D}^\nu q_r) (G_{\mu\rho}^A \widetilde{W}_\nu^{I \rho} - G_{\nu\rho}^A \widetilde{W}_\mu^{I \rho})$ \\
$Q_{q^2GWD}^{(4)}$  &  $i (\bar q_p \gamma^\mu T^A \tau^I \overleftrightarrow{D}^\nu q_r) (G_{\mu\rho}^A \widetilde{W}_\nu^{I \rho} + G_{\nu\rho}^A \widetilde{W}_\mu^{I \rho})$ \\
$Q_{q^2GBD}^{(1)}$  &  $(\bar q_p \gamma^\mu T^A \overleftrightarrow{D}^\nu q_r) (B_{\mu\rho} G_{\nu}^{A\rho} - B_{\nu\rho} G_{\mu}^{A\rho})$ \\
$Q_{q^2GBD}^{(2)}$  &  $i (\bar q_p \gamma^\mu T^A \overleftrightarrow{D}^\nu q_r) (B_{\mu\rho} G_{\nu}^{A\rho} + B_{\nu\rho} G_{\mu}^{A\rho})$ \\
$Q_{q^2GBD}^{(3)}$  &  $(\bar q_p \gamma^\mu T^A \overleftrightarrow{D}^\nu q_r) (B_{\mu\rho} \widetilde{G}_\nu^{A \rho} - B_{\nu\rho} \widetilde{G}_\mu^{A \rho})$ \\
$Q_{q^2GBD}^{(4)}$  &  $i (\bar q_p \gamma^\mu T^A \overleftrightarrow{D}^\nu q_r) (B_{\mu\rho} \widetilde{G}_\nu^{A \rho} + B_{\nu\rho} \widetilde{G}_\mu^{A \rho})$ \\
$Q_{q^2W^2D}^{(3)}$  &  $i \epsilon^{IJK} (\bar q_p \gamma^\mu \tau^I \overleftrightarrow{D}^\nu q_r) (W_{\mu\rho}^J \widetilde{W}_{\nu}^{K\rho} - \widetilde{W}_{\mu\rho}^J W_{\nu}^{K\rho})$ \\
$Q_{q^2W^2D}^{(4)}$  &  $\epsilon^{IJK} (\bar q_p \gamma^\mu \tau^I \overleftrightarrow{D}^\nu q_r) (W_{\mu\rho}^J \widetilde{W}_{\nu}^{K\rho} + \widetilde{W}_{\mu\rho}^J W_{\nu}^{K\rho})$ \\
$Q_{q^2WBD}^{(1)}$  &  $(\bar q_p \gamma^\mu \tau^I \overleftrightarrow{D}^\nu q_r) (B_{\mu\rho} W_{\nu}^{I\rho} - B_{\nu\rho} W_{\mu}^{I\rho})$ \\
$Q_{q^2WBD}^{(2)}$  &  $i (\bar q_p \gamma^\mu \tau^I \overleftrightarrow{D}^\nu q_r) (B_{\mu\rho} W_{\nu}^{I\rho} + B_{\nu\rho} W_{\mu}^{I\rho})$ \\
$Q_{q^2WBD}^{(3)}$  &  $(\bar q_p \gamma^\mu \tau^I \overleftrightarrow{D}^\nu q_r) (B_{\mu\rho}\widetilde{W}_\nu^{I \rho} - B_{\nu\rho} \widetilde W_\mu^{I \rho})$ \\
$Q_{q^2WBD}^{(4)}$  &  $i (\bar q_p \gamma^\mu \tau^I \overleftrightarrow{D}^\nu q_r) (B_{\mu\rho}\widetilde{W}_\nu^{I \rho} + B_{\nu\rho} \widetilde W_\mu^{I \rho})$ \\
$Q_{u^2G^2D}^{(4)}$  &  $i f^{ABC} (\bar u_p \gamma^\mu T^A \overleftrightarrow{D}^\nu u_r) (G_{\mu\rho}^B \widetilde{G}_{\nu}^{C\rho} - \widetilde{G}_{\mu\rho}^B G_{\nu}^{C\rho})$ \\
$Q_{u^2G^2D}^{(5)}$  &  $f^{ABC} (\bar u_p \gamma^\mu T^A \overleftrightarrow{D}^\nu u_r) (G_{\mu\rho}^B \widetilde{G}_{\nu}^{C\rho} + \widetilde{G}_{\mu\rho}^B G_{\nu}^{C\rho})$ \\
$Q_{u^2GBD}^{(1)}$  &  $(\bar u_p \gamma^\mu T^A \overleftrightarrow{D}^\nu u_r) (B_{\mu\rho} G_{\nu}^{A\rho} - B_{\nu\rho} G_{\mu}^{A\rho})$ \\
$Q_{u^2GBD}^{(2)}$  &  $i (\bar u_p \gamma^\mu T^A \overleftrightarrow{D}^\nu u_r) (B_{\mu\rho} G_{\nu}^{A\rho} + B_{\nu\rho} G_{\mu}^{A\rho})$ \\
$Q_{u^2GBD}^{(3)}$  &  $(\bar u_p \gamma^\mu T^A \overleftrightarrow{D}^\nu u_r) (B_{\mu\rho} \widetilde{G}_\nu^{A \rho} - B_{\nu\rho} \widetilde{G}_\mu^{A \rho})$ \\
$Q_{u^2GBD}^{(4)}$  &  $i (\bar u_p \gamma^\mu T^A \overleftrightarrow{D}^\nu u_r) (B_{\mu\rho} \widetilde{G}_\nu^{A \rho} + B_{\nu\rho} \widetilde{G}_\mu^{A \rho})$ \\
$Q_{d^2G^2D}^{(4)}$  &  $i f^{ABC} (\bar d_p \gamma^\mu T^A \overleftrightarrow{D}^\nu d_r) (G_{\mu\rho}^B \widetilde{G}_{\nu}^{C\rho} - \widetilde{G}_{\mu\rho}^B G_{\nu}^{C\rho})$ \\
$Q_{d^2G^2D}^{(5)}$  &  $f^{ABC} (\bar d_p \gamma^\mu T^A \overleftrightarrow{D}^\nu d_r) (G_{\mu\rho}^B \widetilde{G}_{\nu}^{C\rho} + \widetilde{G}_{\mu\rho}^B G_{\nu}^{C\rho})$ \\
$Q_{d^2GBD}^{(1)}$  &  $(\bar d_p \gamma^\mu T^A \overleftrightarrow{D}^\nu d_r) (B_{\mu\rho} G_{\nu}^{A\rho} - B_{\nu\rho} G_{\mu}^{A\rho})$ \\
$Q_{d^2GBD}^{(2)}$  &  $i (\bar d_p \gamma^\mu T^A \overleftrightarrow{D}^\nu d_r) (B_{\mu\rho} G_{\nu}^{A\rho} + B_{\nu\rho} G_{\mu}^{A\rho})$ \\
$Q_{d^2GBD}^{(3)}$  &  $(\bar d_p \gamma^\mu T^A \overleftrightarrow{D}^\nu d_r) (B_{\mu\rho} \widetilde{G}_\nu^{A \rho} - B_{\nu\rho} \widetilde{G}_\mu^{A \rho})$ \\
$Q_{d^2GBD}^{(4)}$  &  $i (\bar d_p \gamma^\mu T^A \overleftrightarrow{D}^\nu d_r) (B_{\mu\rho} \widetilde{G}_\nu^{A \rho} + B_{\nu\rho} \widetilde{G}_\mu^{A \rho})$ 
\end{tabular}
\end{minipage}
\end{adjustbox}
\end{center}
\caption{The hadronic dimension-eight operators in the SMEFT of class-14. The subscripts $p, r$ are weak-eigenstate indices.}
\label{tab:smeft8class_14qud}
\end{table}

%%%%%%%%%%%%%%%%%%%%%%%%%%%%%%%%%%%%%%%%%%
% SMEFT d=8 Class 14 l; Class 15 (RR)
%%%%%%%%%%%%%%%%%%%%%%%%%%%%%%%%%%%%%%%%%%
\begin{table}[H]
\begin{center}
%%%%%%%%%%%%%%%%%%%%%%
\begin{adjustbox}{width=0.95\textwidth,center}
\small
%%%%%%%%%%%%
\begin{minipage}[t]{5.9cm}
\renewcommand{\arraystretch}{1.5}
\begin{tabular}[t]{c|c}
\multicolumn{2}{c}{\boldmath$14:\psi^2X^2D$} \\
\hline
$Q_{l^2G^2D}$  &  $i (\bar l_p \gamma^\mu \overleftrightarrow{D}^\nu l_r) G_{\mu\rho}^A G_\nu^{A\rho}$ \\
$Q_{l^2W^2D}^{(1)}$  &  $i (\bar l_p \gamma^\mu \overleftrightarrow{D}^\nu l_r) W_{\mu\rho}^I W_{\nu}^{I\rho}$ \\
$Q_{l^2W^2D}^{(2)}$  &  $\epsilon^{IJK} (\bar l_p \gamma^\mu \tau^I \overleftrightarrow{D}^\nu l_r) W_{\mu\rho}^J W_{\nu}^{K\rho}$ \\
$Q_{l^2B^2D}$  &  $i (\bar l_p \gamma^\mu \overleftrightarrow{D}^\nu l_r) B_{\mu\rho} B_\nu^{\,\,\,\rho}$ \\
$Q_{e^2G^2D}$  &  $i (\bar e_p \gamma^\mu \overleftrightarrow{D}^\nu e_r) G_{\mu\rho}^A G_\nu^{A\rho}$ \\
$Q_{e^2W^2D}$  &  $i (\bar e_p \gamma^\mu \overleftrightarrow{D}^\nu e_r) W_{\mu\rho}^I W_{\nu}^{I\rho}$ \\
$Q_{e^2B^2D}$  &  $i (\bar e_p \gamma^\mu \overleftrightarrow{D}^\nu e_r) B_{\mu\rho} B_\nu^{\,\,\,\rho}$
\end{tabular}
\end{minipage}
\hspace{1cm}
%%%%%%%%%%%%
\begin{minipage}[t]{7.4cm}
\renewcommand{\arraystretch}{1.5}
\begin{tabular}[t]{c|c}
\multicolumn{2}{c}{\boldmath$14:\psi^2X^2D$} \\
\hline
$Q_{l^2W^2D}^{(3)}$  &  $i \epsilon^{IJK} (\bar l_p \gamma^\mu \tau^I \overleftrightarrow{D}^\nu l_r) (W_{\mu\rho}^J \widetilde{W}_{\nu}^{K\rho} - \widetilde{W}_{\mu\rho}^J W_{\nu}^{K\rho})$ \\
$Q_{l^2W^2D}^{(4)}$  &  $\epsilon^{IJK} (\bar l_p \gamma^\mu \tau^I \overleftrightarrow{D}^\nu l_r) (W_{\mu\rho}^J \widetilde{W}_{\nu}^{K\rho} + \widetilde{W}_{\mu\rho}^J W_{\nu}^{K\rho})$ \\
$Q_{l^2WBD}^{(1)}$  &  $(\bar l_p \gamma^\mu \tau^I \overleftrightarrow{D}^\nu l_r) (B_{\mu\rho} W_{\nu}^{I\rho} - B_{\nu\rho} W_{\mu}^{I\rho})$ \\
$Q_{l^2WBD}^{(2)}$  &  $i (\bar l_p \gamma^\mu \tau^I \overleftrightarrow{D}^\nu l_r) (B_{\mu\rho} W_{\nu}^{I\rho} + B_{\nu\rho} W_{\mu}^{I\rho})$ \\
$Q_{l^2WBD}^{(3)}$  &  $(\bar l_p \gamma^\mu \tau^I \overleftrightarrow{D}^\nu l_r) (B_{\mu\rho}\widetilde{W}_\nu^{I \rho} - B_{\nu\rho} \widetilde W_\mu^{I \rho})$ \\
$Q_{l^2WBD}^{(4)}$  &  $i (\bar l_p \gamma^\mu \tau^I \overleftrightarrow{D}^\nu l_r) (B_{\mu\rho}\widetilde{W}_\nu^{I \rho} + B_{\nu\rho} \widetilde W_\mu^{I \rho})$ 
\end{tabular}
\end{minipage}
\end{adjustbox}
%%%%%%%%%%%%%%%%%%%%%%
\begin{adjustbox}{width=0.87\textwidth,center}
\small
%%%%%%%%%%%%
\begin{minipage}[t]{5.6cm}
\renewcommand{\arraystretch}{1.5}
\begin{tabular}[t]{c|c}
\multicolumn{2}{c}{\boldmath$15:(\bar R R)XH^2D$} \\
\hline
$Q_{e^2WH^2D}^{(1)}$  &  $(\bar{e}_p \gamma^\nu e_r) D^{\mu} (H^\dag \tau^I H) W_{\mu\nu}^I$ \\
$Q_{e^2WH^2D}^{(2)}$  &  $(\bar{e}_p \gamma^\nu e_r) D^{\mu} (H^\dag \tau^I H) \widetilde W_{\mu\nu}^I$ \\
$Q_{e^2WH^2D}^{(3)}$  &  $(\bar{e}_p \gamma^\nu e_r) (H^\dag \overleftrightarrow{D}^{I\mu} H) W_{\mu\nu}^I$ \\
$Q_{e^2WH^2D}^{(4)}$  &  $(\bar{e}_p \gamma^\nu e_r) (H^\dag \overleftrightarrow{D}^{I\mu} H) \widetilde W_{\mu\nu}^I$ \\
$Q_{e^2BH^2D}^{(1)}$  &  $(\bar{e}_p \gamma^\nu e_r) D^{\mu} (H^\dag H) B_{\mu\nu}$ \\
$Q_{e^2BH^2D}^{(2)}$  &  $(\bar{e}_p \gamma^\nu e_r) D^{\mu} (H^\dag H) \widetilde B_{\mu\nu}$ \\
$Q_{e^2BH^2D}^{(3)}$  &  $(\bar{e}_p \gamma^\nu e_r) (H^\dag \overleftrightarrow{D}^\mu H) B_{\mu\nu}$ \\
$Q_{e^2BH^2D}^{(4)}$  &  $(\bar{e}_p \gamma^\nu e_r) (H^\dag \overleftrightarrow{D}^\mu H) \widetilde B_{\mu\nu}$ \\
$Q_{u^2GH^2D}^{(1)}$  &  $(\bar{u}_p \gamma^\nu T^A u_r) D^{\mu} (H^\dag H) G_{\mu\nu}^A$ \\
$Q_{u^2GH^2D}^{(2)}$  &  $(\bar{u}_p \gamma^\nu T^A u_r) D^{\mu} (H^\dag H) \widetilde G_{\mu\nu}^A$ \\
$Q_{u^2GH^2D}^{(3)}$  &  $(\bar{u}_p \gamma^\nu T^A u_r) (H^\dag \overleftrightarrow{D}^\mu H) G_{\mu\nu}^A$ \\
$Q_{u^2GH^2D}^{(4)}$  &  $(\bar{u}_p \gamma^\nu T^A u_r) (H^\dag \overleftrightarrow{D}^\mu H) \widetilde G_{\mu\nu}^A$ \\
$Q_{u^2WH^2D}^{(1)}$  &  $(\bar{u}_p \gamma^\nu u_r) D^{\mu} (H^\dag \tau^I H) W_{\mu\nu}^I$ \\
$Q_{u^2WH^2D}^{(2)}$  &  $(\bar{u}_p \gamma^\nu u_r) D^{\mu} (H^\dag \tau^I H) \widetilde W_{\mu\nu}^I$ \\
$Q_{u^2WH^2D}^{(3)}$  &  $(\bar{u}_p \gamma^\nu u_r) (H^\dag \overleftrightarrow{D}^{I\mu} H) W_{\mu\nu}^I$ \\
$Q_{u^2WH^2D}^{(4)}$  &  $(\bar{u}_p \gamma^\nu u_r) (H^\dag \overleftrightarrow{D}^{I\mu} H) \widetilde W_{\mu\nu}^I$ \\
$Q_{u^2BH^2D}^{(1)}$  &  $(\bar{u}_p \gamma^\nu u_r) D^{\mu} (H^\dag H) B_{\mu\nu}$ \\
$Q_{u^2BH^2D}^{(2)}$  &  $(\bar{u}_p \gamma^\nu u_r) D^{\mu} (H^\dag H) \widetilde B_{\mu\nu}$ \\
$Q_{u^2BH^2D}^{(3)}$  &  $(\bar{u}_p \gamma^\nu u_r) (H^\dag \overleftrightarrow{D}^\mu H) B_{\mu\nu}$ \\
$Q_{u^2BH^2D}^{(4)}$  &  $(\bar{u}_p \gamma^\nu u_r) (H^\dag \overleftrightarrow{D}^\mu H) \widetilde B_{\mu\nu}$
\end{tabular}
\end{minipage}
%%%%%%%%%%%%
\hspace{1cm}
\begin{minipage}[t]{6.4cm}
\renewcommand{\arraystretch}{1.5}
\begin{tabular}[t]{c|c}
\multicolumn{2}{c}{\boldmath$15:(\bar R R)XH^2D$} \\
\hline
$Q_{d^2GH^2D}^{(1)}$  &  $(\bar{d}_p \gamma^\nu T^A d_r) D^{\mu} (H^\dag H) G_{\mu\nu}^A$ \\
$Q_{d^2GH^2D}^{(2)}$  &  $(\bar{d}_p \gamma^\nu T^A d_r) D^{\mu} (H^\dag H) \widetilde G_{\mu\nu}^A$ \\
$Q_{d^2GH^2D}^{(3)}$  &  $(\bar{d}_p \gamma^\nu T^A d_r) (H^\dag \overleftrightarrow{D}^\mu H) G_{\mu\nu}^A$ \\
$Q_{d^2GH^2D}^{(4)}$  &  $(\bar{d}_p \gamma^\nu T^A d_r) (H^\dag \overleftrightarrow{D}^\mu H) \widetilde G_{\mu\nu}^A$ \\
$Q_{d^2WH^2D}^{(1)}$  &  $(\bar{d}_p \gamma^\nu d_r) D^{\mu} (H^\dag \tau^I H) W_{\mu\nu}^I$ \\
$Q_{d^2WH^2D}^{(2)}$  &  $(\bar{d}_p \gamma^\nu d_r) D^{\mu} (H^\dag \tau^I H) \widetilde W_{\mu\nu}^I$ \\
$Q_{d^2WH^2D}^{(3)}$  &  $(\bar{d}_p \gamma^\nu d_r) (H^\dag \overleftrightarrow{D}^{I\mu} H) W_{\mu\nu}^I$ \\
$Q_{d^2WH^2D}^{(4)}$  &  $(\bar{d}_p \gamma^\nu d_r) (H^\dag \overleftrightarrow{D}^{I\mu} H) \widetilde W_{\mu\nu}^I$ \\
$Q_{d^2BH^2D}^{(1)}$  &  $(\bar{d}_p \gamma^\nu d_r) D^{\mu} (H^\dag H) B_{\mu\nu}$ \\
$Q_{d^2BH^2D}^{(2)}$  &  $(\bar{d}_p \gamma^\nu d_r) D^{\mu} (H^\dag H) \widetilde B_{\mu\nu}$ \\
$Q_{d^2BH^2D}^{(3)}$  &  $(\bar{d}_p \gamma^\nu d_r) (H^\dag \overleftrightarrow{D}^\mu H) B_{\mu\nu}$ \\
$Q_{d^2BH^2D}^{(4)}$  &  $(\bar{d}_p \gamma^\nu d_r) (H^\dag \overleftrightarrow{D}^\mu H) \widetilde B_{\mu\nu}$ \\
$Q_{udGH^2}^{(1)} + \hc$  &  $(\bar u_p \gamma^\nu T^A d_r) (\widetilde H^\dag \overleftrightarrow D^\mu H) G_{\mu\nu}^A $ \\
$Q_{udGH^2}^{(2)} + \hc$  &  $(\bar u_p \gamma^\nu T^A d_r) (\widetilde H^\dag \overleftrightarrow D^\mu H) \widetilde G_{\mu\nu}^A $ \\
$Q_{udWH^2}^{(1)} + \hc$  &  $(\bar u_p \gamma^\nu d_r) (\widetilde H^\dag \overleftrightarrow D^{I\mu} H) W_{\mu\nu}^I $ \\
$Q_{udWH^2}^{(2)} + \hc$  &  $(\bar u_p \gamma^\nu d_r) (\widetilde H^\dag \overleftrightarrow D^{I\mu} H) \widetilde W_{\mu\nu}^I $ \\
$Q_{udBH^2}^{(1)} + \hc$  &  $(\bar u_p \gamma^\nu d_r) (\widetilde H^\dag \overleftrightarrow{D}^\mu H) B_{\mu\nu}$ \\
$Q_{udBH^2}^{(2)} + \hc$  &  $(\bar u_p \gamma^\nu d_r) (\widetilde H^\dag \overleftrightarrow{D}^\mu H) \widetilde B_{\mu\nu}$
\end{tabular}
\end{minipage}
%%%%%%%
\end{adjustbox}
\end{center}
\caption{The leptonic dimension-eight operators in the SMEFT of class-14, and the dimension-eight operators of class-15 with field content $(\bar R R)X^2H$. The operators $Q_{udXH^2}$ have distinct Hermitian conjugates. The subscripts $p, r$ are weak-eigenstate indices.}
\label{tab:smeft8class_14le_15RR}
\end{table}

%%%%%%%%%%%%%%%%%%%%%%%%%%%%%%%%%%%%%%%%%%
% SMEFT d=8 Class 14 (LL)
%%%%%%%%%%%%%%%%%%%%%%%%%%%%%%%%%%%%%%%%%%
\begin{table}[H]
\begin{center}
%%%%%%%%%%%%%%%%%%%%%%
\begin{adjustbox}{width=0.91\textwidth,center}
\small
%%%%%%%%%%%%
\begin{minipage}[t]{6.3cm}
\renewcommand{\arraystretch}{1.5}
\begin{tabular}[t]{c|c}
\multicolumn{2}{c}{\boldmath$15:(\bar L L)XH^2D$} \\
\hline
$Q_{l^2WH^2D}^{(1)}$  &  $(\bar{l}_p \gamma^\nu l_r) D^{\mu} (H^\dag \tau^I H) W_{\mu\nu}^I$ \\
$Q_{l^2WH^2D}^{(2)}$  &  $(\bar{l}_p \gamma^\nu l_r) D^{\mu} (H^\dag \tau^I H) \widetilde W_{\mu\nu}^I$ \\
$Q_{l^2WH^2D}^{(3)}$  &  $(\bar{l}_p \gamma^\nu l_r) (H^\dag \overleftrightarrow{D}^{I\mu} H) W_{\mu\nu}^I$ \\
$Q_{l^2WH^2D}^{(4)}$  &  $(\bar{l}_p \gamma^\nu l_r) (H^\dag \overleftrightarrow{D}^{I\mu} H) \widetilde W_{\mu\nu}^I$ \\
$Q_{l^2WH^2D}^{(5)}$  &  $(\bar{l}_p \gamma^\nu \tau^I l_r) D^{\mu} (H^\dag H) W_{\mu\nu}^I$ \\
$Q_{l^2WH^2D}^{(6)}$  &  $(\bar{l}_p \gamma^\nu \tau^I l_r) D^{\mu} (H^\dag H) \widetilde W_{\mu\nu}^I$ \\
$Q_{l^2WH^2D}^{(7)}$  &  $(\bar{l}_p \gamma^\nu \tau^I l_r) (H^\dag \overleftrightarrow{D}^\mu H) W_{\mu\nu}^I$ \\
$Q_{l^2WH^2D}^{(8)}$  &  $(\bar{l}_p \gamma^\nu \tau^I l_r) (H^\dag \overleftrightarrow{D}^\mu H) \widetilde W_{\mu\nu}^I$ \\
$Q_{l^2WH^2D}^{(9)}$  &  $\epsilon^{IJK} (\bar{l}_p \gamma^\nu \tau^I l_r) D^{\mu} (H^\dag \tau^J H) W_{\mu\nu}^K$ \\
$Q_{l^2WH^2D}^{(10)}$  &  $\epsilon^{IJK} (\bar{l}_p \gamma^\nu \tau^I l_r) D^{\mu} (H^\dag \tau^J H) \widetilde W_{\mu\nu}^K$ \\
$Q_{l^2WH^2D}^{(11)}$  &  $\epsilon^{IJK} (\bar{l}_p \gamma^\nu \tau^I l_r) (H^\dag \overleftrightarrow{D}^{J\mu} H) W_{\mu\nu}^K$ \\
$Q_{l^2WH^2D}^{(12)}$  &  $\epsilon^{IJK} (\bar{l}_p \gamma^\nu \tau^I l_r) (H^\dag \overleftrightarrow{D}^{J\mu} H) \widetilde W_{\mu\nu}^K$ \\
$Q_{l^2BH^2D}^{(1)}$  &  $(\bar{l}_p \gamma^\nu \tau^I l_r) D^{\mu} (H^\dag \tau^I H) B_{\mu\nu}$ \\
$Q_{l^2BH^2D}^{(2)}$  &  $(\bar{l}_p \gamma^\nu \tau^I l_r) D^{\mu} (H^\dag \tau^I H) \widetilde B_{\mu\nu}$ \\
$Q_{l^2BH^2D}^{(3)}$  &  $(\bar{l}_p \gamma^\nu \tau^I l_r) (H^\dag \overleftrightarrow{D}^{I\mu} H) B_{\mu\nu}$ \\
$Q_{l^2BH^2D}^{(4)}$  &  $(\bar{l}_p \gamma^\nu \tau^I l_r) (H^\dag \overleftrightarrow{D}^{I\mu} H) \widetilde B_{\mu\nu}$ \\
$Q_{l^2BH^2D}^{(5)}$  &  $(\bar{l}_p \gamma^\nu l_r) D^{\mu} (H^\dag H) B_{\mu\nu}$ \\
$Q_{l^2BH^2D}^{(6)}$  &  $(\bar{l}_p \gamma^\nu l_r) D^{\mu} (H^\dag H) \widetilde B_{\mu\nu}$ \\
$Q_{l^2BH^2D}^{(7)}$  &  $(\bar{l}_p \gamma^\nu l_r) (H^\dag \overleftrightarrow{D}^\mu H) B_{\mu\nu}$ \\
$Q_{l^2BH^2D}^{(8)}$  &  $(\bar{l}_p \gamma^\nu l_r) (H^\dag \overleftrightarrow{D}^\mu H) \widetilde B_{\mu\nu}$ \\
\end{tabular}
\end{minipage}
%%%%%%%%%%%%
\hspace{1cm}
\begin{minipage}[t]{6.3cm}
\renewcommand{\arraystretch}{1.5}
\begin{tabular}[t]{c|c}
\multicolumn{2}{c}{\boldmath$15:(\bar L L)XH^2D$} \\
\hline
$Q_{q^2GH^2D}^{(1)}$  &  $(\bar{q}_p \gamma^\nu T^A \tau^I q_r) D^{\mu} (H^\dag \tau^I H) G_{\mu\nu}^A$ \\
$Q_{q^2GH^2D}^{(2)}$  &  $(\bar{q}_p \gamma^\nu T^A \tau^I q_r) D^{\mu} (H^\dag \tau^I H) \widetilde G_{\mu\nu}^A$ \\
$Q_{q^2GH^2D}^{(3)}$  &  $(\bar{q}_p \gamma^\nu T^A \tau^I q_r) (H^\dag \overleftrightarrow{D}^{I\mu} H) G_{\mu\nu}^A$ \\
$Q_{q^2GH^2D}^{(4)}$  &  $(\bar{q}_p \gamma^\nu T^A \tau^I q_r) (H^\dag \overleftrightarrow{D}^{I\mu} H) \widetilde G_{\mu\nu}^A$ \\
$Q_{q^2GH^2D}^{(5)}$  &  $(\bar{q}_p \gamma^\nu T^A q_r) D^{\mu} (H^\dag H) G_{\mu\nu}^A$ \\
$Q_{q^2GH^2D}^{(6)}$  &  $(\bar{q}_p \gamma^\nu T^A q_r) D^{\mu} (H^\dag H) \widetilde G_{\mu\nu}^A$ \\
$Q_{q^2GH^2D}^{(7)}$  &  $(\bar{q}_p \gamma^\nu T^A q_r) (H^\dag \overleftrightarrow{D}^\mu H) G_{\mu\nu}^A$ \\
$Q_{q^2GH^2D}^{(8)}$  &  $(\bar{q}_p \gamma^\nu T^A q_r) (H^\dag \overleftrightarrow{D}^\mu H) \widetilde G_{\mu\nu}^A$ \\
$Q_{q^2WH^2D}^{(1)}$  &  $(\bar{q}_p \gamma^\nu q_r) D^{\mu} (H^\dag \tau^I H) W_{\mu\nu}^I$ \\
$Q_{q^2WH^2D}^{(2)}$  &  $(\bar{q}_p \gamma^\nu q_r) D^{\mu} (H^\dag \tau^I H) \widetilde W_{\mu\nu}^I$ \\
$Q_{q^2WH^2D}^{(3)}$  &  $(\bar{q}_p \gamma^\nu q_r) (H^\dag \overleftrightarrow{D}^{I\mu} H) W_{\mu\nu}^I$ \\
$Q_{q^2WH^2D}^{(4)}$  &  $(\bar{q}_p \gamma^\nu q_r) (H^\dag \overleftrightarrow{D}^{I\mu} H) \widetilde W_{\mu\nu}^I$ \\
$Q_{q^2WH^2D}^{(5)}$  &  $(\bar{q}_p \gamma^\nu \tau^I q_r) D^{\mu} (H^\dag H) W_{\mu\nu}^I$ \\
$Q_{q^2WH^2D}^{(6)}$  &  $(\bar{q}_p \gamma^\nu \tau^I q_r) D^{\mu} (H^\dag H) \widetilde W_{\mu\nu}^I$ \\
$Q_{q^2WH^2D}^{(7)}$  &  $(\bar{q}_p \gamma^\nu \tau^I q_r) (H^\dag \overleftrightarrow{D}^\mu H) W_{\mu\nu}^I$ \\
$Q_{q^2WH^2D}^{(8)}$  &  $(\bar{q}_p \gamma^\nu \tau^I q_r) (H^\dag \overleftrightarrow{D}^\mu H) \widetilde W_{\mu\nu}^I$ \\
$Q_{q^2WH^2D}^{(9)}$  &  $\epsilon^{IJK} (\bar{q}_p \gamma^\nu \tau^I q_r) D^{\mu} (H^\dag \tau^J H) W_{\mu\nu}^K$ \\
$Q_{q^2WH^2D}^{(10)}$  &  $\epsilon^{IJK} (\bar{q}_p \gamma^\nu \tau^I q_r) D^{\mu} (H^\dag \tau^J H) \widetilde W_{\mu\nu}^K$ \\
$Q_{q^2WH^2D}^{(11)}$  &  $\epsilon^{IJK} (\bar{q}_p \gamma^\nu \tau^I q_r) (H^\dag \overleftrightarrow{D}^{J\mu} H) W_{\mu\nu}^K$ \\
$Q_{q^2WH^2D}^{(12)}$  &  $\epsilon^{IJK} (\bar{q}_p \gamma^\nu \tau^I q_r) (H^\dag \overleftrightarrow{D}^{J\mu} H) \widetilde W_{\mu\nu}^K$ \\
$Q_{q^2BH^2D}^{(1)}$  &  $(\bar{q}_p \gamma^\nu \tau^I q_r) D^{\mu} (H^\dag \tau^I H) B_{\mu\nu}$ \\
$Q_{q^2BH^2D}^{(2)}$  &  $(\bar{q}_p \gamma^\nu \tau^I q_r) D^{\mu} (H^\dag \tau^I H) \widetilde B_{\mu\nu}$ \\
$Q_{q^2BH^2D}^{(3)}$  &  $(\bar{q}_p \gamma^\nu \tau^I q_r) (H^\dag \overleftrightarrow{D}^{I\mu} H) B_{\mu\nu}$ \\
$Q_{q^2BH^2D}^{(4)}$  &  $(\bar{q}_p \gamma^\nu \tau^I q_r) (H^\dag \overleftrightarrow{D}^{I\mu} H) \widetilde B_{\mu\nu}$ \\
$Q_{q^2BH^2D}^{(5)}$  &  $(\bar{q}_p \gamma^\nu q_r) D^{\mu} (H^\dag H) B_{\mu\nu}$ \\
$Q_{q^2BH^2D}^{(6)}$  &  $(\bar{q}_p \gamma^\nu q_r) D^{\mu} (H^\dag H) \widetilde B_{\mu\nu}$ \\
$Q_{q^2BH^2D}^{(7)}$  &  $(\bar{q}_p \gamma^\nu q_r) (H^\dag \overleftrightarrow{D}^\mu H) B_{\mu\nu}$ \\
$Q_{q^2BH^2D}^{(8)}$  &  $(\bar{q}_p \gamma^\nu q_r) (H^\dag \overleftrightarrow{D}^\mu H) \widetilde B_{\mu\nu}$ \\
\end{tabular}
\end{minipage}
%%%%%%%
\end{adjustbox}
\end{center}
\caption{The dimension-eight operators in the SMEFT of class-15 with field content $(\bar L L)X^2H$. The subscripts $p, r$ are weak-eigenstate indices.}
\label{tab:smeft8class_15LL}
\end{table}

%%%%%%%%%%%%%%%%%%%%%%%%%%%%%%%%%%%%%%%%%%
% SMEFT d=8 Classes 16, 17
%%%%%%%%%%%%%%%%%%%%%%%%%%%%%%%%%%%%%%%%%%
\begin{table}[H]
\begin{center}
%%%%%%%%%%%%%%%%%%%%%%
\begin{adjustbox}{width=0.83\textwidth,center}
\small
%%%%%%%%%%%%
\begin{minipage}[t]{5.8cm}
\renewcommand{\arraystretch}{1.5}
\begin{tabular}[t]{c|c}
\multicolumn{2}{c}{\boldmath$16:\psi^2XHD^2 + \hc$} \\
\hline
$Q_{leWHD^2}^{(1)}$  &  $(\bar l_p \sigma^{\mu\nu} D^\rho e_r) \tau^I (D_\nu H) W^I_{\rho\mu}$ \\
$Q_{leWHD^2}^{(2)}$  &  $(\bar l_p D^\rho e_r) \tau^I (D^\nu H) \widetilde W^I_{\rho\nu}$ \\
$Q_{leWHD^2}^{(3)}$  &  $(\bar l_p \sigma^{\mu\nu} e_r) \tau^I (D^\rho H) (D_\rho W^I_{\mu\nu})$ \\
$Q_{leBHD^2}^{(1)}$  &  $(\bar l_p \sigma^{\mu\nu} D^\rho e_r) (D_\nu H) B_{\rho\mu}$ \\
$Q_{leBHD^2}^{(2)}$  &  $(\bar l_p D^\rho e_r) (D^\nu H) \widetilde B_{\rho\nu}$ \\
$Q_{leBHD^2}^{(3)}$  &  $(\bar l_p \sigma^{\mu\nu} e_r) (D^\rho H) (D_\rho B_{\mu\nu})$ \\
$Q_{quGHD^2}^{(1)}$  &  $(\bar q_p \sigma^{\mu\nu} T^A D^\rho u_r) (D_\nu \widetilde H) G^A_{\rho\mu}$ \\
$Q_{quGHD^2}^{(2)}$  &  $(\bar q_p T^A D^\rho u_r) (D^\nu \widetilde H) \widetilde G^A_{\rho\nu}$ \\
$Q_{quGHD^2}^{(3)}$  &  $(\bar q_p \sigma^{\mu\nu} T^A u_r) (D^\rho \widetilde H) (D_\rho G^A_{\mu\nu})$ \\
$Q_{quWHD^2}^{(1)}$  &  $(\bar q_p \sigma^{\mu\nu} D^\rho u_r) \tau^I (D_\nu \widetilde H) W^I_{\rho\mu}$ \\
$Q_{quWHD^2}^{(2)}$  &  $(\bar q_p D^\rho u_r) \tau^I (D^\nu \widetilde H) \widetilde W^I_{\rho\nu}$ \\
$Q_{quWHD^2}^{(3)}$  &  $(\bar q_p \sigma^{\mu\nu} u_r) \tau^I (D^\rho \widetilde H) (D_\rho W^I_{\mu\nu})$ \\
$Q_{quBHD^2}^{(1)}$  &  $(\bar q_p \sigma^{\mu\nu} D^\rho u_r) (D_\nu \widetilde H) B_{\rho\mu}$ \\
$Q_{quBHD^2}^{(2)}$  &  $(\bar q_p D^\rho u_r) (D^\nu \widetilde H) \widetilde B_{\rho\nu}$ \\
$Q_{quBHD^2}^{(3)}$  &  $(\bar q_p \sigma^{\mu\nu} u_r) (D^\rho \widetilde H) (D_\rho B_{\mu\nu})$ \\
$Q_{qdGHD^2}^{(1)}$  &  $(\bar q_p \sigma^{\mu\nu} T^A D^\rho d_r) (D_\nu H) G^A_{\rho\mu}$ \\
$Q_{qdGHD^2}^{(2)}$  &  $(\bar q_p  T^A D^\rho d_r) (D^\nu H) \widetilde G^A_{\rho\nu}$ \\
$Q_{qdGHD^2}^{(3)}$  &  $(\bar q_p \sigma^{\mu\nu} T^A d_r) (D^\rho H) (D_\rho G^A_{\mu\nu})$ \\
$Q_{qdWHD^2}^{(1)}$  &  $(\bar q_p \sigma^{\mu\nu} D^\rho d_r) \tau^I (D_\nu H) W^I_{\rho\mu}$ \\
$Q_{qdWHD^2}^{(2)}$  &  $(\bar q_p  D^\rho d_r) \tau^I (D^\nu H) \widetilde W^I_{\rho\nu}$ \\
$Q_{qdWHD^2}^{(3)}$  &  $(\bar q_p \sigma^{\mu\nu} d_r) \tau^I (D^\rho H) (D_\rho W^I_{\mu\nu})$ \\
$Q_{qdBHD^2}^{(1)}$  &  $(\bar q_p \sigma^{\mu\nu} D^\rho d_r) (D_\nu H) B_{\rho\mu}$ \\
$Q_{qdBHD^2}^{(2)}$  &  $(\bar q_p D^\rho d_r) (D^\nu H) \widetilde B_{\rho\nu}$ \\
$Q_{qdBHD^2}^{(3)}$  &  $(\bar q_p \sigma^{\mu\nu} d_r) (D^\rho H) (D_\rho B_{\mu\nu})$ 
\end{tabular}
\end{minipage}
%%%%%%%%%%%%
\hspace{1cm}
\begin{minipage}[t]{5.7cm}
\renewcommand{\arraystretch}{1.5}
\begin{tabular}[t]{c|c}
\multicolumn{2}{c}{\boldmath$17:\psi^2H^3D^2 + \hc$} \\
\hline
$Q_{leH^3D^2}^{(1)}$  & $(D_\mu H^\dag D^\mu H) (\bar l_p e_r H)$ \\
$Q_{leH^3D^2}^{(2)}$  & $(D_\mu H^\dag \tau^I D^\mu H) (\bar l_p e_r \tau^I H)$ \\
$Q_{leH^3D^2}^{(3)}$  & $(D_\mu H^\dag D_\nu H) (\bar l_p \sigma^{\mu\nu} e_r H)$ \\
$Q_{leH^3D^2}^{(4)}$  & $(D_\mu H^\dag \tau^I D_\nu H) (\bar l_p \sigma^{\mu\nu} e_r \tau^I H)$ \\
$Q_{leH^3D^2}^{(5)}$  & $(H^\dag D_\mu H) (\bar l_p e_r D^\mu H)$ \\
$Q_{leH^3D^2}^{(6)}$  & $(H^\dag D_\mu H) (\bar l_p \sigma^{\mu\nu} e_r D_\nu H)$ \\
$Q_{quH^3D^2}^{(1)}$  & $(D_\mu H^\dag D^\mu H) (\bar q_p u_r \widetilde H)$ \\
$Q_{quH^3D^2}^{(2)}$  & $(D_\mu H^\dag \tau^I D^\mu H) (\bar q_p u_r \tau^I \widetilde H)$ \\
$Q_{quH^3D^2}^{(3)}$  & $(D_\mu H^\dag D_\nu H) (\bar q_p \sigma^{\mu\nu} u_r \widetilde H)$ \\
$Q_{quH^3D^2}^{(4)}$  & $(D_\mu H^\dag \tau^I D_\nu H) (\bar q_p \sigma^{\mu\nu} u_r \tau^I \widetilde H)$ \\
$Q_{quH^3D^2}^{(5)}$  & $(D_\mu H^\dag H) (\bar q_p u_r D^\mu \widetilde H)$ \\
$Q_{quH^3D^2}^{(6)}$  & $(D_\mu H^\dag H) (\bar q_p \sigma^{\mu\nu} u_r D_\nu \widetilde H)$ \\
$Q_{qdH^3D^2}^{(1)}$  & $(D_\mu H^\dag D^\mu H) (\bar q_p d_r H)$ \\
$Q_{qdH^3D^2}^{(2)}$  & $(D_\mu H^\dag \tau^I D^\mu H) (\bar q_p d_r \tau^I H)$ \\
$Q_{qdH^3D^2}^{(3)}$  & $(D_\mu H^\dag D_\nu H) (\bar q_p \sigma^{\mu\nu} d_r H)$ \\
$Q_{qdH^3D^2}^{(4)}$  & $(D_\mu H^\dag \tau^I D_\nu H) (\bar q_p \sigma^{\mu\nu} d_r \tau^I H)$ \\
$Q_{qdH^3D^2}^{(5)}$  & $(H^\dag D_\mu H) (\bar q_p d_r D^\mu H)$ \\
$Q_{qdH^3D^2}^{(6)}$  & $(H^\dag D_\mu H) (\bar q_p \sigma^{\mu\nu} d_r D_\nu H)$ 
\end{tabular}
\end{minipage}
\end{adjustbox}
\end{center}
\caption{The dimension-eight operators in the SMEFT of classes-16, and -17, which have two fermions and two derivates. All of the operators have Hermitian conjugates. The subscripts $p, r$ are weak-eigenstate indices.}
\label{tab:smeft8class_16_17}
\end{table}

%--------------------------------------------------------------------------------------------------------------
\subsection{Results for Four-Fermion Operators}

See Tables~\ref{tab:smeft8class_18_21}, \ref{tab:smeft8class_18},  \ref{tab:smeft8class_19_LL_RR}, \ref{tab:smeft8class_19_LLRR}, \ref{tab:smeft8class_19_LRRL_LRLR}, \ref{tab:smeft8class_19_slashedB}, \ref{tab:smeft8class_20_le_qu}, \ref{tab:smeft8class_20_qd_slashedB}, and \ref{tab:smeft8class_21}.

%%%%%%%%%%%%%%%%%%%%%%%%%%%%%%%%%%%%%%%%%%
% SMEFT d=8 Class 18, 21 (LR)(RL), \slashed B
%%%%%%%%%%%%%%%%%%%%%%%%%%%%%%%%%%%%%%%%%%

\begin{table}[H]
\begin{center}
%%%%%%%%%%%%%%%%%%%%%%
\begin{adjustbox}{width=0.9\textwidth,center}
\small
%%%%%%%%%%%%
\begin{minipage}[t]{4.9cm}
\renewcommand{\arraystretch}{1.5}
\begin{tabular}[t]{c|c}
\multicolumn{2}{c}{\boldmath$18:(\bar L R)(\bar R L)H^2 + \hc$} \\
\hline
$Q_{leqdH^2}^{(1)}$  &  $(\bar l_p^j e_r) (\bar d_s q_{tj}) (H^\dag H)$ \\
$Q_{leqdH^2}^{(2)}$  &  $(\bar l_p e_r) \tau^I (\bar d_s q_t) (H^\dag \tau^I H)$ \\
$Q_{l^2udH^2}$  &  $(\bar l_p d_r H) (\widetilde H^\dag \bar u_s l_t)$ \\
$Q_{lequH^2}^{(5)}$  &  $(\bar l_p e_r H) (\widetilde H^\dag \bar u_s q_t)$ \\
$Q_{q^2udH^2}^{(5)}$  &  $(\bar q_p d_r H) (\widetilde H^\dag \bar u_s q_t)$ \\
$Q_{q^2udH^2}^{(6)}$  &  $(\bar q_p T^A d_r H) (\widetilde H^\dag \bar u_s T^A q_t)$
\end{tabular}
\end{minipage}
\hspace{1cm}
%%%%%%%%%%%%
\begin{minipage}[t]{7.6cm}
\renewcommand{\arraystretch}{1.5}
\begin{tabular}[t]{c|c}
\multicolumn{2}{c}{\boldmath$18(\slashed B):\psi^4H^2 + \hc$} \\
\hline
$Q_{lqudH^2}^{(1)}$  &  $\epsilon_{\alpha\beta\gamma} \epsilon_{jk} (d_p^{\alpha} C u_r^{\beta}) (q_s^{j\gamma} C l_t^k) (H^\dag H)$ \\
$Q_{lqudH^2}^{(2)}$  &  $\epsilon_{\alpha\beta\gamma} (\epsilon \tau^I)_{jk} (d_p^{\alpha} C u_r^{\beta}) (q_s^{j\gamma} C l_t^k) (H^\dag \tau^I H)$ \\
$Q_{eq^2uH^2}$  &  $\epsilon_{\alpha\beta\gamma} \epsilon_{jk}  (q_p^{j\alpha} C q_r^{m\beta}) (u_s^{\gamma} C e_t) (H^\dag_m H^k)$ \\
$Q_{lq^3H^2}^{(1)}$  &  $\epsilon_{\alpha\beta\gamma} \epsilon_{mn} \epsilon_{jk} (q_p^{m\alpha} C q_r^{j\beta}) (q_s^{k\gamma} C l_t^n) (H^\dag H)$ \\
$Q_{lq^3H^2}^{(2)}$  &  $\epsilon_{\alpha\beta\gamma} (\epsilon \tau^I)_{mn} \epsilon_{jk} (q_p^{m\alpha} C q_r^{j\beta}) (q_s^{k\gamma} C l_t^n) (H^\dag \tau^I H)$ \\
$Q_{eu^2dH^2}$  &  $\epsilon_{\alpha\beta\gamma} (d_p^{\alpha} C u_r^{\beta}) (u_s^{\gamma} C e_t) (H^\dag H)$ \\ \hdashline
$Q_{lq^3H^2}^{(3)}$  &  $\epsilon_{\alpha\beta\gamma} \epsilon_{mn} (\epsilon \tau^I)_{jk} (q_p^{m\alpha} C q_r^{j\beta}) (q_s^{k\gamma} C l_t^n) (H^\dag \tau^I H)$ \\
$Q_{lqu^2H^2}$  &  $\epsilon_{\alpha\beta\gamma} \epsilon_{jk} \epsilon_{mn} (l_p^j C q_r^{m\alpha}) (u_s^\beta C u_t^\gamma) \widetilde H^k \widetilde H^n$ \\
$Q_{lqd^2H^2}$  &  $\epsilon_{\alpha\beta\gamma} \epsilon_{jk} \epsilon_{mn} (l_p^j q_r^{m\alpha}) (d_s^\beta C d_t^\gamma) H^k H^n$ \\
$Q_{eq^2dH^2}$  &  $\epsilon_{\alpha\beta\gamma} \epsilon_{jk} \epsilon_{mn} (e_p d_r^\alpha) (q_s^{j\beta} C q_t^{m\gamma}) H^k H^n$
\end{tabular}
\end{minipage}
%%%%%%%
\end{adjustbox}
%%%%%%%%%%%%%%%%%%%%%%
\begin{adjustbox}{width=0.81\textwidth,center}
\small
%%%%%%%%%%%%
\begin{minipage}[t]{4.4cm}
\renewcommand{\arraystretch}{1.5}
\begin{tabular}[t]{c|c}
\multicolumn{2}{c}{\boldmath$21:(\bar L R)(\bar R L)D^2 + \hc$} \\
\hline
$Q_{leqdD^2}^{(1)}$  &  $D_\mu (\bar l_p^j e_r) D^\mu (\bar d_s q_{tj})$  \\
$Q_{leqdD^2}^{(2)}$  &  $(\bar l_p^j \overleftrightarrow{D}_\mu e_r) (\bar d_s \overleftrightarrow{D}^\mu q_{tj})$  
\end{tabular}
\end{minipage}
\hspace{1cm}
%%%%%%%%%%%%
\begin{minipage}[t]{6.7cm}
\renewcommand{\arraystretch}{1.5}
\begin{tabular}[t]{c|c}
\multicolumn{2}{c}{\boldmath$21(\slashed B):\psi^4D^2 + \hc$} \\
\hline
$Q_{lqudD^2}^{(1)}$  &  $\epsilon_{\alpha\beta\gamma} \epsilon_{jk} D_\mu (d_p^{\alpha} C u_r^{\beta}) D^\mu (q_s^{j\gamma} C l_t^k)$  \\
$Q_{lqudD^2}^{(2)}$  &  $\epsilon_{\alpha\beta\gamma} \epsilon_{jk} D_\mu (d_p^{\alpha} C q_r^{j\beta}) D^\mu (u_s^{\gamma} C l_t^k)$  \\
$Q_{eq^2uD^2}$  &  $\epsilon_{\alpha\beta\gamma} \epsilon_{jk} (q_p^{j\alpha} C D_\mu q_r^{k\beta}) D^\mu (u_s^{\gamma} C e_t)$  \\
$Q_{lq^3D^2}$  &  $\epsilon_{\alpha\beta\gamma} \epsilon_{mn} \epsilon_{jk} (q_p^{m\alpha} C D_\mu q_r^{j\beta}) D^\mu (q_s^{k\gamma} C l_t^n)$  \\
$Q_{eu^2dD^2}^{(1)}$  &  $\epsilon_{\alpha\beta\gamma} (u_p^\alpha C D_\mu u_r^\beta) D^\mu (d_s^\gamma C e_t)$  \\ \hdashline
$Q_{eu^2dD^2}^{(2)}$  &  $\epsilon_{\alpha\beta\gamma} (u_p^\alpha C u_r^\beta) (D_\mu d_s^\gamma C D^\mu e_t)$
\end{tabular}
\end{minipage}
%%%%%%%%%
\end{adjustbox}
\end{center}
\caption{The dimension-8 operators of classes-18 and -21 whose fermionic content either has the chiral structure $(\bar L R) (\bar R L)$ or is baryon number violating.
The subscripts $p, r, s, t$ are weak-eigenstate indices. 
Operators below the dashed lines vanish when there is only one generation of fermions.}
\label{tab:smeft8class_18_21}
\end{table}

%%%%%%%%%%%%%%%%%%%%%%%%%%%%%%%%%%%%%%%%%%
% SMEFT d=8 Class 18 (LL)(LL), (RR)(RR), (LL)(RR), (LR(LR)
%%%%%%%%%%%%%%%%%%%%%%%%%%%%%%%%%%%%%%%%%%
\begin{table}[H]
\begin{center}
%%%%%%%%%%%%%%%%%%%%%%
\begin{adjustbox}{width=0.89\textwidth,center}
\small
%%%%%%%%%%%%
\begin{minipage}[t]{6.4cm}
\renewcommand{\arraystretch}{1.5}
\begin{tabular}[t]{c|c}
\multicolumn{2}{c}{\boldmath$18:(\bar L L)(\bar L L)H^2$} \\
\hline
$Q_{l^4H^2}^{(1)}$  &  $(\bar l_p \gamma^\mu l_r) (\bar l_s \gamma_\mu l_t) (H^\dag H)$ \\
$Q_{l^4H^2}^{(2)}$  &  $(\bar l_p \gamma^\mu l_r) (\bar l_s \gamma_\mu \tau^I l_t) (H^\dag \tau^I H)$ \\
$Q_{q^4H^2}^{(1)}$  & $(\bar q_p \gamma^\mu q_r) (\bar q_s \gamma_\mu q_t) (H^\dag H)$ \\
$Q_{q^4H^2}^{(2)}$  & $(\bar q_p \gamma^\mu q_r) (\bar q_s \gamma_\mu \tau^I q_t) (H^\dag \tau^I H)$ \\
$Q_{q^4H^2}^{(3)}$  & $(\bar q_p \gamma^\mu \tau^I q_r) (\bar q_s \gamma_\mu \tau^I q_t) (H^\dag H)$ \\
$Q_{l^2q^2H^2}^{(1)}$  & $(\bar l_p \gamma^\mu l_r) (\bar q_s \gamma_\mu q_t) (H^\dag H)$ \\
$Q_{l^2q^2H^2}^{(2)}$  & $(\bar l_p \gamma^\mu \tau^I l_r) (\bar q_s \gamma_\mu q_t) (H^\dag \tau^I H)$ \\
$Q_{l^2q^2H^2}^{(3)}$  & $(\bar l_p \gamma^\mu \tau^I l_r) (\bar q_s \gamma_\mu \tau^I q_t) (H^\dag H)$ \\
$Q_{l^2q^2H^2}^{(4)}$  & $(\bar l_p \gamma^\mu l_r) (\bar q_s \gamma_\mu \tau^I q_t) (H^\dag \tau^I H)$ \\
$Q_{l^2q^2H^2}^{(5)}$  & $\epsilon^{IJK} (\bar l_p \gamma^\mu \tau^I l_r) (\bar q_s \gamma_\mu \tau^J q_t) (H^\dag \tau^K H)$ \\ \hdashline
$Q_{q^4H^2}^{(5)}$  & $\epsilon^{IJK} (\bar q_p \gamma^\mu \tau^I q_r) (\bar q_s \gamma_\mu \tau^J q_t) (H^\dag \tau^K H)$
\end{tabular}
\end{minipage}
%%%%%%%%%%%%
\hspace{1cm}
\begin{minipage}[t]{5.9cm}
\renewcommand{\arraystretch}{1.5}
\begin{tabular}[t]{c|c}
\multicolumn{2}{c}{\boldmath$18:(\bar R R)(\bar R R)H^2$} \\
\hline
$Q_{e^4H^2}$  &  $(\bar e_p \gamma^\mu e_r) (\bar e_s \gamma_\mu e_t) (H^\dag H)$ \\
$Q_{u^4H^2}$  &  $(\bar u_p \gamma^\mu u_r) (\bar u_s \gamma_\mu u_t) (H^\dag H)$ \\
$Q_{d^4H^2}$  &  $(\bar d_p \gamma^\mu d_r) (\bar d_s \gamma_\mu d_t) (H^\dag H)$ \\
$Q_{e^2u^2H^2}$  &  $(\bar e_p \gamma^\mu e_r) (\bar u_s \gamma_\mu u_t) (H^\dag H)$ \\
$Q_{e^2d^2H^2}$  &  $(\bar e_p \gamma^\mu e_r) (\bar d_s\gamma_\mu d_t) (H^\dag H)$ \\
$Q_{u^2d^2H^2}^{(1)}$  &  $(\bar u_p \gamma^\mu u_r) (\bar d_s \gamma_\mu d_t) (H^\dag H)$ \\
$Q_{u^2d^2H^2}^{(2)}$  &  $(\bar u_p \gamma^\mu T^A u_r) (\bar d_s \gamma_\mu T^A d_t) (H^\dag H)$
\end{tabular}
\end{minipage}
%%%%%%%
\end{adjustbox}
%%%%%%%%%%%%%%%%%%%%%%
\begin{adjustbox}{width=0.94\textwidth,center}
\small
%%%%%%%%%%%%
\begin{minipage}[t]{6.9cm}
\renewcommand{\arraystretch}{1.5}
\begin{tabular}[t]{c|c}
\multicolumn{2}{c}{\boldmath$18:(\bar L L)(\bar R R)H^2$} \\
\hline
$Q_{l^2e^2H^2}^{(1)}$  &  $(\bar l_p \gamma^\mu l_r) (\bar e_s \gamma_\mu e_t) (H^\dag H)$ \\
$Q_{l^2e^2H^2}^{(2)}$  &  $(\bar l_p \gamma^\mu \tau^I l_r) (\bar e_s \gamma_\mu e_t) (H^\dag \tau^I H)$ \\
$Q_{l^2u^2H^2}^{(1)}$  &  $(\bar l_p \gamma^\mu l_r) (\bar u_s \gamma_\mu u_t) (H^\dag H)$ \\
$Q_{l^2u^2H^2}^{(2)}$  &  $(\bar l_p \gamma^\mu \tau^I l_r) (\bar u_s \gamma_\mu u_t) (H^\dag \tau^I H)$ \\
$Q_{l^2d^2H^2}^{(1)}$  &  $(\bar l_p \gamma^\mu l_r) (\bar d_s \gamma_\mu d_t) (H^\dag H)$ \\
$Q_{l^2d^2H^2}^{(2)}$  &  $(\bar l_p \gamma^\mu \tau^I l_r) (\bar d_s \gamma_\mu d_t) (H^\dag \tau^I H)$ \\
$Q_{q^2e^2H^2}^{(1)}$  &  $(\bar q_p \gamma^\mu q_r) (\bar e_s \gamma_\mu e_t) (H^\dag H)$ \\
$Q_{q^2e^2H^2}^{(2)}$  &  $(\bar q_p \gamma^\mu \tau^I q_r) (\bar e_s \gamma_\mu e_t) (H^\dag \tau^I H)$ \\
$Q_{q^2u^2H^2}^{(1)}$  &  $(\bar q_p \gamma^\mu q_r) (\bar u_s \gamma_\mu u_t) (H^\dag H)$ \\ 
$Q_{q^2u^2H^2}^{(2)}$  &  $(\bar q_p \gamma^\mu \tau^I q_r) (\bar u_s \gamma_\mu u_t) (H^\dag \tau^I H)$ \\ 
$Q_{q^2u^2H^2}^{(3)}$  &  $(\bar q_p \gamma^\mu T^A q_r) (\bar u_s \gamma_\mu T^A u_t) (H^\dag H)$ \\ 
$Q_{q^2u^2H^2}^{(4)}$  &  $(\bar q_p \gamma^\mu T^A \tau^I q_r) (\bar u_s \gamma_\mu T^A u_t) (H^\dag \tau^I H)$ \\ 
$Q_{q^2d^2H^2}^{(1)}$  &  $(\bar q_p \gamma^\mu q_r) (\bar d_s \gamma_\mu d_t) (H^\dag H)$ \\
$Q_{q^2d^2H^2}^{(2)}$  &  $(\bar q_p \gamma^\mu \tau^I q_r) (\bar d_s \gamma_\mu d_t) (H^\dag \tau^I H)$ \\
$Q_{q^2d^2H^2}^{(3)}$  &  $(\bar q_p \gamma^\mu T^A q_r) (\bar d_s \gamma_\mu T^A d_t) (H^\dag H)$ \\
$Q_{q^2d^2H^2}^{(4)}$  &  $(\bar q_p \gamma^\mu T^A \tau^I q_r) (\bar d_s \gamma_\mu T^A d_t) (H^\dag \tau^I H)$ 
\end{tabular}
\end{minipage}
%%%%%%%%%%%%
\hspace{1cm}
%%%%%%%%%%%%
\begin{minipage}[t]{6.2cm}
\renewcommand{\arraystretch}{1.5}
\begin{tabular}[t]{c|c}
\multicolumn{2}{c}{\boldmath$18:(\bar L R)(\bar L R)H^2 + \hc$} \\
\hline
$Q_{q^2udH^2}^{(1)}$  &  $(\bar q_p^j u_r) \epsilon_{jk} (\bar q_s^k d_t) (H^\dag H)$ \\
$Q_{q^2udH^2}^{(2)}$  &  $(\bar q_p^j u_r) (\tau^I\epsilon)_{jk} (\bar q_s^k d_t) (H^\dag \tau^I H)$ \\
$Q_{q^2udH^2}^{(3)}$  &  $(\bar q_p^j T^A u_r) \epsilon_{jk} (\bar q_s^k T^A d_t) (H^\dag H)$ \\
$Q_{q^2udH^2}^{(4)}$  &  $(\bar q_p^j T^A u_r) (\tau^I\epsilon)_{jk} (\bar q_s^k T^A d_t) (H^\dag \tau^I H)$ \\
$Q_{lequH^2}^{(1)}$  &  $(\bar l_p^j e_r) \epsilon_{jk} (\bar q_s^k u_t) (H^\dag H)$ \\
$Q_{lequH^2}^{(2)}$  &  $(\bar l_p^j e_r) (\tau^I\epsilon)_{jk} (\bar q_s^k u_t) (H^\dag \tau^I H)$ \\
$Q_{lequH^2}^{(3)}$  &  $(\bar l_p^j \sigma_{\mu\nu} e_r) \epsilon_{jk} (\bar q_s^k \sigma^{\mu\nu} u_t) (H^\dag H)$ \\ 
$Q_{lequH^2}^{(4)}$  &  $(\bar l_p^j \sigma_{\mu\nu} e_r) (\tau^I\epsilon)_{jk} (\bar q_s^k \sigma^{\mu\nu} u_t) (H^\dag \tau^I H)$ \\
$Q_{l^2e^2H^2}^{(3)}$  &  $(\bar l_p e_r H) (\bar l_s e_t H)$ \\
$Q_{leqdH^2}^{(3)}$  &  $(\bar l_p e_r H) (\bar q_s d_t H)$ \\
$Q_{leqdH^2}^{(4)}$  &  $(\bar l_p \sigma_{\mu\nu} e_r H) (\bar q_s \sigma^{\mu\nu} d_t H)$ \\
$Q_{q^2u^2H^2}^{(5)}$  &  $(\bar q_p u_r \widetilde H) (\bar q_s u_t \widetilde H)$ \\
$Q_{q^2u^2H^2}^{(6)}$  &  $(\bar q_p T^A u_r \widetilde H) (\bar q_s T^A u_t \widetilde H)$ \\
$Q_{q^2d^2H^2}^{(5)}$  &  $(\bar q_p d_r H) (\bar q_s d_t H)$ \\
$Q_{q^2d^2H^2}^{(6)}$  &  $(\bar q_p T^A d_r H) (\bar q_s T^A d_t H)$
\end{tabular}
\end{minipage}
%%%%%%%
\end{adjustbox}
\end{center}
\caption{Most of the dimension-eight operators in the SMEFT of class-9, which are further divided into subclasses according to their chiral properties. 
See Table~\ref{tab:smeft8class_18_21} for the remaining class-9 operators.
Operators with ${}+\hc$ have Hermitian conjugates. 
The subscripts $p, r, s, t$ are weak-eigenstate indices.}
\label{tab:smeft8class_18}
\end{table}

%%%%%%%%%%%%%%%%%%%%%%%%%%%%%%%%%%%%%%%%%%
% SMEFT d=8 Class 19
%%%%%%%%%%%%%%%%%%%%%%%%%%%%%%%%%%%%%%%%%%

\begin{table}[H]
\begin{center}
%%%%%%%%%%%%%%%%%%%%%%
\begin{adjustbox}{width=0.85\textwidth,center}
\small
%%%%%%%%%%%%
\begin{minipage}[t]{5.6cm}
\renewcommand{\arraystretch}{1.5}
\begin{tabular}[t]{c|c}
\multicolumn{2}{c}{\boldmath$19:(\bar L L)(\bar L L)X$} \\
\hline
$Q_{l^4W}^{(1)}$  &  $(\bar l_p \gamma^\mu l_r) (\bar l_s \gamma^\nu \tau^I l_t) W_{\mu\nu}^I$ \\
$Q_{l^4W}^{(2)}$  &  $(\bar l_p \gamma^\mu l_r) (\bar l_s \gamma^\nu \tau^I l_t) \widetilde W_{\mu\nu}^I$ \\
$Q_{q^4G}^{(1)}$  & $(\bar q_p \gamma^\mu q_r) (\bar q_s \gamma^\nu T^A q_t) G^A_{\mu\nu}$ \\
$Q_{q^4G}^{(2)}$  & $(\bar q_p \gamma^\mu q_r) (\bar q_s \gamma^\nu T^A q_t) \widetilde G^A_{\mu\nu}$ \\
$Q_{q^4G}^{(3)}$  & $(\bar q_p \gamma^\mu \tau^I q_r) (\bar q_s \gamma^\nu T^A \tau^I q_t) G^A_{\mu\nu}$ \\
$Q_{q^4G}^{(4)}$  & $(\bar q_p \gamma^\mu \tau^I q_r) (\bar q_s \gamma^\nu T^A \tau^I q_t) \widetilde G^A_{\mu\nu}$ \\
$Q_{q^4W}^{(1)}$  & $(\bar q_p \gamma^\mu q_r) (\bar q_s \gamma^\nu \tau^I q_t) W^I_{\mu\nu}$ \\
$Q_{q^4W}^{(2)}$  & $(\bar q_p \gamma^\mu q_r) (\bar q_s \gamma^\nu \tau^I q_t) \widetilde W^I_{\mu\nu}$ \\
$Q_{q^4W}^{(3)}$  & $(\bar q_p \gamma^\mu T^A q_r) (\bar q_s \gamma^\nu T^A \tau^I q_t) W^I_{\mu\nu}$ \\
$Q_{q^4W}^{(4)}$  & $(\bar q_p \gamma^\mu T^A q_r) (\bar q_s \gamma^\nu T^A \tau^I q_t) \widetilde W^I_{\mu\nu}$ \\
$Q_{l^2q^2G}^{(1)}$  & $(\bar l_p \gamma^\mu l_r) (\bar q_s \gamma^\nu T^A q_t) G^A_{\mu\nu}$ \\
$Q_{l^2q^2G}^{(2)}$  & $(\bar l_p \gamma^\mu l_r) (\bar q_s \gamma^\nu T^A q_t) \widetilde G^A_{\mu\nu}$ \\
$Q_{l^2q^2G}^{(3)}$  & $(\bar l_p \gamma^\mu \tau^I l_r) (\bar q_s \gamma^\nu T^A \tau^I q_t) G^A_{\mu\nu}$ \\
$Q_{l^2q^2G}^{(4)}$  & $(\bar l_p \gamma^\mu \tau^I l_r) (\bar q_s \gamma^\nu T^A \tau^I q_t) \widetilde G^A_{\mu\nu}$ \\
$Q_{l^2q^2W}^{(1)}$  &  $(\bar l_p \gamma^\mu l_r)(\bar q_s \gamma^\nu \tau^I q_t) W^I_{\mu\nu}$ \\
$Q_{l^2q^2W}^{(2)}$  &  $(\bar l_p \gamma^\mu l_r)(\bar q_s \gamma^\nu \tau^I q_t) \widetilde W^I_{\mu\nu}$ \\
$Q_{l^2q^2W}^{(3)}$  &  $(\bar l_p \gamma^\mu \tau^I l_r)(\bar q_s \gamma^\nu q_t) W^I_{\mu\nu}$ \\
$Q_{l^2q^2W}^{(4)}$  &  $(\bar l_p \gamma^\mu \tau^I l_r)(\bar q_s \gamma^\nu q_t) \widetilde W^I_{\mu\nu}$ \\
$Q_{l^2q^2W}^{(5)}$  &  $\epsilon^{IJK} (\bar l_p \gamma^\mu \tau^I l_r)(\bar q_s \gamma^\nu \tau^J q_t) W^K_{\mu\nu}$ \\
$Q_{l^2q^2W}^{(6)}$  &  $\epsilon^{IJK} (\bar l_p \gamma^\mu \tau^I l_r)(\bar q_s \gamma^\nu \tau^J q_t) \widetilde W^K_{\mu\nu}$ \\
$Q_{l^2q^2B}^{(1)}$  &  $(\bar l_p \gamma^\mu l_r)(\bar q_s \gamma^\nu q_t) B_{\mu\nu}$ \\
$Q_{l^2q^2B}^{(2)}$  &  $(\bar l_p \gamma^\mu l_r)(\bar q_s \gamma^\nu q_t) \widetilde B_{\mu\nu}$ \\
$Q_{l^2q^2B}^{(3)}$  &  $(\bar l_p \gamma^\mu \tau^I l_r)(\bar q_s \gamma^\nu \tau^I q_t) B_{\mu\nu}$ \\
$Q_{l^2q^2B}^{(4)}$  &  $(\bar l_p \gamma^\mu \tau^I l_r)(\bar q_s \gamma^\nu \tau^I q_t) \widetilde B_{\mu\nu}$ \\ \hdashline
$Q_{l^4B}^{(1)}$  &  $(\bar l_p \gamma^\mu l_r) (\bar l_s \gamma^\nu l_t) B_{\mu\nu}$ \\
$Q_{l^4B}^{(2)}$  &  $(\bar l_p \gamma^\mu l_r) (\bar l_s \gamma^\nu l_t) \widetilde B_{\mu\nu}$ \\
$Q_{q^4B}^{(1)}$  & $(\bar q_p \gamma^\mu q_r) (\bar q_s \gamma^\nu q_t) B_{\mu\nu}$ \\
$Q_{q^4B}^{(2)}$  & $(\bar q_p \gamma^\mu q_r) (\bar q_s \gamma^\nu q_t) \widetilde B_{\mu\nu}$ \\
$Q_{q^4B}^{(3)}$  & $(\bar q_p \gamma^\mu \tau^I q_r) (\bar q_s \gamma^\nu \tau^I q_t) B_{\mu\nu}$ \\
$Q_{q^4B}^{(4)}$  & $(\bar q_p \gamma^\mu \tau^I q_r) (\bar q_s \gamma^\nu \tau^I q_t) \widetilde B_{\mu\nu}$
\end{tabular}
\end{minipage}
\hspace{1cm}
%%%%%%%%%%%%
\begin{minipage}[t]{6.1cm}
\renewcommand{\arraystretch}{1.5}
\begin{tabular}[t]{c|c}
\multicolumn{2}{c}{\boldmath$19:(\bar R R)(\bar R R)X$} \\
\hline
$Q_{u^4G}^{(1)}$  &  $(\bar u_p \gamma^\mu u_r) (\bar u_s \gamma^\nu T^A u_t) G^A_{\mu\nu}$ \\
$Q_{u^4G}^{(2)}$  &  $(\bar u_p \gamma^\mu u_r) (\bar u_s \gamma^\nu T^A u_t) \widetilde G^A_{\mu\nu}$ \\
$Q_{d^4G}^{(1)}$  &  $(\bar d_p \gamma^\mu d_r) (\bar d_s \gamma^\nu T^A d_t) G^A_{\mu\nu}$ \\
$Q_{d^4G}^{(2)}$  &  $(\bar d_p \gamma^\mu d_r) (\bar d_s \gamma^\nu T^A d_t) \widetilde G^A_{\mu\nu}$ \\
$Q_{e^2u^2G}^{(1)}$  &  $(\bar e_p \gamma^\mu e_r)(\bar u_s \gamma^\nu T^A u_t) G_{\mu\nu}^A$ \\
$Q_{e^2u^2G}^{(2)}$  &  $(\bar e_p \gamma^\mu e_r)(\bar u_s \gamma^\nu T^A u_t) \widetilde G_{\mu\nu}^A$ \\
$Q_{e^2u^2B}^{(1)}$  &  $(\bar e_p \gamma^\mu e_r)(\bar u_s \gamma^\nu u_t) B_{\mu\nu}$ \\
$Q_{e^2u^2B}^{(2)}$  &  $(\bar e_p \gamma^\mu e_r)(\bar u_s \gamma^\nu u_t) \widetilde B_{\mu\nu}$ \\
$Q_{e^2d^2G}^{(1)}$  &  $(\bar e_p \gamma^\mu e_r)(\bar d_s \gamma^\nu T^A d_t) G_{\mu\nu}^A$ \\
$Q_{e^2d^2G}^{(2)}$  &  $(\bar e_p \gamma^\mu e_r)(\bar d_s \gamma^\nu T^A d_t) \widetilde G_{\mu\nu}^A$ \\
$Q_{e^2d^2B}^{(1)}$  &  $(\bar e_p \gamma^\mu e_r)(\bar d_s \gamma^\nu d_t) B_{\mu\nu}$ \\
$Q_{e^2d^2B}^{(2)}$  &  $(\bar e_p \gamma^\mu e_r)(\bar d_s \gamma^\nu d_t) \widetilde B_{\mu\nu}$ \\
$Q_{u^2d^2G}^{(1)}$  &  $(\bar u_p \gamma^\mu u_r)(\bar d_s \gamma^\nu T^A d_t) G_{\mu\nu}^A$ \\
$Q_{u^2d^2G}^{(2)}$  &  $(\bar u_p \gamma^\mu u_r)(\bar d_s \gamma^\nu T^A d_t) \widetilde G_{\mu\nu}^A$ \\
$Q_{u^2d^2G}^{(3)}$  &  $(\bar u_p \gamma^\mu T^A u_r)(\bar d_s \gamma^\nu d_t) G_{\mu\nu}^A$ \\
$Q_{u^2d^2G}^{(4)}$  &  $(\bar u_p \gamma^\mu T^A u_r)(\bar d_s \gamma^\nu d_t) \widetilde G_{\mu\nu}^A$ \\
$Q_{u^2d^2G}^{(5)}$  &  $f^{ABC} (\bar u_p \gamma^\mu T^A u_r)(\bar d_s \gamma^\nu T^B d_t) G_{\mu\nu}^C$ \\
$Q_{u^2d^2G}^{(6)}$  &  $f^{ABC} (\bar u_p \gamma^\mu T^A u_r)(\bar d_s \gamma^\nu T^B d_t) \widetilde G_{\mu\nu}^C$ \\
$Q_{u^2d^2G}^{(7)}$  &  $d^{ABC} (\bar u_p \gamma^\mu T^A u_r)(\bar d_s \gamma^\nu T^B d_t) G_{\mu\nu}^C$ \\
$Q_{u^2d^2G}^{(8)}$  &  $d^{ABC} (\bar u_p \gamma^\mu T^A u_r)(\bar d_s \gamma^\nu T^B d_t) \widetilde G_{\mu\nu}^C$ \\
$Q_{u^2d^2B}^{(1)}$  &  $(\bar u_p \gamma^\mu u_r)(\bar d_s \gamma^\nu d_t) B_{\mu\nu}$ \\
$Q_{u^2d^2B}^{(2)}$  &  $(\bar u_p \gamma^\mu u_r)(\bar d_s \gamma^\nu d_t) \widetilde B_{\mu\nu}$ \\
$Q_{u^2d^2B}^{(3)}$  &  $(\bar u_p \gamma^\mu T^A u_r)(\bar d_s \gamma^\nu T^A d_t) B_{\mu\nu}$ \\
$Q_{u^2d^2B}^{(4)}$  &  $(\bar u_p \gamma^\mu T^A u_r)(\bar d_s \gamma^\nu T^A d_t) \widetilde B_{\mu\nu}$ \\ \hdashline
$Q_{e^4B}^{(1)}$  &  $(\bar e_p \gamma^\mu e_r) (\bar e_s \gamma^\nu e_t) B_{\mu\nu}$ \\
$Q_{e^4B}^{(2)}$  &  $(\bar e_p \gamma^\mu e_r) (\bar e_s \gamma^\nu e_t) \widetilde B_{\mu\nu}$ \\
$Q_{u^4B}^{(1)}$  &  $(\bar u_p \gamma^\mu u_r) (\bar u_s \gamma^\nu u_t) B_{\mu\nu}$ \\
$Q_{u^4B}^{(2)}$  &  $(\bar u_p \gamma^\mu u_r) (\bar u_s \gamma^\nu u_t) \widetilde B_{\mu\nu}$ \\
$Q_{d^4B}^{(1)}$  & $(\bar d_p \gamma^\mu d_r) (\bar d_s \gamma^\nu d_t) B_{\mu\nu}$ \\
$Q_{d^4B}^{(2)}$  & $(\bar d_p \gamma^\mu d_r) (\bar d_s \gamma^\nu d_t) \widetilde B_{\mu\nu}$
\end{tabular}
\end{minipage}
%%%%%%%%%
\end{adjustbox}
\end{center}
\caption{The dimension-eight operators in the SMEFT of class-19 with field content $J J X$ with $J = (\bar L L)$ or $(\bar R R)$. 
The subscripts $p, r, s, t$ are weak-eigenstate indices.
Operators below the dashed lines vanish when there is only one generation of fermions.}
\label{tab:smeft8class_19_LL_RR}
\end{table}

%%%%%%%%%%%%%%%%%%%%%%%%%%%%%%

\begin{table}[H]
\begin{center}
%%%%%%%%%%%%%%%%%%%%%%
\begin{adjustbox}{width=0.8\textwidth,center}
\small
%%%%%%%%%%%%
\begin{minipage}[t]{4.9cm}
\renewcommand{\arraystretch}{1.5}
\begin{tabular}[t]{c|c}
\multicolumn{2}{c}{\boldmath$19:(\bar L L)(\bar R R)X$} \\
\hline
$Q_{l^2e^2W}^{(1)}$  &  $(\bar l_p \gamma^\mu \tau^I l_r) (\bar e_s \gamma^\nu e_t) W^I_{\mu\nu}$ \\
$Q_{l^2e^2W}^{(2)}$  &  $(\bar l_p \gamma^\mu \tau^I l_r) (\bar e_s \gamma^\nu e_t) \widetilde W^I_{\mu\nu}$ \\
$Q_{l^2e^2B}^{(1)}$  &  $(\bar l_p \gamma^\mu l_r) (\bar e_s \gamma^\nu e_t) B_{\mu\nu}$ \\
$Q_{l^2e^2B}^{(2)}$  &  $(\bar l_p \gamma^\mu l_r) (\bar e_s \gamma^\nu e_t) \widetilde B_{\mu\nu}$ \\
$Q_{l^2u^2G}^{(1)}$  &  $(\bar l_p \gamma^\mu l_r) (\bar u_s \gamma^\nu T^A u_t) G^A_{\mu\nu}$ \\
$Q_{l^2u^2G}^{(2)}$  &  $(\bar l_p \gamma^\mu l_r) (\bar u_s \gamma^\nu T^A u_t) \widetilde G^A_{\mu\nu}$ \\
$Q_{l^2u^2W}^{(1)}$  &  $(\bar l_p \gamma^\mu \tau^I l_r) (\bar u_s \gamma^\nu u_t) W^I_{\mu\nu}$ \\
$Q_{l^2u^2W}^{(2)}$  &  $(\bar l_p \gamma^\mu \tau^I l_r) (\bar u_s \gamma^\nu u_t) \widetilde W^I_{\mu\nu}$ \\
$Q_{l^2u^2B}^{(1)}$  &  $(\bar l_p \gamma^\mu l_r) (\bar u_s \gamma^\nu u_t) B_{\mu\nu}$ \\
$Q_{l^2u^2B}^{(2)}$  &  $(\bar l_p \gamma^\mu l_r) (\bar u_s \gamma^\nu u_t) \widetilde B_{\mu\nu}$ \\
$Q_{l^2d^2G}^{(1)}$  &  $(\bar l_p \gamma^\mu l_r) (\bar d_s \gamma^\nu T^A d_t) G^A_{\mu\nu}$ \\
$Q_{l^2d^2G}^{(2)}$  &  $(\bar l_p \gamma^\mu l_r) (\bar d_s \gamma^\nu T^A d_t) \widetilde G^A_{\mu\nu}$ \\
$Q_{l^2d^2W}^{(1)}$  &  $(\bar l_p \gamma^\mu \tau^I l_r) (\bar d_s \gamma^\nu d_t) W^I_{\mu\nu}$ \\
$Q_{l^2d^2W}^{(2)}$  &  $(\bar l_p \gamma^\mu \tau^I l_r) (\bar d_s \gamma^\nu d_t) \widetilde W^I_{\mu\nu}$ \\
$Q_{l^2d^2B}^{(1)}$  &  $(\bar l_p \gamma^\mu l_r) (\bar d_s \gamma^\nu d_t) B_{\mu\nu}$ \\
$Q_{l^2d^2B}^{(2)}$  &  $(\bar l_p \gamma^\mu l_r) (\bar d_s \gamma^\nu d_t) \widetilde B_{\mu\nu}$ \\
$Q_{q^2e^2G}^{(1)}$  &  $(\bar q_p \gamma^\mu T^A q_r) (\bar e_s \gamma^\nu e_t) G^A_{\mu\nu}$ \\
$Q_{q^2e^2G}^{(2)}$  &  $(\bar q_p \gamma^\mu T^A q_r) (\bar e_s \gamma^\nu e_t) \widetilde G^A_{\mu\nu}$ \\
$Q_{q^2e^2W}^{(1)}$  &  $(\bar q_p \gamma^\mu \tau^I q_r) (\bar e_s \gamma^\nu e_t) W^I_{\mu\nu}$ \\
$Q_{q^2e^2W}^{(2)}$  &  $(\bar q_p \gamma^\mu \tau^I q_r) (\bar e_s \gamma^\nu e_t) \widetilde W^I_{\mu\nu}$ \\
$Q_{q^2e^2B}^{(1)}$  &  $(\bar q_p \gamma^\mu q_r) (\bar e_s \gamma^\nu e_t) B_{\mu\nu}$ \\
$Q_{q^2e^2B}^{(2)}$  &  $(\bar q_p \gamma^\mu q_r) (\bar e_s \gamma^\nu e_t) \widetilde B_{\mu\nu}$ \\
$Q_{q^2u^2G}^{(1)}$  &  $(\bar q_p \gamma^\mu q_r) (\bar u_s \gamma^\nu T^A u_t) G^A_{\mu\nu}$ \\ 
$Q_{q^2u^2G}^{(2)}$  &  $(\bar q_p \gamma^\mu q_r) (\bar u_s \gamma^\nu T^A u_t) \widetilde G^A_{\mu\nu}$ \\ 
$Q_{q^2u^2G}^{(3)}$  &  $(\bar q_p \gamma^\mu T^A q_r) (\bar u_s \gamma^\nu u_t) G^A_{\mu\nu}$ \\ 
$Q_{q^2u^2G}^{(4)}$  &  $(\bar q_p \gamma^\mu T^A q_r) (\bar u_s \gamma^\nu u_t) \widetilde G^A_{\mu\nu}$ 
\end{tabular}
\end{minipage}
\hspace{1cm}
%%%%%%%%%%%%
\begin{minipage}[t]{6.1cm}
\renewcommand{\arraystretch}{1.5}
\begin{tabular}[t]{c|c}
\multicolumn{2}{c}{\boldmath$19:(\bar L L)(\bar R R)X$} \\
\hline
$Q_{q^2u^2G}^{(5)}$  &  $f^{ABC} (\bar q_p \gamma^\mu T^A q_r) (\bar u_s \gamma^\nu T^B u_t) G^C_{\mu\nu}$ \\ 
$Q_{q^2u^2G}^{(6)}$  &  $f^{ABC} (\bar q_p \gamma^\mu T^A q_r) (\bar u_s \gamma^\nu T^B u_t) \widetilde G^C_{\mu\nu}$ \\ 
$Q_{q^2u^2G}^{(7)}$  &  $d^{ABC} (\bar q_p \gamma^\mu T^A q_r) (\bar u_s \gamma^\nu T^B u_t) G^C_{\mu\nu}$ \\ 
$Q_{q^2u^2G}^{(8)}$  &  $d^{ABC} (\bar q_p \gamma^\mu T^A q_r) (\bar u_s \gamma^\nu T^B u_t) \widetilde G^C_{\mu\nu}$ \\ 
$Q_{q^2u^2W}^{(1)}$  &  $(\bar q_p \gamma^\mu \tau^I q_r) (\bar u_s \gamma^\nu u_t) W^I_{\mu\nu}$ \\ 
$Q_{q^2u^2W}^{(2)}$  &  $(\bar q_p \gamma^\mu \tau^I q_r) (\bar u_s \gamma^\nu u_t) \widetilde W^I_{\mu\nu}$ \\ 
$Q_{q^2u^2W}^{(3)}$  &  $(\bar q_p \gamma^\mu T^A \tau^I q_r) (\bar u_s \gamma^\nu T^A u_t) W^I_{\mu\nu}$ \\ 
$Q_{q^2u^2W}^{(4)}$  &  $(\bar q_p \gamma^\mu T^A \tau^I q_r) (\bar u_s \gamma^\nu T^A u_t) \widetilde W^I_{\mu\nu}$ \\ 
$Q_{q^2u^2B}^{(1)}$  &  $(\bar q_p \gamma^\mu q_r) (\bar u_s \gamma^\nu u_t) B_{\mu\nu}$ \\ 
$Q_{q^2u^2B}^{(2)}$  &  $(\bar q_p \gamma^\mu q_r) (\bar u_s \gamma^\nu u_t) \widetilde B_{\mu\nu}$ \\ 
$Q_{q^2u^2B}^{(3)}$  &  $(\bar q_p \gamma^\mu T^A q_r) (\bar u_s \gamma^\nu T^A u_t) B_{\mu\nu}$ \\ 
$Q_{q^2u^2B}^{(4)}$  &  $(\bar q_p \gamma^\mu T^A q_r) (\bar u_s \gamma^\nu T^A u_t) \widetilde B_{\mu\nu}$ \\ 
$Q_{q^2d^2G}^{(1)}$  &  $(\bar q_p \gamma^\mu q_r) (\bar d_s \gamma^\nu T^A d_t) G^A_{\mu\nu}$ \\ 
$Q_{q^2d^2G}^{(2)}$  &  $(\bar q_p \gamma^\mu q_r) (\bar d_s \gamma^\nu T^A d_t) \widetilde G^A_{\mu\nu}$ \\ 
$Q_{q^2d^2G}^{(3)}$  &  $(\bar q_p \gamma^\mu T^A q_r) (\bar d_s \gamma^\nu d_t) G^A_{\mu\nu}$ \\ 
$Q_{q^2d^2G}^{(4)}$  &  $(\bar q_p \gamma^\mu T^A q_r) (\bar d_s \gamma^\nu d_t) \widetilde G^A_{\mu\nu}$ \\ 
$Q_{q^2d^2G}^{(5)}$  &  $f^{ABC} (\bar q_p \gamma^\mu T^A q_r) (\bar d_s \gamma^\nu T^B d_t) G^C_{\mu\nu}$ \\ 
$Q_{q^2d^2G}^{(6)}$  &  $f^{ABC} (\bar q_p \gamma^\mu T^A q_r) (\bar d_s \gamma^\nu T^B d_t) \widetilde G^C_{\mu\nu}$ \\ 
$Q_{q^2d^2G}^{(7)}$  &  $d^{ABC} (\bar q_p \gamma^\mu T^A q_r) (\bar d_s \gamma^\nu T^B d_t) G^C_{\mu\nu}$ \\ 
$Q_{q^2d^2G}^{(8)}$  &  $d^{ABC} (\bar q_p \gamma^\mu T^A q_r) (\bar d_s \gamma^\nu T^B d_t) \widetilde G^C_{\mu\nu}$ \\ 
$Q_{q^2d^2W}^{(1)}$  &  $(\bar q_p \gamma^\mu \tau^I q_r) (\bar d_s \gamma^\nu d_t) W^I_{\mu\nu}$ \\ 
$Q_{q^2d^2W}^{(2)}$  &  $(\bar q_p \gamma^\mu \tau^I q_r) (\bar d_s \gamma^\nu d_t) \widetilde W^I_{\mu\nu}$ \\ 
$Q_{q^2d^2W}^{(3)}$  &  $(\bar q_p \gamma^\mu T^A \tau^I q_r) (\bar d_s \gamma^\nu T^A d_t) W^I_{\mu\nu}$ \\ 
$Q_{q^2d^2W}^{(4)}$  &  $(\bar q_p \gamma^\mu T^A \tau^I q_r) (\bar d_s \gamma^\nu T^A d_t) \widetilde W^I_{\mu\nu}$ \\ 
$Q_{q^2d^2B}^{(1)}$  &  $(\bar q_p \gamma^\mu q_r) (\bar d_s \gamma^\nu d_t) B_{\mu\nu}$ \\ 
$Q_{q^2d^2B}^{(2)}$  &  $(\bar q_p \gamma^\mu q_r) (\bar d_s \gamma^\nu d_t) \widetilde B_{\mu\nu}$ \\ 
$Q_{q^2d^2B}^{(3)}$  &  $(\bar q_p \gamma^\mu T^A q_r) (\bar d_s \gamma^\nu T^A d_t) B_{\mu\nu}$ \\ 
$Q_{q^2d^2B}^{(4)}$  &  $(\bar q_p \gamma^\mu T^A q_r) (\bar d_s \gamma^\nu T^A d_t) \widetilde B_{\mu\nu}$
\end{tabular}
\end{minipage}
%%%%%%%%%
\end{adjustbox}
\end{center}
\caption{The dimension-eight operators in the SMEFT of class-19 with field content $(\bar L L)(\bar R R) X$. The subscripts $p, r, s, t$ are weak-eigenstate indices.}
\label{tab:smeft8class_19_LLRR}
\end{table}

%%%%%%%%%%%%%%%%%%%%%%%%%%%%%%
%%%%%%%%%%%%%%%%%%%%%%%%%%%%%%

\begin{table}[H]
\begin{center}
%%%%%%%%%%%%%%%%%%%%%%
\begin{adjustbox}{width=0.95\textwidth,center}
\small
%%%%%%%%%%%%
\begin{minipage}[t]{5.7cm}
\renewcommand{\arraystretch}{1.5}
\begin{tabular}[t]{c|c}
\multicolumn{2}{c}{\boldmath$19:(\bar L R)(\bar R L)X + \hc$} \\
\hline
$Q_{ledqG}^{(1)}$ & $(\bar l_p^j \sigma^{\mu\nu} e_r) (\bar d_s T^A q_{tj}) G^A_{\mu\nu}$ \\
$Q_{ledqG}^{(2)}$ & $(\bar l_p^j e_r) (\bar d_s \sigma^{\mu\nu} T^A q_{tj}) G^A_{\mu\nu}$ \\
$Q_{ledqW}^{(1)}$ & $(\bar l_p \sigma^{\mu\nu} e_r) \tau^I (\bar d_s q_t)  W^I_{\mu\nu}$ \\
$Q_{ledqW}^{(2)}$ & $(\bar l_p e_r) \tau^I (\bar d_s \sigma^{\mu\nu} q_t) W^I_{\mu\nu}$ \\
$Q_{ledqB}^{(1)}$ & $(\bar l_p^j \sigma^{\mu\nu} e_r) (\bar d_s q_{tj}) B_{\mu\nu}$ \\
$Q_{ledqB}^{(2)}$ & $(\bar l_p^j e_r) (\bar d_s \sigma^{\mu\nu} q_{tj}) B_{\mu\nu}$
\end{tabular}
\end{minipage}
\hspace{1cm}
%%%%%%%%%%%%
\begin{minipage}[t]{7.6cm}
\renewcommand{\arraystretch}{1.5}
\begin{tabular}[t]{c|c}
\multicolumn{2}{c}{\boldmath$19:(\bar L R)(\bar L R)X + \hc$} \\
\hline
$Q_{q^2udG}^{(1)}$  &  $(\bar q_p^j \sigma^{\mu\nu} T^A u_r) \epsilon_{jk} (\bar q_s^k d_t) G^A_{\mu\nu}$ \\
$Q_{q^2udG}^{(2)}$  &  $(\bar q_p^j \sigma^{\mu\nu} u_r) \epsilon_{jk} (\bar q_s^k T^A d_t) G^A_{\mu\nu}$ \\
$Q_{q^2udG}^{(3)}$  &  $(\bar q_p^j T^A u_r) \epsilon_{jk} (\bar q_s^k \sigma^{\mu\nu} d_t) G^A_{\mu\nu}$ \\
$Q_{q^2udG}^{(4)}$  &  $(\bar q_p^j u_r) \epsilon_{jk} (\bar q_s^k \sigma^{\mu\nu} T^A d_t) G^A_{\mu\nu}$ \\
$Q_{q^2udG}^{(5)}$  &  $(\bar q_p^j \sigma^{\mu\rho} T^A u_r) \epsilon_{jk} (\bar q_s^k \sigma_{\rho\nu} d_t) G_\mu^{A\nu}$ \\
$Q_{q^2udG}^{(6)}$  &  $(\bar q_p^j \sigma^{\mu\rho} u_r) \epsilon_{jk} (\bar q_s^k \sigma_{\rho\nu} T^A d_t) G_\mu^{A\nu}$ \\
$Q_{q^2udW}^{(1)}$  &  $(\bar q_p^j \sigma^{\mu\nu} u_r) (\tau^I \epsilon)_{jk} (\bar q_s^k d_t) W^I_{\mu\nu}$ \\
$Q_{q^2udW}^{(2)}$  &  $(\bar q_p^j u_r) (\tau^I \epsilon)_{jk} (\bar q_s^k \sigma^{\mu\nu} d_t) W^I_{\mu\nu}$ \\
$Q_{q^2udW}^{(3)}$  &  $(\bar q_p^j \sigma^{\mu\rho} u_r) (\tau^I \epsilon)_{jk} (\bar q_s^k \sigma_{\rho\nu} d_t) W^{I\nu}_\mu$ \\
$Q_{q^2udB}^{(1)}$  &  $(\bar q_p^j \sigma^{\mu\nu} u_r) \epsilon_{jk} (\bar q_s^k d_t) B_{\mu\nu}$ \\
$Q_{q^2udB}^{(2)}$  &  $(\bar q_p^j u_r) \epsilon_{jk} (\bar q_s^k \sigma^{\mu\nu} d_t) B_{\mu\nu}$ \\
$Q_{q^2udB}^{(3)}$  &  $(\bar q_p^j \sigma^{\mu\rho} u_r) \epsilon_{jk} (\bar q_s^k \sigma_{\rho\nu} d_t) B_\mu^\nu$ \\
$Q_{lequG}^{(1)}$  &  $(\bar l_p^j \sigma^{\mu\nu} e_r) \epsilon_{jk} (\bar q_s^k T^A u_t) G^A_{\mu\nu}$ \\
$Q_{lequG}^{(2)}$  &  $(\bar l_p^j e_r) \epsilon_{jk} (\bar q_s^k \sigma^{\mu\nu} T^A u_t) G^A_{\mu\nu}$ \\
$Q_{lequG}^{(3)}$  &  $(\bar l_p^j \sigma^{\mu\rho} e_r) \epsilon_{jk} (\bar q_s^k \sigma_{\rho\nu} T^A u_t) G_\mu^{A\nu}$ \\
$Q_{lequW}^{(1)}$  &  $(\bar l_p^j \sigma^{\mu\nu} e_r) (\tau^I \epsilon)_{jk} (\bar q_s^k u_t) W^I_{\mu\nu}$ \\
$Q_{lequW}^{(2)}$  &  $(\bar l_p^j e_r) (\tau^I \epsilon)_{jk} (\bar q_s^k \sigma^{\mu\nu} u_t) W^I_{\mu\nu}$ \\
$Q_{lequW}^{(3)}$  &  $(\bar l_p^j \sigma^{\mu\rho} e_r) (\tau^I \epsilon)_{jk} (\bar q_s^k \sigma_{\rho\nu} u_t) W^{I\nu}_\mu$ \\
$Q_{lequB}^{(1)}$  &  $(\bar l_p^j \sigma^{\mu\nu} e_r) \epsilon_{jk} (\bar q_s^k u_t) B_{\mu\nu}$ \\
$Q_{lequB}^{(2)}$  &  $(\bar l_p^j e_r) \epsilon_{jk} (\bar q_s^k \sigma^{\mu\nu} u_t) B_{\mu\nu}$ \\
$Q_{lequB}^{(3)}$  &  $(\bar l_p^j \sigma^{\mu\rho} e_r) \epsilon_{jk} (\bar q_s^k \sigma_{\rho\nu} u_t) B_\mu^\nu$
\end{tabular}
\end{minipage}
%%%%%%%%%
\end{adjustbox}
\end{center}
\caption{The dimension-eight operators in the SMEFT of class-19 with field content $(\bar L R) (\bar R L) X$ or $(\bar L R) (\bar L R) X$. 
All of the operators have Hermitian conjugates. 
The subscripts $p, r, s, t$ are weak-eigenstate indices. }
\label{tab:smeft8class_19_LRRL_LRLR}
\end{table}

\begin{table}[H]
\begin{center}
%%%%%%%%%%%%%%%%%%%%%%
\begin{adjustbox}{width=0.52\textwidth,center}
\small
%%%%%%%%%%%%
\begin{minipage}[t]{7.8cm}
\renewcommand{\arraystretch}{1.5}
\begin{tabular}[t]{c|c}
\multicolumn{2}{c}{\boldmath$19(\slashed B):\psi^4X + \hc$} \\
\hline
$Q_{lqudG}^{(1)}$  &  $(T^A)_\gamma^\delta \epsilon_{\delta\alpha\beta}   \epsilon_{jk} (d_p^{\alpha} C \sigma^{\mu\nu} u_r^{\beta}) (q_s^{j\gamma} C l_t^k) G^A_{\mu\nu}$ \\
$Q_{lqudG}^{(2)}$  &  $(T^A)_\gamma^\delta \epsilon_{\delta\alpha\beta} \epsilon_{jk} (d_p^{\alpha} C u_r^{\beta}) (q_s^{j\gamma} C \sigma^{\mu\nu} l_t^k) G^A_{\mu\nu}$ \\
$Q_{lqudG}^{(3)}$  &  $(T^A)_{(\alpha}^\delta \epsilon_{\beta)\gamma\delta}  \epsilon_{jk}   (d_p^{\alpha} C \sigma^{\mu\nu} u_r^{\beta}) (q_s^{j\gamma} C l_t^k) G^A_{\mu\nu}$ \\
$Q_{lqudG}^{(4)}$  &  $(T^A)_{(\alpha}^\delta \epsilon_{\beta)\gamma\delta} \epsilon_{jk} (d_p^{\alpha} C u_r^{\beta}) (q_s^{j\gamma} C \sigma^{\mu\nu} l_t^k) G^A_{\mu\nu}$ \\
$Q_{lqudW}^{(1)}$  &  $\epsilon_{\alpha\beta\gamma} (\epsilon \tau^I)_{jk} (d_p^{\alpha} C \sigma^{\mu\nu} u_r^{\beta}) (q_s^{j\gamma} C l_t^k) W^I_{\mu\nu}$ \\
$Q_{lqudW}^{(2)}$  &  $\epsilon_{\alpha\beta\gamma} (\epsilon \tau^I)_{jk} (d_p^{\alpha} C u_r^{\beta}) (q_s^{j\gamma} C \sigma^{\mu\nu} l_t^k) W^I_{\mu\nu}$ \\
$Q_{lqudB}^{(1)}$  &  $\epsilon_{\alpha\beta\gamma} \epsilon_{jk} (d_p^{\alpha} C \sigma^{\mu\nu} u_r^{\beta}) (q_s^{j\gamma} C l_t^k)  B_{\mu\nu}$ \\
$Q_{lqudB}^{(2)}$  &  $\epsilon_{\alpha\beta\gamma} \epsilon_{jk} (d_p^{\alpha} C u_r^{\beta}) (q_s^{j\gamma} C \sigma^{\mu\nu} l_t^k) B_{\mu\nu}$ \\
$Q_{eq^2uG}^{(1)}$  &  $(T^A)_\gamma^\delta \epsilon_{\delta\alpha\beta}  \epsilon_{jk} (q_p^{j\alpha} C \sigma^{\mu\nu} q_r^{k\beta}) (u_s^{\gamma}  C e_t) G^A_{\mu\nu}$ \\
$Q_{eq^2uG}^{(2)}$  &  $(T^A)_{(\alpha}^\delta \epsilon_{\beta)\gamma\delta}   \epsilon_{jk} (q_p^{j\alpha} C q_r^{k\beta}) (u_s^{\gamma} C \sigma^{\mu\nu} e_t) G^A_{\mu\nu}$ \\
$Q_{eq^2uW}^{(1)}$  &  $\epsilon_{\alpha\beta\gamma} (\epsilon \tau^I)_{jk} (q_p^{j\alpha} C \sigma^{\mu\nu} q_r^{k\beta}) (u_s^{\gamma} C e_t) W^I_{\mu\nu}$ \\
$Q_{eq^2uB}^{(1)}$  &  $\epsilon_{\alpha\beta\gamma} \epsilon_{jk} (q_p^{j\alpha} C q_r^{k\beta}) (u_s^{\gamma} C \sigma^{\mu\nu} e_t) B_{\mu\nu}$ \\
$Q_{lq^3G}^{(1)}$ & $(T^A)_\gamma^\delta \epsilon_{\delta\alpha\beta}  \epsilon_{mn} \epsilon_{jk} (q_p^{m\alpha} C \sigma^{\mu\nu} q_r^{j\beta}) (q_s^{k\gamma} C l_t^n) G^A_{\mu\nu}$ \\ 
$Q_{lq^3G}^{(2)}$ & $(T^A)_{(\alpha}^\delta \epsilon_{\beta)\gamma\delta} \epsilon_{mn} \epsilon_{jk} (q_p^{m\alpha} C q_r^{j\beta}) (q_s^{k\gamma} C \sigma^{\mu\nu} l_t^n) G^A_{\mu\nu}$ \\
$Q_{lq^3W}^{(1)}$  &  $\epsilon_{\alpha\beta\gamma} (\epsilon \tau^I)_{mn} \epsilon_{jk} (q_p^{m\alpha} C q_r^{j\beta}) (q_s^{k\gamma} C \sigma^{\mu\nu} l_t^n) W^I_{\mu\nu}$ \\
$Q_{lq^3W}^{(2)}$  & $\epsilon_{\alpha\beta\gamma} (\epsilon \tau^I)_{mj} \epsilon_{kn} (q_p^{m\alpha} C \sigma^{\mu\nu}q_r^{j\beta}) (q_s^{k\gamma} C  l_t^n) W^I_{\mu\nu}$  \\
$Q_{lq^3B}^{(1)}$ & $\epsilon_{\alpha\beta\gamma} \epsilon_{mn} \epsilon_{jk} (q_p^{m\alpha} C q_r^{j\beta}) (q_s^{k\gamma} C \sigma^{\mu\nu} l_t^n) B_{\mu\nu}$ \\ 
$Q_{eu^2dG}^{(1)}$ & $(T^A)_\gamma^\delta \epsilon_{\delta\alpha\beta} (d_p^{\alpha} C \sigma^{\mu\nu} u_r^{\beta}) (u_s^{\gamma} C e_t) G^A_{\mu\nu}$ \\
$Q_{eu^2dG}^{(2)}$ & $(T^A)_\gamma^\delta \epsilon_{\delta\alpha\beta} (u_p^{\alpha} C \sigma^{\mu\nu} u_r^{\beta}) (d_s^{\gamma} C e_t) G^A_{\mu\nu}$ \\
$Q_{eu^2dG}^{(3)}$ & $(T^A)_{(\alpha}^\delta \epsilon_{\beta)\gamma\delta}  (u_p^{\alpha} C u_r^{\beta}) (d_s^{\gamma} C \sigma^{\mu\nu} e_t) G^A_{\mu\nu}$ \\
$Q_{eu^2dB}^{(1)}$ &  $\epsilon_{\alpha\beta\gamma} (d_p^{\alpha} C \sigma^{\mu\nu} u_r^{\beta}) (u_s^{\gamma} C e_t) B_{\mu\nu}$ \\
$Q_{eu^2dB}^{(2)}$ & $\epsilon_{\alpha\beta\gamma} (u_p^{\alpha} C \sigma^{\mu\nu} u_r^{\beta}) (d_s^{\gamma} C e_t) B_{\mu\nu}$ \\ \hdashline
$Q_{eq^2uW}^{(2)}$  &  $\epsilon_{\alpha\beta\gamma} (\epsilon \tau^I)_{jk} (q_p^{j\alpha} C q_r^{k\beta}) (u_s^{\gamma} C \sigma^{\mu\nu} e_t) W^I_{\mu\nu}$ \\
$Q_{eq^2uB}^{(2)}$  &  $\epsilon_{\alpha\beta\gamma} \epsilon_{jk} (q_p^{j\alpha} C \sigma^{\mu\nu} q_r^{k\beta}) (u_s^{\gamma} C e_t) B_{\mu\nu}$ \\
$Q_{lq^3G}^{(3)}$ & $(T^A)_{(\alpha}^\delta \epsilon_{\beta)\gamma\delta} \epsilon_{mn} \epsilon_{jk} (q_p^{m\alpha} C \sigma^{\mu\nu} q_r^{j\beta}) (q_s^{k\gamma} C l_t^n) G^A_{\mu\nu}$ \\ 
$Q_{lq^3G}^{(4)}$ & $(T^A)_\gamma^\delta \epsilon_{\delta\alpha\beta}  \epsilon_{mn} \epsilon_{jk} (q_p^{m\alpha} C q_r^{j\beta}) (q_s^{k\gamma} C \sigma^{\mu\nu}  l_t^n) G^A_{\mu\nu}$ \\
$Q_{lq^3W}^{(3)}$  &  $\epsilon_{\alpha\beta\gamma} \epsilon_{mn} (\epsilon \tau^I)_{jk} (q_p^{m\alpha} C q_r^{j\beta}) (q_s^{k\gamma} C \sigma^{\mu\nu} l_t^n) W^I_{\mu\nu}$ \\
$Q_{lq^3B}^{(2)}$ & $\epsilon_{\alpha\beta\gamma} \epsilon_{mn} \epsilon_{jk} (q_p^{m\alpha} C \sigma^{\mu\nu} q_r^{j\beta}) (q_s^{k\gamma} C l_t^n) B_{\mu\nu}$ 
\end{tabular}
\end{minipage}
%%%%%%%%%
\end{adjustbox}
\end{center}
\caption{The baryon number violating dimension-eight operators of class-19. 
All of the operators have Hermitian conjugates. 
The subscripts $p, r, s, t$ are weak-eigenstate indices. 
Operators below the dashed line vanish when there is only one generation of fermions.}
\label{tab:smeft8class_19_slashedB}
\end{table}

%%%%%%%%%%%%%%%%%%%%%%%%%%%%%%%%%%%%%%%%%%
% SMEFT d=8 Class 20
%%%%%%%%%%%%%%%%%%%%%%%%%%%%%%%%%%%%%%%%%%

\begin{table}[H]
\begin{center}
%%%%%%%%%%%%%%%%%%%%%%
\begin{adjustbox}{width=0.83\textwidth,center}
\small
%%%%%%%%%%%%
\begin{minipage}[t]{5.4cm}
\renewcommand{\arraystretch}{1.5}
\begin{tabular}[t]{c|c}
\multicolumn{2}{c}{\boldmath$20:\psi^4HD + \hc$} \\
\hline
$Q_{l^3eHD}^{(1)}$  &  $i (\bar l_p \gamma^\mu l_r) [(\bar l_s e_t) D_\mu H]$ \\
$Q_{l^3eHD}^{(2)}$  &  $i (\bar l_p \gamma^\mu \tau^I l_r) [(\bar l_s e_t) \tau^I D_\mu H]$ \\
$Q_{l^3eHD}^{(3)}$  &  $i (\bar l_p \gamma^\mu l_r) [(D_\mu \bar l_s e_t) H]$ \\
$Q_{le^3HD}^{(1)}$  &  $i (\bar e_p \gamma^\mu e_r) [(\bar l_s D_\mu e_t) H]$ \\
$Q_{leq^2HD}^{(1)}$  &  $i (\bar q_p \gamma^\mu q_r) [(\bar l_s e_t) D_\mu H]$ \\
$Q_{leq^2HD}^{(2)}$  &  $i (\bar l_p \gamma^\mu q_r^\alpha) [(\bar q_{s\alpha} e_t) D_\mu H]$ \\
$Q_{leq^2HD}^{(3)}$  &  $i (\bar q_p \gamma^\mu \tau^I q_r) [(\bar l_s e_t) \tau^I D_\mu H]$ \\
$Q_{leq^2HD}^{(4)}$  &  $i (\bar l_p \gamma^\mu \tau^I q_r^\alpha) [(\bar q_{s\alpha} e_t) \tau^I D_\mu H]$ \\
$Q_{leq^2HD}^{(5)}$  &  $i (\bar q_p \gamma^\mu q_r) [(\bar l_s D_\mu e_t) H]$ \\
$Q_{leq^2HD}^{(6)}$  &  $i (\bar q_p \gamma^\nu \tau^I q_r) [(\bar l_s D_\mu e_t) \tau^I H]$ \\
$Q_{leu^2HD}^{(1)}$  &  $i (\bar u_p \gamma^\mu u_r) [(\bar l_s e_t) D_\mu H]$ \\
$Q_{leu^2HD}^{(2)}$  &  $i (\bar u_{p\alpha} \gamma^\mu e_r) [(\bar l_s u_t^\alpha) D_\mu H]$ \\
$Q_{leu^2HD}^{(3)}$  &  $i (\bar u_p \gamma^\mu u_r) [(D_\mu \bar l_s e_t) H]$\\
$Q_{led^2HD}^{(1)}$  &  $i (\bar d_p \gamma^\mu d_r) [(\bar l_s e_t) D_\mu H]$ \\
$Q_{led^2HD}^{(2)}$  &  $i (\bar d_{p\alpha} \gamma^\mu e_r) [(\bar l_s d_t^\alpha) D_\mu H]$ \\
$Q_{led^2HD}^{(3)}$  &  $i (\bar d_p \gamma^\mu d_r) [(D_\mu \bar l_s e_t) H]$ \\
$Q_{leudHD}^{(1)}$  &  $i \epsilon_{jk} (\bar u_p \gamma^\mu d_r) (\bar e_s l_t^j) D_\mu H^k $ \\
$Q_{leudHD}^{(2)}$  &  $i \epsilon_{jk}  (\bar e_p \gamma^\mu d_r^\alpha) (\bar u_{s\alpha} l_t^j) D_\mu H^k$ \\
$Q_{leudHD}^{(3)}$  &  $i \epsilon_{jk} (\bar u_p \gamma^\mu d_r) (\bar e_s D_\mu l_t^j) H^k $ \\ \hdashline
$Q_{le^3HD}^{(2)}$  &  $i (\bar e_p \gamma^\mu e_r) [(\bar l_s e_t) D_\mu H]$
\end{tabular}
\end{minipage}
\hspace{1cm}
%%%%%%%%%%%%
\begin{minipage}[t]{6.1cm}
\renewcommand{\arraystretch}{1.5}
\begin{tabular}[t]{c|c}
\multicolumn{2}{c}{\boldmath$20:\psi^4HD + \hc$} \\
\hline
$Q_{l^2quHD}^{(1)}$  &  $i (\bar l_p \gamma^\mu l_r) [(\bar q_s u_t) D_\mu \widetilde H]$ \\
$Q_{l^2quHD}^{(2)}$  &  $i (\bar q_{p\alpha} \gamma^\mu l_r) [(\bar l_s u_t^\alpha) D_\mu \widetilde H]$ \\
$Q_{l^2quHD}^{(3)}$  &  $i (\bar l_p \gamma^\mu \tau^I l_r) [(\bar q_s u_t) \tau^I D_\mu \widetilde H]$ \\
$Q_{l^2quHD}^{(4)}$  &  $i (\bar q_{p\alpha} \gamma^\mu \tau^I l_r) [(\bar l_s u_t^\alpha) \tau^I D_\mu \widetilde H]$ \\
$Q_{l^2quHD}^{(5)}$  &  $i (\bar l_p \gamma^\mu l_r) [(\bar q_s D_\mu u_t) \widetilde H]$ \\
$Q_{l^2quHD}^{(6)}$  &  $i (\bar l_p \gamma^\mu \tau^I l_r) [(\bar q_s D_\mu u_t) \tau^I \widetilde H]$ \\
$Q_{e^2quHD}^{(1)}$  &  $i (\bar e_p \gamma^\mu e_r) [(\bar q_s u_t) D_\mu \widetilde H]$ \\
$Q_{e^2quHD}^{(2)}$  &  $i (\bar e_p \gamma^\mu u_r^\alpha) [(\bar q_{s\alpha} e_t) D_\mu \widetilde H]$ \\
$Q_{e^2quHD}^{(3)}$  &  $i (\bar e_p \gamma^\mu e_r) [(D_\mu \bar q_s u_t) \widetilde H]$ \\
$Q_{q^3uHD}^{(1)}$  &  $i (\bar q_p \gamma^\mu q_r) [(\bar q_s u_t) D_\mu \widetilde H]$ \\
$Q_{q^3uHD}^{(2)}$  &  $i (\bar q_p \gamma^\mu \tau^I q_r) [(\bar q_s u_t) \tau^I D_\mu \widetilde H]$ \\
$Q_{q^3uHD}^{(3)}$  &  $i (\bar q_p \gamma^\mu T^A q_r) [(\bar q_s T^A u_t) D_\mu \widetilde H]$ \\
$Q_{q^3uHD}^{(4)}$  &  $i (\bar q_p \gamma^\mu T^A \tau^I q_r) [(\bar q_s T^A u_t) \tau^I D_\mu \widetilde H]$ \\
$Q_{q^3uHD}^{(5)}$  &  $i (\bar q_p \gamma^\mu q_r) [(D_\mu \bar q_s u_t) \widetilde H]$ \\
$Q_{q^3uHD}^{(6)}$  &  $i (\bar q_p \gamma^\mu \tau^I q_r) [(D_\mu \bar q_s u_t) \tau^I  \widetilde H]$ \\
$Q_{qu^3HD}^{(1)}$  &  $i (\bar u_p \gamma^\mu u_r) [(\bar q_s u_t) D_\mu \widetilde H]$ \\
$Q_{qu^3HD}^{(2)}$  &  $i (\bar u_p \gamma^\mu T^A u_r) [(\bar q_s T^A u_t) D_\mu \widetilde H]$ \\
$Q_{qu^3HD}^{(3)}$  &  $i (\bar u_p \gamma^\mu u_r) [(\bar q_s  D_\mu u_t) \widetilde H]$ \\
$Q_{qud^2HD}^{(1)}$  &  $i (\bar d_p \gamma^\mu d_r) [(\bar q_s u_t) D_\mu \widetilde H]$ \\
$Q_{qud^2HD}^{(2)}$  &  $i (\bar d_p \gamma^\mu u_r) [(\bar q_s d_t) D_\mu \widetilde H]$ \\
$Q_{qud^2HD}^{(3)}$  &  $i (\bar d_p \gamma^\mu T^A d_r) [(\bar q_s T^A u_t) D_\mu \widetilde H]$ \\
$Q_{qud^2HD}^{(4)}$  &  $i (\bar d_p \gamma^\mu T^A u_r) [(\bar q_s T^A d_t)  D_\mu \widetilde H$] \\
$Q_{qud^2HD}^{(5)}$  &  $i (\bar d_p \gamma^\mu d_r) [(D_\mu \bar q_s u_t) \widetilde H]$ \\
$Q_{qud^2HD}^{(6)}$  &  $i (\bar d_p \gamma^\mu T^A d_r) [(D_\mu \bar q_s T^A u_t) \widetilde H]$ \
\end{tabular}
\end{minipage}
%%%%%%%%%
\end{adjustbox}
\end{center}
\caption{The dimension-eight operators in the SMEFT of class-20 whose field content superficially includes either an electron-type or up-quark-type Yukawa interaction.  
All of the operators have Hermitian conjugates. 
The subscripts $p, r, s, t$ are weak-eigenstate indices.
The operator below the dashed line is redundant when there is only one generation of fermions.}
\label{tab:smeft8class_20_le_qu}
\end{table}

%%%%%%%%%%%%%%%%%%%%%%%%%%%%%%%%%%%%%

\begin{table}[H]
\begin{center}
%%%%%%%%%%%%%%%%%%%%%%
\begin{adjustbox}{width=0.95\textwidth,center}
\small
%%%%%%%%%%%%
\begin{minipage}[t]{6.1cm}
\renewcommand{\arraystretch}{1.5}
\begin{tabular}[t]{c|c}
\multicolumn{2}{c}{\boldmath$20:\psi^4HD + \hc$} \\
\hline
$Q_{l^2qdHD}^{(1)}$  &  $i (\bar l_p \gamma^\mu l_r) [(\bar q_s d_t) D_\mu H]$ \\
$Q_{l^2qdHD}^{(2)}$  &  $i (\bar q_{p\alpha} \gamma^\mu l_r) [(\bar l_s d_t^\alpha) D_\mu H]$ \\
$Q_{l^2qdHD}^{(3)}$  &  $i (\bar l_p \gamma^\mu \tau^I l_r) [(\bar q_s d_t) \tau^I D_\mu H]$ \\
$Q_{l^2qdHD}^{(4)}$  &  $i (\bar q_{p\alpha} \gamma^\mu \tau^I l_r) [(\bar l_s d_t^\alpha) \tau^I D_\mu H]$ \\
$Q_{l^2qdHD}^{(5)}$  &  $i (\bar l_p \gamma^\mu l_r) [(\bar q_s D_\mu d_t) H]$ \\
$Q_{l^2qdHD}^{(6)}$  &  $i (\bar l_p \gamma^\mu \tau^I l_r) [(\bar q_s D_\mu d_t) \tau^I H]$ \\
$Q_{e^2qdHD}^{(1)}$  &  $i (\bar e_p \gamma^\mu e_r) [(\bar q_s d_t) D_\mu H]$ \\
$Q_{e^2qdHD}^{(2)}$  &  $i (\bar e_p \gamma^\mu d_r^\alpha) [(\bar q_{s\alpha} e_t) D_\mu H]$ \\
$Q_{e^2qdHD}^{(3)}$  &  $i (\bar e_p \gamma^\mu e_r) [(D_\mu \bar q_s d_t) H]$ \\
$Q_{q^3dHD}^{(1)}$  &  $i (\bar q_p \gamma^\mu q_r) [(\bar q_s d_t) D_\mu H]$ \\
$Q_{q^3dHD}^{(2)}$  &  $i (\bar q_p \gamma^\mu \tau^I q_r) [(\bar q_s d_t) \tau^I D_\mu H]$ \\
$Q_{q^3dHD}^{(3)}$  &  $i (\bar q_p \gamma^\mu T^A q_r) [(\bar q_s T^A d_t) D_\mu H]$ \\
$Q_{q^3dHD}^{(4)}$  &  $i (\bar q_p \gamma^\mu T^A \tau^I q_r) [(\bar q_s T^A d_t) \tau^I D_\mu H]$ \\
$Q_{q^3dHD}^{(5)}$  &  $i (\bar q_p \gamma^\mu q_r) [(D_\mu \bar q_s d_t) H]$ \\
$Q_{q^3dHD}^{(6)}$  &  $i (\bar q_p \gamma^\mu \tau^I q_r) [(D_\mu \bar q_s d_t) \tau^I H]$ \\
$Q_{qu^2dHD}^{(1)}$  &  $i (\bar u_p \gamma^\mu u_r) [(\bar q_s d_t) D_\mu H]$ \\
$Q_{qu^2dHD}^{(2)}$  &  $i (\bar u_p \gamma^\mu d_r) [(\bar q_s u_t) D_\mu H]$ \\
$Q_{qu^2dHD}^{(3)}$  &  $i (\bar u_p \gamma^\mu T^A u_r) [(\bar q_s T^A d_t) D_\mu H]$ \\
$Q_{qu^2dHD}^{(4)}$  &  $i (\bar u_p \gamma^\mu T^A d_r) [(\bar q_s T^A u_t)  D_\mu H]$ \\
$Q_{qu^2dHD}^{(5)}$  &  $i (\bar u_p \gamma^\mu u_r) [(D_\mu \bar q_s d_t) H]$ \\
$Q_{qu^2dHD}^{(6)}$  &  $i (\bar u_p \gamma^\mu T^A u_r) [(D_\mu \bar q_s T^A d_t) H]$ \\
$Q_{qd^3HD}^{(1)}$  &  $i (\bar d_p \gamma^\mu d_r) [(\bar q_s d_t) D_\mu H]$ \\
$Q_{qd^3HD}^{(2)}$  &  $i (\bar d_p \gamma^\mu T^A d_r) [(\bar q_s T^A d_t) D_\mu H]$ \\
$Q_{qd^3HD}^{(3)}$  &  $i (\bar d_p \gamma^\mu d_r) [(\bar q_s D_\mu d_t) H]$ \\
\end{tabular}
\end{minipage}
\hspace{1cm}
%%%%%%%%%%%%
\begin{minipage}[t]{7.1cm}
\renewcommand{\arraystretch}{1.5}
\begin{tabular}[t]{c|c}
\multicolumn{2}{c}{\boldmath$20(\slashed B):\psi^4HD + \hc$} \\
\hline
$Q_{lu^2dHD}^{(1)}$  &  $i \epsilon_{\alpha\beta\gamma} [D_\mu H^\dag (u_p^\alpha C \gamma^\mu l_r)] (u_s^\beta C d_t^\gamma)$ \\
$Q_{lu^2dHD}^{(2)}$  &  $i \epsilon_{\alpha\beta\gamma} [H^\dag (u_p^\alpha C \gamma^\mu l_r)] (D_\mu u_s^\beta C d_t^\gamma)$ \\
$Q_{lud^2HD}^{(1)}$  &  $i \epsilon_{\alpha\beta\gamma} \epsilon_{jk} (d_p^\alpha C \gamma^\mu l_r^j) (d_s^\beta C u_t^\gamma) D_\mu H^k$ \\
$Q_{lud^2HD}^{(2)}$  &  $i \epsilon_{\alpha\beta\gamma} \epsilon_{jk} (d_p^\alpha C \gamma^\mu l_r^j) (D_\mu d_s^\beta C u_t^\gamma) H^k$ \\
$Q_{lq^2uHD}^{(1)}$  &  $i \epsilon_{\alpha\beta\gamma} \epsilon_{jn} \epsilon_{km} D_\mu H^{n\dag} (q_p^{m\alpha} C \gamma^\mu u_r^\beta) (q_s^{j\gamma} C l_t^k)$ \\
$Q_{lq^2uHD}^{(2)}$  &  $i \epsilon_{\alpha\beta\gamma} \epsilon_{kn} \epsilon_{jm} D_\mu H^{n\dag} (q_p^{m\alpha} C \gamma^\mu u_r^\beta) (q_s^{j\gamma} C l_t^k)$ \\
$Q_{lq^2uHD}^{(3)}$  &  $i \epsilon_{\alpha\beta\gamma} \epsilon_{jn} \epsilon_{km} H^{n\dag} (q_p^{m\alpha} C \gamma^\mu u_r^\beta) (D_\mu q_s^{j\gamma} C l_t^k)$ \\
$Q_{lq^2dHD}^{(1)}$  &  $i \epsilon_{\alpha\beta\gamma} \epsilon_{jn} \epsilon_{km} (q_p^{m\alpha} C \gamma^\mu d_r^\beta) (q_s^{j\gamma} C l_t^k) D_\mu H^n$ \\
$Q_{lq^2dHD}^{(2)}$  &  $i \epsilon_{\alpha\beta\gamma} \epsilon_{kn} \epsilon_{jm} (q_p^{m\alpha} C \gamma^\mu d_r^\beta) (q_s^{j\gamma} C l_t^k) D_\mu H^n$ \\
$Q_{lq^2dHD}^{(3)}$  &  $i \epsilon_{\alpha\beta\gamma} \epsilon_{jn} \epsilon_{km} (q_p^{m\alpha} C \gamma^\mu d_r^\beta) (D_\mu q_s^{j\gamma} C l_t^k) H^n$ \\
$Q_{eq^3HD}$  &  $i \epsilon_{\alpha\beta\gamma} \epsilon_{mn} \epsilon_{jk} (q_p^{j\alpha} C \gamma^\mu e_r) (D_\mu q_s^{k\beta} C q_t^{m\gamma}) H^n$ \\
$Q_{equ^2HD}^{(1)}$  &  $i \epsilon_{\alpha\beta\gamma} [D_\mu H^\dag (u_p^\alpha C \gamma^\mu q_r^\beta)] (u_s^\gamma C e_t)$ \\
$Q_{equ^2HD}^{(2)}$  &  $i \epsilon_{\alpha\beta\gamma} [H^\dag (u_p^\alpha C \gamma^\mu q_r^\beta)] (D_\mu u_s^\gamma C e_t)$ \\
$Q_{equdHD}^{(1)}$  &  $i \epsilon_{\alpha\beta\gamma} \epsilon_{jk} (q_p^{j\alpha} C \gamma^\mu u_r^\beta) (d_s^\gamma C e_t) D_\mu H^k$ \\
$Q_{equdHD}^{(2)}$  &  $i \epsilon_{\alpha\beta\gamma} \epsilon_{jk} (q_p^{j\alpha} C \gamma^\mu d_r^\beta) (u_s^\gamma C e_t) D_\mu H^k$ \\
$Q_{equdHD}^{(3)}$  &  $i \epsilon_{\alpha\beta\gamma} \epsilon_{jk} (q_p^{j\alpha} C \gamma^\mu u_r^\beta) (d_s^\gamma C D_\mu e_t) H^k$ 
\end{tabular}
\end{minipage}
%%%%%%%%%
\end{adjustbox}
\end{center}
\caption{The dimension-eight operators in the SMEFT of class-20 whose field content superficially includes a down-quark-type Yukawa interaction or is baryon number violating.  
All of the operators have Hermitian conjugates. 
The subscripts $p, r, s, t$ are weak-eigenstate indices.}
\label{tab:smeft8class_20_qd_slashedB}
\end{table}

%%%%%%%%%%%%%%%%%%%%%%%%%%%%%%%%%%%%%%%%%%
% SMEFT d=8 Class 21
%%%%%%%%%%%%%%%%%%%%%%%%%%%%%%%%%%%%%%%%%%

\begin{table}[H]
\begin{center}
%%%%%%%%%%%%%%%%%%%%%%
\begin{adjustbox}{width=0.81\textwidth,center}
\small
%%%%%%%%%%%%
\begin{minipage}[t]{5.1cm}
\renewcommand{\arraystretch}{1.5}
\begin{tabular}[t]{c|c}
\multicolumn{2}{c}{\boldmath$21:(\bar L L)(\bar L L)D^2$} \\
\hline
$Q_{l^4D^2}^{(1)}$  &  $D^\nu (\bar l_p \gamma^\mu l_r) D_\nu (\bar l_s \gamma_\mu l_t)$  \\
$Q_{l^4D^2}^{(2)}$  &  $(\bar l_p \gamma^\mu \overleftrightarrow{D}^\nu l_r) (\bar l_s \gamma_\mu \overleftrightarrow{D}_\nu l_t)$ \\
$Q_{q^4D^2}^{(1)}$  &  $D^\nu (\bar q_p \gamma^\mu q_r) D_\nu (\bar q_s \gamma_\mu q_t)$  \\
$Q_{q^4D^2}^{(2)}$  &  $(\bar q_p \gamma^\mu \overleftrightarrow{D}^\nu q_r) (\bar q_s \gamma_\mu \overleftrightarrow{D}_\nu q_t)$ \\
$Q_{q^4D^2}^{(3)}$  &  $D^\nu (\bar q_p \gamma^\mu \tau^I q_r) D_\nu (\bar q_s \gamma_\mu \tau^I q_t)$  \\
$Q_{q^4D^2}^{(4)}$  &  $(\bar q_p \gamma^\mu \overleftrightarrow{D}^{I\nu} q_r) (\bar q_s \gamma_\mu \overleftrightarrow{D}^I_\nu q_t)$ \\
$Q_{l^2q^2D^2}^{(1)}$  &  $D^\nu (\bar l_p \gamma^\mu l_r) D_\nu (\bar q_s \gamma_\mu q_t)$  \\
$Q_{l^2q^2D^2}^{(2)}$  &  $(\bar l_p \gamma^\mu \overleftrightarrow{D}^\nu l_r) (\bar q_s \gamma_\mu \overleftrightarrow{D}_\nu q_t)$ \\
$Q_{l^2q^2D^2}^{(3)}$  &  $D^\nu (\bar l_p \gamma^\mu \tau^I l_r) D_\nu (\bar q_s \gamma_\mu \tau^I q_t)$  \\
$Q_{l^2q^2D^2}^{(4)}$  &  $(\bar l_p \gamma^\mu \overleftrightarrow{D}^{I\nu} l_r) (\bar q_s \gamma_\mu \overleftrightarrow{D}^I_\nu q_t)$
\end{tabular}
\end{minipage}
\hspace{1cm}
%%%%%%%%%%%%
\begin{minipage}[t]{6.1cm}
\renewcommand{\arraystretch}{1.5}
\begin{tabular}[t]{c|c}
\multicolumn{2}{c}{\boldmath$21:(\bar R R)(\bar R R)D^2$} \\
\hline
$Q_{e^4D^2}$  &  $D^\nu (\bar e_p \gamma^\mu e_r) D_\nu (\bar e_s \gamma_\mu e_t)$  \\
$Q_{u^4D^2}^{(1)}$  &  $D^\nu (\bar u_p \gamma^\mu u_r) D_\nu (\bar u_s \gamma_\mu u_t)$  \\
$Q_{u^4D^2}^{(2)}$  &  $(\bar u_p \gamma^\mu \overleftrightarrow{D}^\nu u_r) (\bar u_s \gamma_\mu \overleftrightarrow{D}_\nu u_t)$ \\
$Q_{d^4D^2}^{(1)}$  &  $D^\nu (\bar d_p \gamma^\mu d_r) D_\nu (\bar d_s \gamma_\mu d_t)$  \\
$Q_{d^4D^2}^{(2)}$  &  $(\bar d_p \gamma^\mu \overleftrightarrow{D}^\nu d_r) (\bar d_s \gamma_\mu \overleftrightarrow{D}_\nu d_t)$ \\
$Q_{e^2u^2D^2}^{(1)}$  &  $D^\nu (\bar e_p \gamma^\mu e_r) D_\nu (\bar u_s \gamma_\mu u_t)$  \\
$Q_{e^2u^2D^2}^{(2)}$  &  $(\bar e_p \gamma^\mu \overleftrightarrow{D}^\nu e_r) (\bar u_s \gamma_\mu \overleftrightarrow{D}_\nu u_t)$ \\
$Q_{e^2d^2D^2}^{(1)}$  &  $D^\nu (\bar e_p \gamma^\mu e_r) D_\nu (\bar d_s \gamma_\mu d_t)$  \\
$Q_{e^2d^2D^2}^{(2)}$  &  $(\bar e_p \gamma^\mu \overleftrightarrow{D}^\nu e_r) (\bar d_s \gamma_\mu \overleftrightarrow{D}_\nu d_t)$ \\
$Q_{u^2d^2D^2}^{(1)}$  &  $D^\nu (\bar u_p \gamma^\mu u_r) D_\nu (\bar d_s \gamma_\mu d_t)$  \\
$Q_{u^2d^2D^2}^{(2)}$  &  $(\bar u_p \gamma^\mu \overleftrightarrow{D}^\nu u_r) (\bar d_s \gamma_\mu \overleftrightarrow{D}_\nu d_t)$ \\
$Q_{u^2d^2D^2}^{(3)}$  &  $D^\nu (\bar u_p \gamma^\mu T^A u_r) D_\nu (\bar d_s \gamma_\mu T^A d_t)$  \\
$Q_{u^2d^2D^2}^{(4)}$  &  $(\bar u_p \gamma^\mu T^A \overleftrightarrow{D}^\nu u_r) (\bar d_s \gamma_\mu T^A \overleftrightarrow{D}_\nu d_t)$
\end{tabular}
\end{minipage}
%%%%%%%%%
\end{adjustbox}
%%%%%%%%%%%%%%%%%%%%%%
\begin{adjustbox}{width=0.84\textwidth,center}
\small
%%%%%%%%%%%%
\begin{minipage}[t]{5.9cm}
\renewcommand{\arraystretch}{1.5}
\begin{tabular}[t]{c|c}
\multicolumn{2}{c}{\boldmath$21:(\bar L L)(\bar R R)D^2$} \\
\hline
$Q_{l^2e^2D^2}^{(1)}$  &  $D^\nu (\bar l_p \gamma^\mu l_r) D_\nu (\bar e_s \gamma_\mu e_t)$  \\
$Q_{l^2e^2D^2}^{(2)}$  &  $(\bar l_p \gamma^\mu \overleftrightarrow{D}^\nu l_r) (\bar e_s \gamma_\mu \overleftrightarrow{D}_\nu e_t)$ \\
$Q_{l^2u^2D^2}^{(1)}$  &  $D^\nu (\bar l_p \gamma^\mu l_r) D_\nu (\bar u_s \gamma_\mu u_t)$  \\
$Q_{l^2u^2D^2}^{(2)}$  &  $(\bar l_p \gamma^\mu \overleftrightarrow{D}^\nu l_r) (\bar u_s \gamma_\mu \overleftrightarrow{D}_\nu u_t)$ \\
$Q_{l^2d^2D^2}^{(1)}$  &  $D^\nu (\bar l_p \gamma^\mu l_r) D_\nu (\bar d_s \gamma_\mu d_t)$  \\
$Q_{l^2d^2D^2}^{(2)}$  &  $(\bar l_p \gamma^\mu \overleftrightarrow{D}^\nu l_r) (\bar d_s \gamma_\mu \overleftrightarrow{D}_\nu d_t)$ \\
$Q_{q^2e^2D^2}^{(1)}$  &  $D^\nu (\bar q_p \gamma^\mu q_r) D_\nu (\bar e_s \gamma_\mu e_t)$  \\
$Q_{q^2e^2D^2}^{(2)}$  &  $(\bar q_p \gamma^\mu \overleftrightarrow{D}^\nu q_r) (\bar e_s \gamma_\mu \overleftrightarrow{D}_\nu e_t)$ \\
$Q_{q^2u^2D^2}^{(1)}$  &  $D^\nu (\bar q_p \gamma^\mu q_r) D_\nu (\bar u_s \gamma_\mu u_t)$  \\
$Q_{q^2u^2D^2}^{(2)}$  &  $(\bar q_p \gamma^\mu \overleftrightarrow{D}^\nu q_r) (\bar u_s \gamma_\mu \overleftrightarrow{D}_\nu u_t)$ \\
$Q_{q^2u^2D^2}^{(3)}$  &  $D^\nu (\bar q_p \gamma^\mu T^A q_r) D_\nu (\bar u_s \gamma_\mu T^A u_t)$  \\
$Q_{q^2u^2D^2}^{(4)}$  &  $(\bar q_p \gamma^\mu T^A  \overleftrightarrow{D}^\nu q_r) (\bar u_s \gamma_\mu T^A  \overleftrightarrow{D}_\nu u_t)$ \\
$Q_{q^2d^2D^2}^{(1)}$  &  $D^\nu (\bar q_p \gamma^\mu q_r) D_\nu (\bar d_s \gamma_\mu d_t)$  \\
$Q_{q^2d^2D^2}^{(2)}$  &  $(\bar q_p \gamma^\mu \overleftrightarrow{D}^\nu q_r) (\bar d_s \gamma_\mu \overleftrightarrow{D}_\nu d_t)$ \\
$Q_{q^2d^2D^2}^{(3)}$  &  $D^\nu (\bar q_p \gamma^\mu T^A q_r) D_\nu (\bar d_s \gamma_\mu T^A d_t)$  \\
$Q_{q^2d^2D^2}^{(4)}$  &  $(\bar q_p \gamma^\mu T^A  \overleftrightarrow{D}^\nu q_r) (\bar d_s \gamma_\mu T^A  \overleftrightarrow{D}_\nu d_t)$
\end{tabular}
\end{minipage}
\hspace{1cm}
%%%%%%%%%%%%
\begin{minipage}[t]{5.7cm}
\renewcommand{\arraystretch}{1.5}
\begin{tabular}[t]{c|c}
\multicolumn{2}{c}{\boldmath$21:(\bar L R)(\bar L R)D^2 + \hc$} \\
\hline

$Q_{q^2udD^2}^{(1)}$  &  $D_\mu (\bar q_p^j u_r) \epsilon_{jk} D^\mu (\bar q_s^k d_t)$  \\
$Q_{q^2udD^2}^{(2)}$  &  $D_\mu (\bar q_p^j T^A u_r) \epsilon_{jk} D^\mu (\bar q_s^k T^A d_t)$   \\
$Q_{q^2udD^2}^{(3)}$  &  $(\bar q_p^j \overleftrightarrow{D}^\mu u_r) \epsilon_{jk} (\bar q_s^k \overleftrightarrow{D}_\mu  d_t)$   \\
$Q_{lequD^2}^{(1)}$  &  $D_\mu (\bar l_p^j e_r) \epsilon_{jk} D^\mu (\bar q_s^k u_t)$ \\
$Q_{lequD^2}^{(2)}$  &  $D_\mu (\bar l_p^j u_r^\alpha) \epsilon_{jk} D^\mu (\bar q_{s\alpha}^k e_t)$ \\
$Q_{lequD^2}^{(3)}$  &  $(\bar l_p^j  \overleftrightarrow{D}^\mu e_r) \epsilon_{jk} (\bar q_s^k \overleftrightarrow{D}_\mu u_t) $  
\end{tabular}
\end{minipage}
%%%%%%%%%
\end{adjustbox}
\end{center}
\caption{Most of the dimension-eight operators in the SMEFT of class-21, which are further divided into subclasses according to their chiral properties. 
See Table~\ref{tab:smeft8class_18_21} for the remaining class-21 operators.
Operators with ${}+\hc$ have Hermitian conjugates.
The subscripts $p, r, s, t$ are weak-eigenstate indices. }
\label{tab:smeft8class_21}
\end{table}

%%----------------------------------------------------------------------------------------------------------------------------------------------------------------

\section{Phenomenology}
\label{sec:pheno}

In this Section we briefly discuss a few aspects of phenomenology involving dimension-8 operators, focusing on processes that first start at dimension-8 and/or involve interplay between dimension-6 and -8 effects.
In particular, we discuss light-by-light scattering and electroweak precision data.
We also present a model where dimension-8 effects are arguably more important than dimension-6 effects.

Note that sometimes we will explicitly write factors of the cutoff scale of the effective theory, $C_i \to c_i / \Lambda^{d-4}$, to better highlight the different orders in the EFT expansion in our phenomenological studies.

%--------------------------------------------------------------------------------------------------------------
\subsection{Light-by-Light Scattering}

The possibility of non-linear processes involving solely photons had been discussed back in the 1930s~\cite{Halpern:1933dya, Euler:1935zz, Heisenberg:1935qt, Euler:1936oxn}.
Later in the 1950s the cross section for elastic light-by-light scattering was computed in QED~\cite{Karplus:1950zza, Karplus:1950zz}.
Almost 70 years would pass before this process was finally observed in vacuum in 2019 by the ATLAS collaboration at the LHC~\cite{Aad:2019ock}.
Additionally, the CMS collaboration reports evidence for elastic light-by-light scattering in vacuum~\cite{Sirunyan:2018fhl}.

Interactions in the SMEFT involving four photons first start at dimension-8
\begin{equation}
\label{eq:lbl}
\mathcal{L}_{LbL} = \frac{\alpha^2}{90 M_e^4} \left[ \mathscr{C}_{LbL}^{(1)} \left(F_{\mu\nu} F^{\mu\nu}\right)^2 + \mathscr{C}_{LbL}^{(2)} \left(F_{\mu\nu} \widetilde F^{\mu\nu}\right)^2 + \widetilde{\mathscr{C}}_{LbL} \left(F_{\mu\nu} F^{\mu\nu}\right) \left(F_{\rho\sigma} \widetilde F^{\rho\sigma}\right) \right] ,
\end{equation}
where $F_{\mu\nu}$ is the photon field strength, $\alpha$ is the fine structure constant, and $M_e$ is the mass of the electron.
The normalization is such that order one coefficients are generated when the electron is integrated out with $E \ll M_e$.

On the other hand, for a less restrictive energy range, $E \ll \Lambda$, the SMEFT Wilson coefficients from Table~\ref{tab:smeft8class_1_2_3_4} appearing in Eq.~\eqref{eq:lbl} are
\begin{align}
\label{eq:lblco}
\mathscr{C}_{LbL}^{(1)} &= \frac{90 M_e^4}{\Lambda^4} 16 \pi^2 \left[ \frac{1}{g_2^4} \left( c_{W^4}^{(1)} + c_{W^4}^{(3)} \right) + \frac{1}{g_1^4} c_{B^4}^{(1)} + \frac{1}{g_2^2 g_1^2} \left( c_{W^2B^2}^{(1)} + c_{W^2B^2}^{(3)} \right) \right] , \nn
\mathscr{C}_{LbL}^{(2)} &= \frac{90 M_e^4}{\Lambda^4} 16 \pi^2 \left[ \frac{1}{g_2^4} \left( c_{W^4}^{(2)} + c_{W^4}^{(4)} \right) + \frac{1}{g_1^4} c_{B^4}^{(2)} + \frac{1}{g_2^2 g_1^2} \left( c_{W^2B^2}^{(2)} + c_{W^2B^2}^{(4)} \right)  \right] , \nn
\widetilde{\mathscr{C}}_{LbL} &= \frac{90 M_e^4}{\Lambda^4} 16 \pi^2 \left[ \frac{1}{g_2^4} \left( c_{W^4}^{(5)} + c_{W^4}^{(6)} \right) + \frac{1}{g_1^4} c_{B^4}^{(3)} + \frac{1}{g_2^2 g_1^2} \left( c_{W^2B^2}^{(5)} + c_{W^2B^2}^{(6)} + c_{W^2B^2}^{(7)} \right)  \right] ,
\end{align}
where we have dropped terms that are not enhanced by $\O(16 \pi^2)$.
If the physics beyond the SM (BSM) that generates~\eqref{eq:lblco} is loop suppressed then these additional terms must be included.

Axiomatic principles of quantum field theory such as unitarity, analyticity, and crossing symmetry yield constraints on the parameters of a theory, see \textit{e.g.}~\cite{Manohar:2008tc, Bellazzini:2016xrt}. 
For example, Refs.~\cite{Gunion:1990kf, Grinstein:2013fia} used a once-subtracted dispersion relation to derive sum rules for the couplings of an extended Higgs boson sector.
Dimension-8 SMEFT operators with four derivatives are constrained by twice-subtracted dispersion relations, which lead to positivity bounds on Wilson coefficients~\cite{Bellazzini:2018paj, Zhang:2018shp, Bi:2019phv, Remmen:2019cyz, Remmen:2020vts}.
This follows from the contour at infinity vanishing due to the Froissart bound~\cite{Froissart:1961ux}, and the crossed-channel branch-cut having the same sign the original-channel making the sum of non-vanishing integrals positive-definite by the optical theorem.

The positivity bounds on the coefficients in~\eqref{eq:lblco} are~\cite{Remmen:2019cyz}
\begin{align}
\label{eq:posi}
\mathscr{C}_{LbL}^{(1)} &> 0 , \nn
\mathscr{C}_{LbL}^{(2)} &> 0 , \nn
4 \mathscr{C}_{LbL}^{(1)} \mathscr{C}_{LbL}^{(2)} &> (\widetilde{\mathscr{C}}_{LbL})^2 .
\end{align}
Unsurprisingly QED satisfies these bounds, $\mathscr{C}_{LbL}^{(1)} = 1$ and $\mathscr{C}_{LbL}^{(2)} = 7 / 4$ when $E \ll M_e$ and the electron is integrated out at one-loop~\cite{Schwinger:1951nm}.
The coefficient $\widetilde{\mathscr{C}}_{LbL}$ is not generated in QED as electromagnetic interactions conserve parity, which also satisfies~\eqref{eq:posi}.
The bounds~\ref{eq:posi} can be combined with other positivity bounds coming from different initial scattering states to further constrain the $X^4$ SMEFT Wilson coefficients.

%--------------------------------------------------------------------------------------------------------------
\subsection{Electroweak Precision Data}

Historically, the constraints on BSM physics from electroweak precision data were frequently summarized in terms of the parameters $S$, $T$, and $U$~\cite{Peskin:1990zt, Marciano:1990dp, Kennedy:1990ib, Holdom:1990tc, Golden:1990ig, Altarelli:1990zd}.
The leading contributions to $S$ and $T$ come from dimension-6 operators, whereas $U$ first arises from a dimension-8 operator~\cite{Grinstein:1991cd} motivating its discussion here.
Additionally this discussion will help frame the results of Sec.~\ref{sec:quartets}.
However, for heavy BSM physics, it is important to keep in mind that the SMEFT is the preferred framework to use to describe electroweak precision data as it is completely general, unlike an $STU$ analysis.
For example, the dimension-6 operators that contribute to $S$ and $T$ also contribute to Higgs and diboson processes, see \textit{e.g.}~\cite{Ellis:2018gqa}.
To put a modern twist on this analysis we use the geometric interpretation of the SMEFT (geoSMEFT)~\cite{Helset:2018fgq, Helset:2020yio, Hays:2020scx} to formulate expressions for $S$, $T$, and $U$ to all orders in $v_T / \Lambda$, with $v_T$ defined in Eq.~\eqref{eq:vT}.

Higher-dimensional operators change the definitions of SM parameters in a variety of ways.
Field redefinitions are needed to relate these combinations of inputs to measured quantities.
To start, consider the potential for the Higgs field in the SMEFT through dimension-8
\begin{equation}
\label{eq:V}
V_{\rm SMEFT} = \lambda \left(H^\dag H - \frac{v^2}{2}\right)^2 - \frac{c_H}{\Lambda^2} (H^\dag H)^3 - \frac{c_{H^8}}{\Lambda^4} (H^\dag H)^4 .
\end{equation}
The dimension-8 operators relevant for EWPD are given in Tables~\ref{tab:smeft8class_1_2_3_4} and~\ref{tab:smeft8class_5_6_7_8}.
Due to the presence of the higher-dimensional operators in Eq.~\eqref{eq:V}, the minimum of Higgs boson potential is shifted~\cite{Hays:2018zze, Helset:2020yio}
\begin{equation}
\label{eq:vT}
\langle H^\dag H \rangle \equiv \frac{v_T^2}{2} = \frac{v^2}{2}\left(1 + \frac{v^2}{\Lambda^2} \frac{3 c_H}{4 \lambda} + \frac{v^4}{\Lambda^4} \frac{9 c_H^2 + 4 \lambda c_{H^8}}{8 \lambda^2} \right)
\end{equation}
with $v_T \approx 246$~GeV.

Now we turn to canonically normalizing the electroweak gauge, Higgs, and Goldstone bosons.
Care must be taken as the field redefinitions are matrix equations.
The geometric interpretation provides an elegant way to perform these transformations to all-orders in $v_T / \Lambda$.
The Higgs-derivative operators through dimension-8 are
\begin{align}
\label{eq:Hkin}
\mathcal{L}_{H, \rm kin} &= (D_\mu H^\dag) (D^\mu H) + \frac{c_{H\Box}}{\Lambda^2} (H^\dag H) \Box (H^\dag H) + \frac{c_{HD}}{\Lambda^2} [(D_\mu H^\dag) H] [H^\dag (D^\mu H)] \nn
&+\frac{c_{H^6D^2}^{(1)}}{\Lambda^4} (H^\dag H)^2 (D_\mu H^\dag D^\mu H) + \frac{c_{H^6D^2}^{(2)}}{\Lambda^4} (H^\dag H) (H^\dag \tau^I H) (D_\mu H^\dag \tau^I D^\mu H) .
\end{align}
Defining
\begin{equation}
H = \frac{1}{\sqrt{2}} 
\begin{pmatrix}
\phi_2 + i \phi_1 \\
\phi_4 - i \phi_3
\end{pmatrix} ,
\end{equation}
we can write Eq.~\eqref{eq:Hkin} in the language of the geoSMEFT as 
\begin{equation}
\mathcal{L}_{H, \rm kin} = \frac{1}{2} h_{\mathcal{I} \mathcal{J}}(\phi) \phi^\mathcal{I} \phi^\mathcal{J}  ,
\end{equation}
where $\phi_\mathcal{I} = (\phi_1, \phi_2, \phi_3, \phi_4)$ with $\langle \phi_\mathcal{I} \rangle = v_T \delta_{\mathcal{I} 4}$.
The weak eigenstates are related to the mass eigenstates through a metric on Higgs field space, $h_{\mathcal{I} \mathcal{J}}$, and a unitary matrix not considered here.
We are working with the weak eigenstates and have not introduced the mass eigenstates here as all we need is the metric $h_{\mathcal{I} \mathcal{J}}$ to define the $T$ parameter.
See Ref.~\cite{Helset:2020yio} for expressions involving mass eigenstates.
The gauge kinetic terms to all-orders in $v_T / \Lambda$ are
\begin{align}
\label{eq:lew}
\mathcal{L}_{\rm EW,\, kin} &= - \frac{1}{4} W^I_{\mu\nu} W^{I\mu\nu} - \frac{1}{4}  B_{\mu\nu} B^{\mu\nu} \\
&+ \frac{c_{HW}}{\Lambda^2} (H^\dag H) W^I_{\mu\nu} W^{I\mu\nu} + \frac{c_{HB}}{\Lambda^2} (H^\dag H) B_{\mu\nu} B^{\mu\nu} + \frac{c_{HWB}}{\Lambda^2} (H^\dag \tau^I H) W^I_{\mu\nu} B^{\mu\nu}  \nn
&+ \sum_{n=0}^\infty \frac{H^{2n}}{\Lambda^{4+2n}} \left(c_{B^2H^{4+2n}}^{(1)}(H^\dag H)^2 B_{\mu\nu} B^{\mu\nu} + c_{W^2H^{4+2n}}^{(1)} (H^\dag H)^2 W^I_{\mu\nu} W^{I\mu\nu} \right. \nn
&\left. + c_{WBH^{4+2n}}^{(1)} (H^\dag H) (H^\dag \tau^I H) W^I_{\mu\nu} B^{\mu\nu} + c_{W^2H^{4+2n}}^{(3)} (H^\dag \tau^I H) (H^\dag \tau^J H) W^I_{\mu\nu} W^{J\mu\nu} \right)\nonumber
\end{align}
In the geoSMEFT Eq.~\eqref{eq:lew} takes the form
\begin{equation}
\mathcal{L}_{\rm EW,\, kin} = - \frac{1}{4} g_{\mathcal{A} \mathcal{B}}(H) \mathcal{W}^\mathcal{A} \mathcal{W}^\mathcal{B} ,
\end{equation}
where $\mathcal{W}^\mathcal{A} = (W^I, B) = (W^1, W^2, W^3, B)$ and $g_{\mathcal{A} \mathcal{B}}$ is another metric on Higgs field space
This metric, $g_{\mathcal{A} \mathcal{B}}$, again along with a unitary matrix not considered here, relates the weak eigenstate gauge bosons to the mass eigenstates.

The geometric definitions of $S$, $T$, and $U$ are
\begin{align}
\label{eq:geostu}
\frac{1}{16\pi} S &= \langle g_{34} \rangle + \langle g_{43} \rangle , \nn
\bar \alpha\, T &= \langle h_{11} \rangle - \langle h_{33} \rangle = \langle h_{22} \rangle - \langle h_{33} \rangle , \nn
\frac{1}{16\pi} U &= \langle g_{11} \rangle - \langle g_{33} \rangle = \langle g_{22} \rangle - \langle g_{33} \rangle
\end{align}
where $\bar \alpha$ takes into account shift in the definition of the fine structure constant due to the presence of higher-dimensional operators, see Appendix A of~\cite{Hays:2020scx} for an explicit expression.
Note this shift does not affect our analysis of light-by-light scattering; since all of those effects start at dimension-8 we can freely trade $\alpha$ for $\bar\alpha$.
Given~\eqref{eq:geostu} it straightforward to work out the contributions to $S$, $T$, and $U$
\begin{align}
\label{eq:stu}
\frac{1}{16\pi} S &= \frac{v_T^2}{\Lambda^2} c_{HWB} + \sum_{n=0}^\infty \frac{v_T^{4+2n}}{2^n \Lambda^{4+2n}} c_{WBH^{4+2n}}^{(1)}, \nn
\bar\alpha\, T &= - \frac{v_T^2}{2 \Lambda^2} c_{HD} - \frac{v_T^4}{2 \Lambda^4} c_{H^6D^2}^{(2)} , \nn
\frac{1}{16\pi} U &= \sum_{n=0}^\infty \frac{v_T^{4+2n}}{2^n \Lambda^{4+2n}} c_{W^2H^{4+2n}}^{(3)} ,
\end{align}
where we give $T$ to $\mathcal{O}(v_T^4 / \Lambda^4)$ and $S$ and $U$ to all-orders in $v_T / \Lambda$.

%--------------------------------------------------------------------------------------------------------------
\subsection{Scalar $SU(2)_w$ Quartets}
\label{sec:quartets}

Although the dimension-6 operators are formally the leading terms in the EFT expansion, there are various reasons why it may be necessary to consider dimension-8 effects.
Here we explore a scenario where the difference in experimental precision in different classes of measurements causes dimension-8 effects to be important.
The measurements are double Higgs boson production for which only upper limits exists on the cross section exist~\cite{Sirunyan:2018ayu, Aad:2019yxi}, and EWPD, which as the name implies, is precisely measured.
For example, one of the more precisely measured observables in this class is the width of the $Z$ boson, which has a relative precision of $9 \cdot 10^{-4}$~\cite{ALEPH:2005ab}.

The models under consideration add to the SM field content a new scalar field, $\Theta$, that is an $SU(2)_w$ quartet with either $\hyp = 3/2$ or $1/2$. 
The Lagrangian for the $\hyp = 3/2$ case is
\begin{align}
\label{eq:quar}
\mathcal{L}_\Theta &= D_\mu \Theta^\dag D^\mu \Theta - M^2 \Theta^\dag \Theta  + [\lambda_1 \Theta_{jkm}^\dag H^j H^k H^m + \hc] \nn
&- \lambda_{\alpha1} (H^\dag H) (\Theta^\dag \Theta)  - \lambda_{\alpha2} (H_n^\dag H^m) (\Theta_{jkm}^\dag \Theta^{jkn})  + \O(\Theta^4) .
\end{align}
For the $\hyp = 1/2$ case the term linear in $\Theta$ is instead $[\lambda_1 \Theta_{jkm}^\dag H^j H^k \widetilde H^m + \hc]$.
Note that $\Theta$ is completely symmetric in its $SU(2)_w$ indices.
Assuming $M \gg v_T$ we integrate out $\Theta$ and match to the SMEFT.
In both cases only one dimension-6 operator, $(H^\dag H)^3$, is generated  at tree level~\cite{Henning:2014wua, Dawson:2017vgm, deBlas:2017xtg}
\begin{equation}
C_H = \frac{2\hyp}{3} \frac{|\lambda_1|^2}{M^2} 
\end{equation}
while at one-loop the triple $W$ operator, $Q_W$, is also generated~\cite{Henning:2014wua, Bakshi:2018ics}.
At this stage the scalar quartets look like great candidates to enhance the double Higgs boson production rate.

However things are not so simple.
When electroweak symmetry is broken the term linear in $\Theta$ in Eq.~\eqref{eq:quar} will force $\Theta$ to have a non-zero vacuum expectation value (vev).
If $\Theta$ gets a vev, $v_\Theta$, its quantum numbers dictate that it contributes to the $T$ parameter~\cite{Dawson:2017vgm}
\begin{equation}
1 + \bar\alpha\, T = \frac{v_T^2}{v_T^2 - 2 v_\Theta^2 [\tfrac{3}{2} (\tfrac{3}{2} + 1) - 3 \hyp^2]} 
\end{equation}

With this in mind we extend the matching to dimension-8 starting with the $\hyp = 3/2$ case.
We find
\begin{align}
\label{eq:quar8}
\mathcal{L}_\Theta^{(d=8)} &= - \frac{\lambda_\alpha |\lambda_1|^2}{M^4} (H^\dag H)^4 + \frac{3 |\lambda_1|^2}{M^4} (H^\dag H)^2 (D_\mu H^\dag) (D^\mu H) \nn
&+ \frac{6 |\lambda_1|^2}{M^4} (H^\dag H) H^\dag_j H^k (D_\mu H^\dag_k) (D^\mu H^j)
\end{align}
with $\lambda_\alpha =  \lambda_{\alpha1} +  \lambda_{\alpha2}$.
The last term on the right-hand side of Eq.~\eqref{eq:quar8} is not in our operator basis.
We use the Fierz identity for Pauli matrices, Eq.~\eqref{eq:fierz}, to convert that operator into our basis, yielding for the $\hyp = 3/2$ case
\begin{equation}
C_{H^8} = - \frac{\lambda_\alpha |\lambda_1|^2}{M^4}, \quad C_{H^6D^2}^{(1)} = \frac{6 |\lambda_1|^2}{M^4}, \quad C_{H^6D^2}^{(2)} = \frac{3 |\lambda_1|^2}{M^4} .
\end{equation}
The $\hyp = 1/2$ case is slightly more complicated
\begin{align}
\mathcal{L}_\Theta^{(d=8)} &= - \frac{\lambda_\alpha |\lambda_1|^2}{3M^4} (H^\dag H)^4 + \frac{7|\lambda_1|^2}{3M^4} (H^\dag H)^2 (D_\mu H^\dag) (D^\mu H) \nn
&+ \frac{2 |\lambda_1|^2}{3M^4} (H^\dag H) \left[-H^\dag_j H^k (D_\mu H^\dag_k) (D^\mu H^j)\right. \nn
&\left. + H^\dag_j H^\dag_k (D_\mu H^k) (D^\mu H^j) + H^j H^k (D_\mu H^\dag_k) (D^\mu H^\dag_j) \right]
\end{align}
Integrating the last two terms by parts, applying the Higgs boson EOM to fields with two derivatives acting on them, and using Eq.~\eqref{eq:fierz} we find
\begin{align}
&C_H = \left(1 - \frac{4 \lambda v^2}{3M^2}\right) \frac{|\lambda_1|^2}{M^2},\quad C_{H^8} = \left(8 \lambda - \lambda_\alpha \right) \frac{|\lambda_1|^2}{3M^4},\nn
&C_{H^6D^2}^{(1)} = \frac{2 |\lambda_1|^2}{3M^4},\quad C_{H^6D^2}^{(2)} = - \frac{|\lambda_1|^2}{M^4}, \nn
&C_{leH^5} =  \frac{2}{3} Y_e^\dag \frac{ |\lambda_1|^2}{ M^4},\quad C_{quH^5} = \frac{2}{3} Y_u^\dag \frac{ |\lambda_1|^2}{ M^4},\quad C_{qdH^5} =  \frac{2}{3} Y_d^\dag \frac{ |\lambda_1|^2}{M^4} .
\end{align}
We included the dimension-6 coefficient $C_H$ here as it receives another contribution from mass term in the Higgs EOM, which was used to reduce a dimension-8 operator into operators in our basis. 
This completes the matching of the scalar quartet models to the SMEFT at dimension-8.

Using~\eqref{eq:stu} we see that $T_{\hyp=3/2} = - 3 T_{\hyp=1/2} = - \frac{3}{2\bar\alpha} |\lambda_1|^2 (\tfrac{v_T}{M})^4$.
Our result for the contributions of these models to the $T$ parameter agrees with what was found in Ref.~\cite{Dawson:2017vgm}.
The implication of our matching results is that these model cannot in fact provide a large enhancement to double Higgs boson production.
This was unclear, from an EFT perspective at least, until the matching was done at dimension-8.

%%----------------------------------------------------------------------------------------------------------------------------------------------------------------

\section{Comments on Renormalization Group Evolution}
\label{sec:rge}

By construction measurements in an effective field theory take place at energies below the cutoff scale of the EFT.
As such, to link measurements to parameters of possible physics in the UV, it is crucial to understand how the coefficients of the EFT operators evolve from one scale to another.
This behavior is described by the renormalization group evolution equations
\begin{equation}
\gamma_{ij} C_j = 16 \pi^2 \mu \frac{d C_i}{d \mu} \equiv \dot C_i ,
\end{equation}
where $\gamma$ is the anomalous dimension matrix and $\mu$ is the renormalization scale.

The renormalization group evolution of the SMEFT operators at one-loop is known at dimension-5~\cite{Babu:1993qv}, dimension-6~\cite{Jenkins:2013zja, Jenkins:2013wua, Alonso:2013hga, Alonso:2014zka}, and dimension-7~\cite{Liao:2016hru, Liao:2019tep}.
The loop diagrams in the preceding computations were built from one dimension-$d$ SMEFT contact amplitude and one tree level SM amplitude.
For the dimension-6 RGE there are also one-loop amplitudes resulting from two insertions of dimension-5 operators, and Ref.~\cite{Davidson:2018zuo} computed these contributions to the RGE equations.
Typically, the two dimension-5 operator contribution to the dimension-6 RGE will be suppressed with respect to the one dimension-6 contribution because all odd mass dimension operators in the SMEFT violate at least one of baryon and lepton number~\cite{Kobach:2016ami, Helset:2019eyc}.

It is well beyond the scope of this work to compute the RGE equations for the dimension-8 operators.
Instead we content ourselves to make a few comments about the structure of the associated anomalous dimension matrix. 
The renormalization of a dimension-8 operator can happen due to one insertion of a dimension-8 operator in a loop amplitude, two insertions of dimension-6 operators, or one insertion of dimension-5 operator and one insertion of a dimension-7 operator.
The lattermost type of amplitude will typically be suppressed as was the case with the two dimension-5 operator contribution to the dimension-6 RGE.
However, the two dimension-6 operator contribution to the dimension-8 RGE will generally be comparable in magnitude to the one dimension-8 operator contribution.
Dimension-8 is the lowest mass dimension where there are two co-leading contributions to the RGE.

Many of the entries in the anomalous dimension matrix vanish~\cite{Cheung:2015aba} including beyond one-loop~\cite{Bern:2020ikv}.
One way to understand these zeroes in the anomalous dimension matrix is through non-renormalization theorem of Ref.~\cite{Cheung:2015aba}, which applies to loop amplitudes with one dimension-$d$ SMEFT insertion and one SM tree level amplitude.
To start define holomorphic and anti-holomorphic weights as $w = n - h$ and $\overline w = n + h$, respectively, where $n$ is the number of particles created by an operator and $h$ is the sum of the helicities of the particles created.
Then the theorem can be stated as operators belonging to subclass $j$ can renormalize operators of another subclass $i$ if $w_i \geq w_j$ and $\overline w_i \geq \overline w_j$.
The theorem can only be violated when the SM tree amplitude contains two Yukawa interactions.
The results of the non-renormalization theorem applied to dimension-8 operators are visualized in Table~\ref{tab:rge}.
Visually the theorem transitions down or to the left in the weight lattice.
Furthermore, if the SMEFT is UV-completed by a weak-coupled and renormalizable theory, then only the operator classes in red can be generated via tree level matching, whereas the black and blue operator classes are only generated at loop level in this scenario.
However, the operator classes in blue can be renormalized by the red tree level operator classes even though the blue operator classes are not generated by these types of UV theories at tree level~\cite{Craig:2019wmo}.

A contact amplitude resulting from an insertion of dimension-$d$ SMEFT operator (with $d$ even) obeys the relation
\begin{equation}
\label{eq:smeft_n}
2 d \geq w_k + \overline w_k \geq d .
\end{equation}
Furthermore, restricting to particles of spin-1 or less we also have
\begin{equation}
\label{eq:smeft_h}
d \geq |w_k - \overline w_k| .
\end{equation}
The relations~\eqref{eq:smeft_n} and~\eqref{eq:smeft_h} do not, in general, respect the weight bounds of SM tree level amplitudes, $w_k \geq 4$ and $\overline w_k \geq 4$ (which apply to non-exceptional amplitudes).
Therefore the theorem needs to be modified to determine vanishing entries among the two dimension-6 contributions to the dimension-8 anomalous dimension matrix as the theorem, as currently formulated, relies on the SM tree level weight bounds.
It would be interesting to examine in detail the structure of the anomalous dimension matrix resulting from two insertion of dimension-6 operators.

Inverting the logic of the non-renormalization theorem, operators of the same type will generically mix under RGE evolution.
An interesting restriction on how general the mixing can be comes from Lagrangian terms with multiple flavor structures.
Examples of such terms include $Q_{l^4H^2}$, $Q_{q^4H^2}^{(1,3)}$, $Q_{u^4H^2}$ and $Q_{d^4H^2}$.
The different flavor structures in these terms can only be mixed by Yukawa contributions to the RGE, \textit{i.e.} gauge and $\lambda$ contributions do not mix different flavor structures of a given term.
See Ref.~\cite{Alonso:2014zka} for further discussion on this point.
On the other hand, at the dimension-6, there are plenty of example of mixing amongst different terms of the same type.
For example, the gauge contribution to the dimension-6 RGE mixes the bosonic operators $C_{H\Box}$ and $C_{HD}$~\cite{Alonso:2013hga}.
Along these lines we expect the mixing of $Q_{q^2H^4D}^{(2)}$ to be closely related to that of $Q_{q^2H^4D}^{(3)}$ because, as shown in Sec.~\ref{sec:cls_ops}, these terms contain orthogonal linear combinations of operators.
A similar relation is expected for $Q_{l^2H^4D}^{(2, 3)}$.

%--------------------------------------------------------------------------------------------------------------
\begin{table}[H]
\begin{center}
\renewcommand{\arraystretch}{1.2}
\begin{tabular}[t]{c c *5{| p{2.35cm}} }
%-----------------
\multirow{5}{*}{$\overline w$} & 8 &
\cellcolor[gray]{0.94}$X_L^4$ & 
\cellcolor[gray]{0.88}$X_L^3 H^2$, ${\color{blue} X_L^2 \psi^2 H}$, ${\color{red} X_L \psi^4}$        & 
\cellcolor[gray]{0.82}${\color{red} X_L^2 H^4}$, ${\color{red} X_L \psi^2 H^3}$, ${\color{red}  \psi^4 H^2}$       &    
\cellcolor[gray]{0.76}${\color{red} \psi^2 H^5}$      & 
\cellcolor[gray]{0.70}${\color{red} H^8}$            \\ \cline{2-7}
%-----------------
& 6 &
           & 
 \cellcolor[gray]{0.94}$X_L^2 H^2 D^2$, $X_L^2 \psi \bar \psi D$, ${\color{blue} X_L \psi^2 H D^2}$, ${\color{red} \psi^4 D^2}$ & 
 \cellcolor[gray]{0.88}${\color{red}  X_L H^4 D^2}$, ${\color{blue} X_L^2 \bar \psi^2 H}$, ${\color{red}  X_L \psi \bar \psi H^2 D}$, ${\color{red}  \psi^2 H^3 D^2}$, ${\color{red}  X_L \psi^2 \bar \psi^2}$, ${\color{red}  \psi^3 \bar \psi H D}$   & 
 \cellcolor[gray]{0.82}${\color{red} H^6 D^2}$, ${\color{red} \psi \bar \psi H^4 D}$, ${\color{red} \psi^2 \bar \psi^2 H^2}$             &  
 \cellcolor[gray]{0.76}$ {\color{red} \bar \psi^2 H^5}$                  \\ \cline{2-7}
%-----------------
& 4 &           
           &                           
           & 
\cellcolor[gray]{0.94}$X_L^2 X_R^2$, ${\color{blue} X_L X_R H^2 D^2}$, ${\color{red} H^4 D^4}$, ${\color{blue} X_L X_R \psi \bar \psi D}$, ${\color{blue} X_R \psi^2 H D^2}$, ${\color{blue} {\color{blue} X_L \bar \psi^2 H D^2}}$, ${\color{red} \psi \bar \psi H^2 D^3}$, ${\color{red} \psi^2 \bar \psi^2 D^2}$ & 
\cellcolor[gray]{0.88}${\color{red}  X_R H^4 D^2}$, ${\color{blue} X_R^2 \psi^2 H}$, ${\color{red}  X_R \psi \bar \psi H^2 D}$, ${\color{red} \bar  \psi^2 H^3 D^2}$, ${\color{red}  X_R \psi^2 \bar \psi^2}$, ${\color{red}  \psi \bar \psi^3 H D}$     & 
\cellcolor[gray]{0.82}${\color{red} X_R^2 H^4}$, ${\color{red} X_R \bar  \psi^2 H^3}$, ${\color{red} \bar \psi^4 H^2}$ \\ \cline{2-7}
%-----------------
& 2 &          
           &                           
           &                         
           & 
\cellcolor[gray]{0.94}$X_R^2 H^2 D^2$, $X_R^2 \psi \bar \psi D$, ${\color{blue} X_R \bar \psi^2 H D^2}$, $ {\color{red} \bar \psi^4 D^2}$ & 
\cellcolor[gray]{0.88}$X_R^3 H^2$, ${\color{blue} X_R^2 \bar \psi^2 H}$, ${\color{red} X_R \bar \psi^4}$ \\ \cline{2-7}
%-----------------     
& 0 &      
           &                           
           &                         
           &                           
           & 
\cellcolor[gray]{0.94}$X_R^4$ \\ \cline{2-7}
%-----------------  
& & 0 & 2 & 4 & 6 & 8 \\
\multicolumn{2}{c}{} & \multicolumn{5}{c}{$w$}
\end{tabular}
\end{center}
\caption{Weight lattice for dimension-8 operators in the SMEFT. 
By the non-renormalization theorem of Ref.~\cite{Cheung:2015aba} a subclass of operators, $j$, can renormalize another subclass, $i$, at one-loop if $w_i \geq w_j$ and $\overline w_i \geq \overline w_j$. 
Visually this prevents transitions down or to the left.
Additionally, if the SMEFT is UV-completed by a weak-coupled and renormalizable theory, then only the operator classes in red can be generated via tree level matching, whereas the black and blue operator classes are only generated at loop level in this scenario.
However, the operator classes in blue can be renormalized by the red tree level operator classes even though the blue operator classes are not generated by these types of UV theories at tree level~\cite{Craig:2019wmo}.}
\label{tab:rge}
\end{table}

%%----------------------------------------------------------------------------------------------------------------------------------------------------------------

\section{Conclusions}
\label{sec:con}

In this work we presented a complete basis of dimension-8 operators in the Standard Model Effective Field Theory.
Multiple checks have been passed including that no operator in the basis can be removed completely using the equations of motion.
As a sample of what can be done with the a complete basis of dimension-8 operators, we briefly considered the phenomenology of light-by-light scattering, electroweak precision data, and commented on the structure of the dimension-8 RGE.
Additionally, we matched theories of $SU(2)_w$ quartets onto the SMEFT up to dimension-8 allowing to us showcase the interplay between dimension-6 effects and dimension-8 effects, the latter of which cannot be neglected in those models.

%%----------------------------------------------------------------------------------------------------------------------------------------------------------------

\section*{Acknowledgements}
\addcontentsline{toc}{section}{Acknowledgements}

We thank Renato Fonseca, Adam Martin, Grant Remmen, Nicholas Rodd, Ver\'onica Sanz, Michael Trott for useful discussions.
The author is grateful for the privilege of being able to stay home and work on a project like this.
Ref.~\cite{Li:2020gnx} appeared on the arXiv simultaneously with this work.
It also presents a complete basis of dimension-8 SMEFT operators.
We are especially grateful to Jing Shu and Jiang-Hao Yu for discussions of Ref.~\cite{Li:2020gnx}.

Additionally, there were a few issues in original version of the paper.
(v2 on the arXiv is the JHEP version.)
We would also like to thank:
\begin{itemize}
\item [v3] Marco Ardu for finding a mistake in $Q_{leqdH^2}^{(4)}$
\item [v4] Mikael Chala for finding a couple of mistakes in the matching in Section~\ref{sec:quartets}
\item [v5] Serge Hamoudou, Jacky Kumar, and David London for demonstrating the flavor indices were incorrect on the right-hand side of the bottommost equation in~\eqref{eq:l5} and that the operator $Q_{q^4H^2}^{(5)}$ was originally missing from the basis
\item [v6] Serge Hamoudou for helping identify that the combination $\epsilon \tau^I$ ($\tau^I \epsilon$) should go with a pair of $SU(2)_w$ (anti-)fermions.
\end{itemize}

%%----------------------------------------------------------------------------------------------------------------------------------------------------------------

\begin{appendix}

\section{Dimension-6 and -7 Operators}
\label{sec:app}

For the sake of convenience we reproduce the tables of dimension-6 and -7 operators here.
Table~\ref{tab:smeft7ops} contains the dimension-7 operators, and is adapted from Ref.~\cite{Liao:2016hru}.
The classification scheme we use for the dimension-7 comes from labelling the classes in Table~2 of Ref.~\cite{Henning:2015alf} in descending order.
Table~\ref{tab:smeft6ops} contains the dimension-6 operators, and is adapted from Ref.~\cite{Alonso:2013hga}.
In contrast with~\cite{Alonso:2013hga} we also include in Table~\ref{tab:smeft6ops} the baryon number violating operators as listed in Ref.~\cite{Alonso:2014zka}.

%%%%%%%%%%%%%%%%%%%%%%%%%%%%%%%%%%%%%%%%%%%%%%%%%%%%%%%%%%%%%%%%%%%%%%%
% SMEFT d=7 Operators
%%%%%%%%%%%%%%%%%%%%%%%%%%%%%%%%%%%%%%%%%%%%%%%%%%%%%%%%%%%%%%%%%%%%%%% 

\begin{table}[H]
\hspace{-0.5cm}
\begin{center}
\begin{adjustbox}{width=0.83\textwidth,center}
\small
\begin{minipage}[t]{6.3cm}
\renewcommand{\arraystretch}{1.5}
\begin{tabular}[t]{c|c}
\multicolumn{2}{c}{\boldmath$1:\psi^2XH^2+\hc$} \\
\hline
$Q_{l^2WH^2}$  &  $\epsilon_{mn} (\epsilon \tau^I)_{jk} (l_p^m C \sigma^{\mu\nu} l_r^j) H^n H^k W_{\mu\nu}^I$  \\ \hdashline
$Q_{l^2BH^2}$   &  $\epsilon_{mn} \epsilon_{jk} (l_p^m C \sigma^{\mu\nu} l_r^j) H^n H^k B_{\mu\nu}$ 
\end{tabular}
\end{minipage}
\hspace{1cm}
\begin{minipage}[t]{5.1cm}
\renewcommand{\arraystretch}{1.5}
\begin{tabular}[t]{c|c}
\multicolumn{2}{c}{\boldmath$2:\psi^2H^4+\hc$} \\
\hline
$Q_{l^2H^4}$  &  $\epsilon_{mn} \epsilon_{jk} (l_p^m C l_r^j)H^n H^k (H^\dag H)$ 
\end{tabular}
\end{minipage}
\end{adjustbox}
\begin{adjustbox}{width=0.73\textwidth,center}
\small
\begin{minipage}[t]{4.8cm}
\renewcommand{\arraystretch}{1.5}
\begin{tabular}[t]{c|c}
\multicolumn{2}{c}{\boldmath$3(B):\psi^4H+\hc$} \\
\hline
$Q_{l^3eH}$  &  $\epsilon_{jk} \epsilon_{mn} (\bar{e}_p l_r^j) (l_s^k C l_t^m) H^n $ \\
$Q_{leudH}$  &  $\epsilon_{jk} (\bar{d}_p l_r^j) (u_s C e_t) H^k $ \\
$Q_{l^2qdH}^{(1)}$  &  $\epsilon_{jk} \epsilon_{mn} (\bar{d}_p l_r^j) (q_s^k C l_t^m) H^n $ \\
$Q_{l^2qdH}^{(2)}$  &  $\epsilon_{jm} \epsilon_{kn} (\bar{d}_p l_r^j) (q_s^k C l_t^m) H^n $ \\
$Q_{l^2quH}$  &  $\epsilon_{jk} (\bar{q}_p^m u_r) (l_{sm} C l_t^j) H^k $ 
\end{tabular}
\end{minipage}
\hspace{1cm}
\begin{minipage}[t]{5.1cm}
\renewcommand{\arraystretch}{1.5}
\begin{tabular}[t]{c|c}
\multicolumn{2}{c}{\boldmath$3(\slashed{B}):\psi^4H+\hc$} \\
\hline
$Q_{lud^2H}$  &  $\epsilon_{\alpha\beta\gamma} (\bar{l}_p d_r^\alpha) (u_s^\beta C d_t^\gamma) \widetilde{H} $ \\
$Q_{lq^2dH}$  &  $\epsilon_{\alpha\beta\gamma} (\bar{l}_p^m d_r^\alpha) (q_{sm}^\beta C q_t^{j\gamma}) H^\dag_j $ \\ \hdashline
$Q_{ld^3H}$  &  $\epsilon_{\alpha\beta\gamma} (\bar{l}_p d_r^\alpha) (d_s^\beta C d_t^\gamma) H$ \\
$Q_{eqd^2H}$  &  $\epsilon_{\alpha\beta\gamma} (\bar{e}_p q_r^{j\alpha}) (d_s^\beta C d_t^\gamma) H^\dag_j$
\end{tabular}
\end{minipage}
\end{adjustbox}
\begin{adjustbox}{width=0.76\textwidth,center}
\small
\begin{minipage}[t]{5.8cm}
\renewcommand{\arraystretch}{1.5}
\begin{tabular}[t]{c|c}
\multicolumn{2}{c}{\boldmath$4:\psi^2H^3D+\hc$} \\
\hline
$Q_{leH^3D}$     & $\epsilon_{mn} \epsilon_{jk} (l_p^m C \gamma^\mu e_r) H^n H^j i D_{\mu} H^k$ 
\end{tabular}
\end{minipage}
\hspace{1cm}
\begin{minipage}[t]{4.6cm}
\renewcommand{\arraystretch}{1.5}
\begin{tabular}[t]{c|c}
\multicolumn{2}{c}{\boldmath$5(B):\psi^4D+\hc$} \\
\hline
$Q_{l^2udD}$    & $\epsilon_{jk} (\bar{d}_p \gamma^\mu u_r) (l_s^j C i D_\mu l_t^k)$ 
\end{tabular}
\end{minipage}
\end{adjustbox}
\begin{adjustbox}{width=0.77\textwidth,center}
\small
\begin{minipage}[t]{5.6cm}
\renewcommand{\arraystretch}{1.5}
\begin{tabular}[t]{c|c}
\multicolumn{2}{c}{\boldmath$6:\psi^2H^2D^2+\hc$} \\
\hline
$Q_{l^2H^2D^2}^{(1)}$ & $\epsilon_{jk} \epsilon_{mn} (l_p^j C D^{\mu} l_r^k) H^m (D_{\mu} H^n)$ \\
$Q_{l^2H^2D^2}^{(2)}$ & $\epsilon_{jm} \epsilon_{kn} (l_p^j C D^{\mu} l_r^k) H^m (D_{\mu} H^n)$ 
\end{tabular}
\end{minipage}
\hspace{1cm}
\begin{minipage}[t]{5.0cm}
\renewcommand{\arraystretch}{1.5}
\begin{tabular}[t]{c|c}
\multicolumn{2}{c}{\boldmath$5(\slashed{B}):\psi^4D+\hc$} \\
\hline
$Q_{lqd^2D}$   & $\epsilon_{\alpha\beta\gamma} (\bar{l}_p \gamma^\mu q_r^\alpha) (d_s^\beta C i D_\mu d_t^\gamma)$ \\
$Q_{ed^3D}$  & $\epsilon_{\alpha\beta\gamma} (\bar{e}_p \gamma^\mu d_r^\alpha) (d_s^\beta C i D_\mu d_t^\gamma)$ 
\end{tabular}
\end{minipage}
\end{adjustbox}
\end{center}
\caption{The dimension-seven operators in the SMEFT. The operators are divided into six classes according to their field content. The classes-3 and -5 are further divided into subclasses according to their baryon number. All of the operators have Hermitian conjugates. The subscripts $p, r, s, t$ are weak-eigenstate indices. Operators below the dashed lines in classes-1 and -3 vanish when there is only one generation of fermions.}
\label{tab:smeft7ops}
\end{table}

%%%%%%%%%%%%%%%%%%%%%%%%%%%%%%%%%%%%%%%%%%%%%%%%%%%%%%%%%%%%%%%%%%%%%%%
% SMEFT d=6 Operators
%%%%%%%%%%%%%%%%%%%%%%%%%%%%%%%%%%%%%%%%%%%%%%%%%%%%%%%%%%%%%%%%%%%%%%% 
\begin{table}[H]
\hspace{-0.5cm}
\begin{center}
\begin{adjustbox}{width=1\textwidth,center}
\small
\begin{minipage}[t]{4.45cm}
\renewcommand{\arraystretch}{1.5}
\begin{tabular}[t]{c|c}
\multicolumn{2}{c}{\boldmath$1:X^3$} \\
\hline
$Q_G$                & $f^{ABC} G_\mu^{A\nu} G_\nu^{B\rho} G_\rho^{C\mu} $ \\
$Q_{\widetilde G}$          & $f^{ABC} \widetilde G_\mu^{A\nu} G_\nu^{B\rho} G_\rho^{C\mu} $ \\
$Q_W$                & $\epsilon^{IJK} W_\mu^{I\nu} W_\nu^{J\rho} W_\rho^{K\mu}$ \\ 
$Q_{\widetilde W}$          & $\epsilon^{IJK} \widetilde W_\mu^{I\nu} W_\nu^{J\rho} W_\rho^{K\mu}$ \\
\end{tabular}
\end{minipage}
\begin{minipage}[t]{2.7cm}
\renewcommand{\arraystretch}{1.5}
\begin{tabular}[t]{c|c}
\multicolumn{2}{c}{\boldmath$2:H^6$} \\
\hline
$Q_H$       & $(H^\dag H)^3$ 
\end{tabular}
\end{minipage}
\begin{minipage}[t]{5.1cm}
\renewcommand{\arraystretch}{1.5}
\begin{tabular}[t]{c|c}
\multicolumn{2}{c}{\boldmath$3:H^4 D^2$} \\
\hline
$Q_{H\Box}$ & $(H^\dag H)\Box(H^\dag H)$ \\
$Q_{H D}$   & $\ \left(H^\dag D^\mu H\right)^* \left(H^\dag D_\mu H\right)$ 
\end{tabular}
\end{minipage}
\begin{minipage}[t]{2.7cm}
\renewcommand{\arraystretch}{1.5}
\begin{tabular}[t]{c|c}
\multicolumn{2}{c}{\boldmath$5: \psi^2H^3 + \hc$} \\
\hline
$Q_{eH}$           & $(H^\dag H)(\bar l_p e_r H)$ \\
$Q_{uH}$          & $(H^\dag H)(\bar q_p u_r \widetilde H )$ \\
$Q_{dH}$           & $(H^\dag H)(\bar q_p d_r H)$\\
\end{tabular}
\end{minipage}

\vspace{0.25cm}

\end{adjustbox}

\begin{adjustbox}{width=0.9\textwidth,center}

\begin{minipage}[t]{4.7cm}
\renewcommand{\arraystretch}{1.5}
\begin{tabular}[t]{c|c}
\multicolumn{2}{c}{\boldmath$4:X^2H^2$} \\
\hline
$Q_{H G}$     & $H^\dag H\, G^A_{\mu\nu} G^{A\mu\nu}$ \\
$Q_{H\widetilde G}$         & $H^\dag H\, \widetilde G^A_{\mu\nu} G^{A\mu\nu}$ \\
$Q_{H W}$     & $H^\dag H\, W^I_{\mu\nu} W^{I\mu\nu}$ \\
$Q_{H\widetilde W}$         & $H^\dag H\, \widetilde W^I_{\mu\nu} W^{I\mu\nu}$ \\
$Q_{H B}$     & $ H^\dag H\, B_{\mu\nu} B^{\mu\nu}$ \\
$Q_{H\widetilde B}$         & $H^\dag H\, \widetilde B_{\mu\nu} B^{\mu\nu}$ \\
$Q_{H WB}$     & $ H^\dag \tau^I H\, W^I_{\mu\nu} B^{\mu\nu}$ \\
$Q_{H\widetilde W B}$         & $H^\dag \tau^I H\, \widetilde W^I_{\mu\nu} B^{\mu\nu}$ 
\end{tabular}
\end{minipage}
\begin{minipage}[t]{5.2cm}
\renewcommand{\arraystretch}{1.5}
\begin{tabular}[t]{c|c}
\multicolumn{2}{c}{\boldmath$6:\psi^2 XH+\hc$} \\
\hline
$Q_{eW}$      & $(\bar l_p \sigma^{\mu\nu} e_r) \tau^I H W_{\mu\nu}^I$ \\
$Q_{eB}$        & $(\bar l_p \sigma^{\mu\nu} e_r) H B_{\mu\nu}$ \\
$Q_{uG}$        & $(\bar q_p \sigma^{\mu\nu} T^A u_r) \widetilde H \, G_{\mu\nu}^A$ \\
$Q_{uW}$        & $(\bar q_p \sigma^{\mu\nu} u_r) \tau^I \widetilde H \, W_{\mu\nu}^I$ \\
$Q_{uB}$        & $(\bar q_p \sigma^{\mu\nu} u_r) \widetilde H \, B_{\mu\nu}$ \\
$Q_{dG}$        & $(\bar q_p \sigma^{\mu\nu} T^A d_r) H\, G_{\mu\nu}^A$ \\
$Q_{dW}$         & $(\bar q_p \sigma^{\mu\nu} d_r) \tau^I H\, W_{\mu\nu}^I$ \\
$Q_{dB}$        & $(\bar q_p \sigma^{\mu\nu} d_r) H\, B_{\mu\nu}$ 
\end{tabular}
\end{minipage}
\begin{minipage}[t]{5.4cm}
\renewcommand{\arraystretch}{1.5}
\begin{tabular}[t]{c|c}
\multicolumn{2}{c}{\boldmath$7:\psi^2H^2 D$} \\
\hline
$Q_{H l}^{(1)}$      & $(H^\dag i\overleftrightarrow{D}_\mu H)(\bar l_p \gamma^\mu l_r)$\\
$Q_{H l}^{(3)}$      & $(H^\dag i\overleftrightarrow{D}^I_\mu H)(\bar l_p \tau^I \gamma^\mu l_r)$\\
$Q_{H e}$            & $(H^\dag i\overleftrightarrow{D}_\mu H)(\bar e_p \gamma^\mu e_r)$\\
$Q_{H q}^{(1)}$      & $(H^\dag i\overleftrightarrow{D}_\mu H)(\bar q_p \gamma^\mu q_r)$\\
$Q_{H q}^{(3)}$      & $(H^\dag i\overleftrightarrow{D}^I_\mu H)(\bar q_p \tau^I \gamma^\mu q_r)$\\
$Q_{H u}$            & $(H^\dag i\overleftrightarrow{D}_\mu H)(\bar u_p \gamma^\mu u_r)$\\
$Q_{H d}$            & $(H^\dag i\overleftrightarrow{D}_\mu H)(\bar d_p \gamma^\mu d_r)$\\
$Q_{H u d}$ + h.c.   & $i(\widetilde H ^\dag D_\mu H)(\bar u_p \gamma^\mu d_r)$\\
\end{tabular}
\end{minipage}
\end{adjustbox}
\end{center}

\begin{center}
\begin{adjustbox}{width=0.9\textwidth,center}
\begin{minipage}[t]{4.75cm}
\renewcommand{\arraystretch}{1.5}
\begin{tabular}[t]{c|c}
\multicolumn{2}{c}{\boldmath$8:(\bar L L)(\bar L L)$} \\
\hline
$Q_{ll}$        & $(\bar l_p \gamma^\mu l_r)(\bar l_s \gamma_\mu l_t)$ \\
$Q_{qq}^{(1)}$  & $(\bar q_p \gamma^\mu q_r)(\bar q_s \gamma_\mu q_t)$ \\
$Q_{qq}^{(3)}$  & $(\bar q_p \gamma^\mu \tau^I q_r)(\bar q_s \gamma_\mu \tau^I q_t)$ \\
$Q_{lq}^{(1)}$                & $(\bar l_p \gamma^\mu l_r)(\bar q_s \gamma_\mu q_t)$ \\
$Q_{lq}^{(3)}$                & $(\bar l_p \gamma^\mu \tau^I l_r)(\bar q_s \gamma_\mu \tau^I q_t)$ 
\end{tabular}
\end{minipage}
\hspace{0.2cm}
\begin{minipage}[t]{5.25cm}
\renewcommand{\arraystretch}{1.5}
\begin{tabular}[t]{c|c}
\multicolumn{2}{c}{\boldmath$8:(\bar R R)(\bar R R)$} \\
\hline
$Q_{ee}$               & $(\bar e_p \gamma^\mu e_r)(\bar e_s \gamma_\mu e_t)$ \\
$Q_{uu}$        & $(\bar u_p \gamma^\mu u_r)(\bar u_s \gamma_\mu u_t)$ \\
$Q_{dd}$        & $(\bar d_p \gamma^\mu d_r)(\bar d_s \gamma_\mu d_t)$ \\
$Q_{eu}$                      & $(\bar e_p \gamma^\mu e_r)(\bar u_s \gamma_\mu u_t)$ \\
$Q_{ed}$                      & $(\bar e_p \gamma^\mu e_r)(\bar d_s\gamma_\mu d_t)$ \\
$Q_{ud}^{(1)}$                & $(\bar u_p \gamma^\mu u_r)(\bar d_s \gamma_\mu d_t)$ \\
$Q_{ud}^{(8)}$                & $(\bar u_p \gamma^\mu T^A u_r)(\bar d_s \gamma_\mu T^A d_t)$ \\
\end{tabular}
\end{minipage}
\hspace{0.2cm}
\begin{minipage}[t]{4.75cm}
\renewcommand{\arraystretch}{1.5}
\begin{tabular}[t]{c|c}
\multicolumn{2}{c}{\boldmath$8:(\bar L L)(\bar R R)$} \\
\hline
$Q_{le}$               & $(\bar l_p \gamma^\mu l_r)(\bar e_s \gamma_\mu e_t)$ \\
$Q_{lu}$               & $(\bar l_p \gamma^\mu l_r)(\bar u_s \gamma_\mu u_t)$ \\
$Q_{ld}$               & $(\bar l_p \gamma^\mu l_r)(\bar d_s \gamma_\mu d_t)$ \\
$Q_{qe}$               & $(\bar q_p \gamma^\mu q_r)(\bar e_s \gamma_\mu e_t)$ \\
$Q_{qu}^{(1)}$         & $(\bar q_p \gamma^\mu q_r)(\bar u_s \gamma_\mu u_t)$ \\ 
$Q_{qu}^{(8)}$         & $(\bar q_p \gamma^\mu T^A q_r)(\bar u_s \gamma_\mu T^A u_t)$ \\ 
$Q_{qd}^{(1)}$ & $(\bar q_p \gamma^\mu q_r)(\bar d_s \gamma_\mu d_t)$ \\
$Q_{qd}^{(8)}$ & $(\bar q_p \gamma^\mu T^A q_r)(\bar d_s \gamma_\mu T^A d_t)$\\
\end{tabular}
\end{minipage}

\end{adjustbox}

\vspace{0.25cm}

\begin{adjustbox}{width=0.9\textwidth,center}

\begin{minipage}[t]{3.75cm}
\renewcommand{\arraystretch}{1.5}
\begin{tabular}[t]{c|c}
\multicolumn{2}{c}{\boldmath$8:(\bar LR)(\bar RL)+\hc$} \\
\hline
$Q_{ledq}$ & $(\bar l_p^j e_r)(\bar d_s q_{tj})$ 
\end{tabular}
\end{minipage}
\hspace{0.4cm}
\begin{minipage}[t]{5.5cm}
\renewcommand{\arraystretch}{1.5}
\begin{tabular}[t]{c|c}
\multicolumn{2}{c}{\boldmath$8:(\bar LR)(\bar L R)+\hc$} \\
\hline
$Q_{quqd}^{(1)}$ & $(\bar q_p^j u_r) \epsilon_{jk} (\bar q_s^k d_t)$ \\
$Q_{quqd}^{(8)}$ & $(\bar q_p^j T^A u_r) \epsilon_{jk} (\bar q_s^k T^A d_t)$ \\
$Q_{lequ}^{(1)}$ & $(\bar l_p^j e_r) \epsilon_{jk} (\bar q_s^k u_t)$ \\
$Q_{lequ}^{(3)}$ & $(\bar l_p^j \sigma_{\mu\nu} e_r) \epsilon_{jk} (\bar q_s^k \sigma^{\mu\nu} u_t)$
\end{tabular}
\end{minipage}
\begin{minipage}[t]{5.6cm}
\renewcommand{\arraystretch}{1.5}
\begin{tabular}[t]{c|c}
\multicolumn{2}{c}{\boldmath$8:(\slashed{B})+\hc$} \\
\hline
$Q_{duql}$ & $\epsilon_{\alpha\beta\gamma} \epsilon_{jk} (d_p^{\alpha} C u_r^{\beta}) (q_s^{j\gamma} C l_t^k)$ \\
$Q_{qque}$ & $\epsilon_{\alpha\beta\gamma} \epsilon_{jk} (q_p^{j\alpha} C q_r^{k\beta}) (u_s^{\gamma} C e_t)$ \\
$Q_{qqql}$ & $\epsilon_{\alpha\beta\gamma} \epsilon_{mn} \epsilon_{jk} (q_p^{m\alpha} C q_r^{j\beta}) (q_s^{k\gamma} C l_t^n)$ \\
$Q_{duue}$ & $\epsilon_{\alpha\beta\gamma} (d_p^{\alpha} C u_r^{\beta}) (u_s^{\gamma} C e_t)$
\end{tabular}
\end{minipage}
\end{adjustbox}
\end{center}
\caption{The dimension-six operators in the SMEFT. The operators are divided into eight classes according to their field content. The class-8 $\psi^4$ four-fermion operators are further divided into subclasses according to their chiral and baryonic properties. Operators with ${}+\hc$ have Hermitian conjugates, as does the $\psi^2 H^2 D$ operator $Q_{Hud}$. The subscripts $p, r, s, t$ are weak-eigenstate indices.}
\label{tab:smeft6ops}
\end{table}

\end{appendix}

\clearpage

%%----------------------------------------------------------------------------------------------------------------------------------------------------------------

\bibliographystyle{JHEP}
\bibliography{dim8}

%%----------------------------------------------------------------------------------------------------------------------------------------------------------------

\end{document}